\documentclass[a4paper,11pt]{article}
\pdfoutput=1 
\usepackage{tikz}
\usetikzlibrary{svg.path}
\usepackage{scalerel}
\usepackage{jheppub} 

\reversemarginpar
\usepackage{graphicx}
\usepackage{amsmath}
\usepackage{booktabs,tabularx,array,colortbl}
\allowdisplaybreaks        
\usepackage{hyperref}
\usepackage{pdfpages}
\usepackage[section]{placeins}
\usepackage{newunicodechar}
\newunicodechar{−}{-}
\usepackage{epstopdf}
\usepackage{xcolor}
\usepackage{empheq}
\usepackage{tensor}
\usepackage[colorinlistoftodos]{todonotes}
\usepackage{epsfig}
\usepackage{graphicx,color}
\usepackage{cancel}
\usepackage[utf8]{inputenc}
\usepackage{simpler-wick}
\usepackage{comment}
\usepackage{bm}
\usepackage{mathtools}
\usepackage{slashed}
\usepackage{bbm}
\usepackage{etoolbox}
\pretocmd{\thebibliography}{\clearpage}{}{}

\newcommand{\veuv}{\varepsilon_{\rm {UV}}}

\newcommand{\gT}{g_T}
\newcommand{\eT}{\epsilon_T}
\newcommand{\kT}{k_T}
\newcommand{\kt}{\tilde k_T}
\newcommand{\kTsq}{k_T^2}
\newcommand{\STT}{S_{TT}}
\newcommand{\SLT}{S_{LT}}
\newcommand{\Skk}{S_{TT}^{kk}}
\newcommand{\Skt}{S_{TT}^{k\tilde k}}
\newcommand{\SSq}{S_{TT}^{2}}

\newcommand{\be}{\begin{equation}}
\newcommand{\ee}{\end{equation}}
\newcommand{\bea}{\begin{eqnarray}}
\newcommand{\eea}{\end{eqnarray}}
\newcommand{\balign}{\begin{align}}
\newcommand{\ealign}{\end{align}}
\newcommand{\as}{\alpha_s}

\newcommand{\bg}{\begin{gather}}
\newcommand{\foma}{\end{gather}}

\newcommand{\noopsort}[1]{}

\def\ve{\varepsilon}

\def\<{\langle}
\def\>{\rangle}

\def\({\left(}
\def\[{\left[}
\def\){\right)}
\def\]{\right]}

\def\ln{\hbox{ln}}

\def\nbar{\bar{n}}

\newcommand{\ben}{\begin{eqnarray}}
\newcommand{\een}{\end{eqnarray}}

\newcommand{\bef}{\begin{figure}[htb]\centering}
\newcommand{\eef}{\end{figure}}

\newcommand{\kperp}{k_\perp}

\usepackage[normalem]{ulem} 
\DeclareRobustCommand{\old}[1]{\ifmmode{\color[rgb]{1,0,0}\cancel{#1}}\else{\color[rgb]{1,0,0}\sout{#1}}\fi}
\DeclareRobustCommand{\neww}[1]{\ifmmode{\color[rgb]{0,0,1}#1}\else{\color[rgb]{0,0,1}\uwave{#1}}\fi}

\newcommand{\Sone}{\ensuremath{{}^3S_1^{[8]}}}
\newcommand{\Szero}{\ensuremath{{}^1S_0^{[8]}}}
\newcommand{\PJ}{\ensuremath{{}^3P_J^{[8]}}}
\newcommand{\Pzero}{\ensuremath{{}^3P_0^{[8]}}}

\providecommand{\nbar}{\bar n}
\providecommand{\eT}{\epsilon_T}
\providecommand{\gT}{g_T}

\definecolor{orcidlogocol}{HTML}{A6CE39}
\tikzset{
  orcidlogo/.pic={
    \fill[orcidlogocol] svg{M256,128c0,70.7-57.3,128-128,128C57.3,256,0,198.7,0,128C0,57.3,57.3,0,128,0C198.7,0,256,57.3,256,128z};
    \fill[white] svg{M86.3,186.2H70.9V79.1h15.4v48.4V186.2z}
                 svg{M108.9,79.1h41.6c39.6,0,57,28.3,57,53.6c0,27.5-21.5,53.6-56.8,53.6h-41.8V79.1z M124.3,172.4h24.5c34.9,0,42.9-26.5,42.9-39.7c0-21.5-13.7-39.7-43.7-39.7h-23.7V172.4z}
                 svg{M88.7,56.8c0,5.5-4.5,10.1-10.1,10.1c-5.6,0-10.1-4.6-10.1-10.1c0-5.6,4.5-10.1,10.1-10.1C84.2,46.7,88.7,51.3,88.7,56.8z};
  }
}

\newcommand\orcidicon[1]{\href{https://orcid.org/#1}{\mbox{\scalerel*{
\begin{tikzpicture}[yscale=-1,transform shape]
\pic{orcidlogo};
\end{tikzpicture}}{|}}}}

\title{Polarized gluon TMD fragmentation into spin-1 quarkonium at NLO}
\author[a]{Alessia Bongallino\orcidicon{0009-0008-9832-2222},}
\author[a]{Miguel G. Echevarria\orcidicon{0000-0003-2888-8526}}
\author[a]{and Samuel F. Romera\orcidicon{0009-0007-0853-6212}}
\affiliation[a]{Department of Physics and EHU Quantum Center,\\ University of the Basque Country (EHU), 48940, Leioa, Spain}

\emailAdd{alessia.bongallino@ehu.eus}
\emailAdd{miguel.garciae@ehu.eus}
\emailAdd{samuel.fernandez@ehu.eus}

\date{\today}

\abstract{
We compute the complete set of gluon transverse-momentum-dependent fragmentation functions (TMDFFs) into quarkonium through next-to-leading order in perturbative QCD.
We account for all polarization configurations of both the fragmenting gluon and the produced $J^{\rm PC}=1^{--}$ quarkonium.
Within NRQCD, we include all color-octet channels that contribute through $\mathcal O(\alpha_s^2)$ at the considered order in the heavy-quark relative velocity $v$, namely ${}^3S_1^{[8]}$, ${}^1S_0^{[8]}$ and ${}^3P_J^{[8]}$.
Several structures of the calculation follow only from the spin-1 nature of the final-state hadron, so our framework applies more generally to the fragmentation of a gluon into any spin-1 hadron.
Our results agree with all known collinear fragmentation functions, up to differences in $\delta(1-z)$ terms that we discuss, and with the TMDFFs at LO, and we cross-check the unpolarized TMDFFs at NLO against previous calculations.
Most of the polarized TMDFFs, and their full transverse-momentum dependence, are presented here for the first time.
They provide the perturbative input for studies of quarkonium polarization in jets, at the EIC and in $e^+e^-$ annihilation.
}

\begin{document}
\maketitle



\section{Introduction}
\label{sec:intro}

Heavy quarkonium production is a privileged laboratory for the interplay between perturbative and non-perturbative QCD. The presence of the heavy-quark mass $m_Q$ provides a perturbative scale for the creation of the heavy quark--antiquark pair, while its subsequent hadronization is described, within non-relativistic QCD (NRQCD)~\cite{Bodwin:1994jh}, by an expansion in the relative velocity $v$ of the pair in terms of universal long-distance matrix elements (LDMEs). Despite decades of effort, a consistent description of the yields, transverse-momentum spectra and polarization of quarkonia across different processes remains elusive, and the relative weight of  color-singlet and  color-octet mechanisms is still debated~\cite{Brambilla:2010cs, Bodwin:2014gia, Lansberg:2019adr}.

Two qualitatively different kinematic regimes must be distinguished in quarkonium production.
When the quarkonium mass $M$ is of the order of the hard scale of the process $Q$, $M \sim Q$, the pair is produced directly in the hard process.
In the TMD regime, when its transverse momentum $q_T$ is small compared to its mass, $q_T\ll M\sim Q$, TMD factorization has to be supplemented by soft radiation from the heavy-quark pair, encoded in TMD shape functions and TMD soft functions~\cite{Echevarria:2019ynx,Fleming:2019pzj,Boer:2023zit,Echevarria:2024idp,Copeland:2025vop,Echevarria:2025shf}.
When instead a hard scale much larger than the quarkonium mass is present, $Q \gg M$, production through the fragmentation of a single parton becomes important and, in the region $q_T \sim M\ll Q$, the relevant non-perturbative objects are the TMD fragmentation functions (TMDFFs) of a light quark or a gluon into the quarkonium. 
These TMDFFs can in turn be matched onto NRQCD LDMEs with perturbatively calculable coefficients~\cite{Echevarria:2020qjk,Echevarria:2023dme,Copeland:2023wbu}.

The study of quarkonium TMDFFs is comparatively recent. The light-quark TMDFFs were first analyzed within NRQCD in  Ref.~\cite{Echevarria:2020qjk}. 
The full set of polarized TMDFFs for light quarks and gluons fragmenting into a $J/\psi$ was obtained at leading order (LO) in  Ref.~\cite{Copeland:2023wbu}, and used to show that, in semi-inclusive deep-inelastic scattering (SIDIS) at large $Q^2$, light-quark fragmentation can compete with, and in some regions dominate over, photon--gluon 
fusion~\cite{Copeland:2023qed}. The gluon TMDFF into quarkonium in the ${}^3S_1^{[8]}$ channel was computed at next-to-leading order (NLO) in  Ref.~\cite{Echevarria:2023dme}, where the cancellation of rapidity divergences in the properly defined TMDFF, including the soft factor, was shown explicitly, together with the consistency of its large-$q_T$ limit with the collinear fragmentation function (FF).
On the collinear side, the gluon FF into ${}^3S_1^{[8]}$ is known at LO~\cite{Braaten:1994vv} and at NLO~\mbox{\cite{Braaten:2000pc,Ma:2013yla}}, and NNLO corrections have recently become available~\cite{Feng:2026owq}.
The ${}^1S_0^{[8]}$ and ${}^3P_J^{[8]}$ channels start at $\mathcal O(\alpha_s^2)$; they can be obtained from the color-singlet $g\to\eta_Q$~\cite{Braaten:1993rw} and $g\to\chi_{QJ}$~\cite{Braaten:1994kd} calculations, are given in Refs.~\cite{Ma:2013yla,Ma:2014eja}, and are known at NLO~\cite{Feng:2018ulg,Zhang:2018mlo,Zhang:2020atv}.
Much less is known for polarized quarkonium.
The fragmentation functions of an unpolarized gluon into a transversely or longitudinally polarized quarkonium were obtained in Refs.~\mbox{\cite{Beneke:1995yb,Braaten:2000pc,Ma:2015yka}}, and recently at NNLO for the ${}^3S_1^{[8]}$ channel~\mbox{\cite{Feng:2026owq}}, while the polarized TMDFFs, including the polarization of the fragmenting gluon, are known only at LO~\mbox{\cite{Copeland:2023wbu}}.

The gluon TMDFF computed here enters a broad set of processes. 
One of the most direct experimental handles at present is the quarkonium production inside jets, where the momentum fraction $z$ carried by the quarkonium and its transverse momentum $j_\perp$ with respect to the jet axis are described by fragmenting jet functions whose matching coefficients involve the gluon FF and, at $j_\perp \sim M$, the matching onto gluon TMDFFs is expected~\cite{Baumgart:2014upa, Bain:2016rrv}.
Different color-octet channels lead to distinct $z$ and $j_\perp$ shapes~\cite{Baumgart:2014upa,Bain:2016rrv, Bain:2017wvk, Kang:2017yde}, and the polarization of the quarkonium inside the jet provides an additional handle.
Prompt $J/\psi$ in jets has been measured by LHCb~\cite{LHCb:2017llq} and CMS~\cite{CMS:2019ebt,CMS:2021puf}, and LHCb has recently extended these measurements to $\psi(2S)$ and $\chi_{c1}(3872)$, finding prompt distributions that differ significantly from NRQCD predictions supplemented by a parton shower~\cite{LHCb:2024ybz}; the extension to bottomonium is being pursued~\cite{Ha:2026mct}.
At the EIC, $J/\psi$ production in SIDIS at large $Q^2$ receives contributions from quark and gluon fragmentation~\cite{Copeland:2023qed}.
Since the gluon TMDFF enters first at $\mathcal{O}(\alpha_s)$ through the hard process, it is needed at NLO to separate these mechanisms, a prerequisite for extracting the color-octet LDMEs~\cite{Bacchetta:2018ivt, Copeland:2023wbu}.

Many of the structures that appear in the calculation are not specific to quarkonium, but follow from the spin-1 nature of the final-state hadron.
The general decomposition of quark and gluon TMD correlators for spin-1 hadrons, including the tensor-polarized functions that have no counterpart for spin-$\tfrac{1}{2}$ hadrons, was developed in  Refs.~\cite{Ji:1993vw,Bacchetta:2000jk,Mulders:2000sh,Boer:2016xqr}.

Our results therefore also bear on the spin alignment of light vector mesons, which is governed by the tensor-polarized fragmentation function $D_{1LL}$~\cite{Chen:2020pty}. It has been measured in $e^+e^-$ annihilation at LEP~\cite{DELPHI:1997ruo,OPAL:1997vmw} and in $pp$ collisions at RHIC~\cite{STAR:2008lcm} and the LHC~\cite{ALICE:2019aid}, and the gluon contribution enters at NLO in $e^+e^-$ annihilation and at LO in hadronic collisions. In that case the NRQCD LDMEs are replaced by genuinely non-perturbative inputs, but the tensor decomposition, the projectors, the evolution and the perturbative matching onto collinear functions carry over, whereas the NRQCD short-distance coefficients do not.

In this work, we present the complete set of gluon TMDFFs into $J^{\rm PC} = 1^{--}$ quarkonium -- and, more generally, into a spin-1 hadron -- through NLO in perturbative QCD, accounting for all polarization configurations of both the fragmenting gluon and the produced hadron.
Within the NRQCD framework, for quarkonium production, we include all relevant angular-momentum channels at the considered orders in $\alpha_s$ and the heavy-quark relative velocity $v$, namely the color-octet ${}^3S_1^{[8]}$, ${}^1S_0^{[8]}$, and ${}^3P_J^{[8]}$ states.

In order to validate our results we perform a number of independent checks against the literature: the reproduction of the LO collinear FF~\cite{Braaten:1994vv}, the comparison of the large-$q_T$ limit with the NLO collinear FF~\mbox{\cite{Braaten:2000pc,Ma:2013yla,Feng:2026owq}}, the cancellation of rapidity divergences and the agreement with the unpolarized NLO result of Ref.~\cite{Echevarria:2023dme} (up to a constant term in the virtual correction that we correct here), the matching onto LDMEs at $q_T \gg M$~\cite{Echevarria:2020qjk}, and the consistency of the spin decomposition with the LO polarized results of  Ref.~\cite{Copeland:2023wbu}.

The paper is organized as follows. In section~\ref{sec:theory} we introduce the definition of the gluon TMDFF for a spin-1 hadron and its decomposition into polarized structures. Section~\ref{sec:results} contains the NLO calculation, including the treatment of ultraviolet, infrared and rapidity divergences. We also discuss the matching onto NRQCD LDMEs, onto the collinear FF and the large-$q_T$ limit. We conclude in section~\ref{sec:conclusions}. Conventions are collected in App.~\mbox{\ref{sec:conventions}}, and the heavy-quark spinor structures and polarization projectors in App.~\mbox{\ref{app:polarisationandexpansion}}.



\section{Theory}
\label{sec:theory}

In this section we set up the operator definition of the gluon TMD fragmentation correlator for a spin-1 hadron, decompose it into eighteen leading-twist TMDFFs, give a compact construction of the projectors that isolate each of them, and discuss their factorization into short-distance coefficients and long-distance matrix elements within NRQCD.

\subsection{Gluon TMD fragmentation correlators}
\label{sec:tmd_correlators}

The gauge-invariant building block for the gluon is given by
\begin{equation}
  \mathcal{B}_{n\perp}^{\mu}
  = \frac{1}{g_s}\Big[W_n^\dagger\, iD_{n\perp}^{\mu}\, W_n\Big] ,
  \qquad
  W_n(y) = P\exp\!\Big[ig_s\int_{-\infty}^{0}\!ds\;\nbar\cdot A_n(y+\nbar s)\Big] ,
  \label{eq:Bfield}
\end{equation}
which at leading power is equivalent to the field-strength formulation with staple-like Wilson lines used in  Refs.~\cite{Mulders:2000sh,Boer:2016xqr,Copeland:2023wbu}.
The bare gluon TMD fragmentation correlator for a hadron in the spin state $S$ with momentum $P$ is then~\cite{Mulders:2000sh}
\begin{align}
  \Delta^{\lambda\sigma}_{g\to H}(z,b_T;S)
  &= \frac{P^+}{N_c^2-1}\sum_X\int\frac{d\xi^-}{2\pi}\,
     e^{-iP^+\xi^-/z}
  \nonumber\\
  &\quad\times
  \langle 0|\, T\big[\mathcal{B}^{\lambda,a}_{n\perp}\big]\!\Big(\frac{\xi}{2}\Big)
  |H(P,S),X\rangle\,
  \langle H(P,S),X|\,\bar T\big[\mathcal{B}^{\sigma,a}_{n\perp}\big]\!\Big(\!-\frac{\xi}{2}\Big)|0\rangle \, ,
  \label{eq:corr_b}
\end{align}
where $\xi^\mu=\xi^-\bar n^\mu+b_T^\mu$.
We work in the light-cone basis (defined in the Appendix~\ref{sec:conventions}) and in the hadron frame where the produced hadron $H$ (of mass $M$) carries no transverse momentum:
\begin{equation}
    \label{eq:hadron_frame}
    P^\mu = P^+ {n}^\mu + \frac{M^2}{2 P^+} \nbar^\mu \, , \qquad g_T^{\mu \nu} P_\nu = 0 \, , 
\end{equation}
and the fragmenting gluon carries longitudinal momentum $P^+/z$ and transverse momentum $k_T$, such that its transverse-momentum-space counterpart is defined through the Fourier transform,
\begin{equation}
    \label{eq:ft}
    \Delta^{\lambda\sigma}(z,\boldsymbol b_T)=\int d^{d-2}\boldsymbol k_\perp\,
    e^{i\boldsymbol k_\perp \cdot\boldsymbol b_T}\,\Delta^{\lambda\sigma}(z,\boldsymbol k_\perp) \, .
\end{equation}

Moreover, the renormalized correlator is
\begin{equation}
  \Delta^{\lambda\sigma}_{g\to H}(z,b_T;S;\mu,\zeta)
  = Z_g(\mu,\zeta)\,R_g(\mu,\zeta)\,
    \Delta^{\lambda\sigma}_{g\to H}(z,b_T;S) ,
  \label{eq:renorm}
\end{equation}
where $Z_g$ removes the ultraviolet divergences and $R_g$ the rapidity divergences.
With the $\delta$-regulator the zero-bin subtraction~\cite{Manohar:2006nz} coincides with the soft factor, and
\begin{equation}
    R_g = S^{-1/2}(b_T;\mu,\zeta) \, ,
\end{equation}
with $S(b_T;\mu,\zeta)$ the adjoint soft function~\cite{Echevarria:2015byo,Echevarria:2016scs}.
Both factors depend only on the operator in Eq.~\eqref{eq:corr_b}, and not on the projection onto a given gluon or hadron polarization.
Therefore, all the TMD fragmentation functions defined below share the same ultraviolet anomalous dimension and the rapidity anomalous dimension.

The spin state of the vector hadron is described by a $3\times3$ density matrix, parametrized by a vector polarization $(S_L,S_T^\mu)$ and a tensor polarization $(S_{LL},S_{LT}^\mu,S_{TT}^{\mu\nu})$~\cite{Bacchetta:2000jk}.
Its explicit form is given in Sec.~\ref{sec:nrqcd}. Moreover, at leading twist, the correlator is a $2\times2$ tensor in the transverse plane; it splits into a trace part, which describes unpolarized gluons, an antisymmetric part, which describes circularly polarized gluons (a helicity asymmetry), and a symmetric traceless part, which describes linearly polarized gluons (helicity flip $|\Delta h|=2$)~\cite{Mulders:2000sh,Boer:2016xqr,Copeland:2023wbu}.
We write
\begin{equation}
  \Delta^{\lambda\sigma}_{g\to H}(z,k_T;S)
  = \frac{1}{3}\sum_{X}\Gamma^{\lambda\sigma}_{X}(z,k_T) ,
  \qquad X\in\{U,L,T,LL,LT,TT\} ,
  \label{eq:decomp}
\end{equation}
where each $\Gamma_X$ is linear in the polarization parameter labeled by
$X$:
\begin{align}
  \Gamma^{\lambda\sigma}_{U}
  &= -g_T^{\lambda\sigma} D_1
     + \frac{k_T^{\lambda\sigma}}{M^2}\, H_1^\perp ,
  \label{eq:GU}\\
  \Gamma^{\lambda\sigma}_{L}
  &= S_L\Big[\, i\epsilon_T^{\lambda\sigma} G_{1L}
     + \frac{\epsilon_{T \alpha}^{\{ \lambda} k_T^{\sigma \} \alpha} }{2M^2}\, H_{1L}^\perp\Big] ,
  \label{eq:GL}\\
  \Gamma^{\lambda\sigma}_{T}
  &= -g_T^{\lambda\sigma}\frac{\epsilon_T^{S_T k_T}}{M} D_{1T}^\perp
     + i\epsilon_T^{\lambda\sigma}\frac{k_T\cdot S_T}{M} G_{1T}
     \nonumber\\
     & - \frac{\epsilon_T^{k_T\{\lambda}S_T^{\sigma\}}
            +\epsilon_T^{S_T\{\lambda}k_T^{\sigma\}}}{4M}\, H_{1T}
     - \frac{\epsilon_{T \alpha}^{\{ \lambda} k_T^{\sigma \} \alpha S_T } }{2M^3} \, H_{1T}^\perp ,
  \label{eq:GT}\\
  \Gamma^{\lambda\sigma}_{LL}
  &= S_{LL}\Big[-g_T^{\lambda\sigma} D_{1LL}
     + \frac{k_T^{\lambda\sigma}}{M^2}\, H_{1LL}^\perp\Big] ,
  \label{eq:GLL}\\
  \Gamma^{\lambda\sigma}_{LT}
  &= -g_T^{\lambda\sigma}\frac{k_T\cdot S_{LT}}{M} D_{1LT}
     + i\epsilon_T^{\lambda\sigma}\frac{\epsilon_T^{S_{LT} k_T}}{M} G_{1LT} \nonumber\\
     & + \frac{S_{LT}^{\{\lambda}k_T^{\sigma\}}}{M}\, H_{1LT}
     + \frac{k_T^{\lambda\sigma \alpha}\,S_{LT \alpha}}{M^3}\, H_{1LT}^\perp ,
  \label{eq:GLT}\\
  \Gamma^{\lambda\sigma}_{TT}
  &= -g_T^{\lambda\sigma}\frac{k_T^{\alpha \beta} S_{TT \alpha \beta}}{M^2} D_{1TT}
     + i\epsilon_T^{\lambda\sigma}\frac{\epsilon_{T\gamma}^\beta k_T^{\gamma \alpha} S_{TT \alpha \beta}}{M^2} G_{1TT} \nonumber\\
     &+ S_{TT}^{\lambda\sigma}\, H_{1TT} + \frac{S_{TT \alpha}^{\{\lambda} k_T^{\sigma \}\alpha} }{M^2} H_{1TT}^{\perp} 
     + \frac{k_T^{\lambda\sigma\rho\tau}S_{TT\,\rho\tau}}{M^4}\, H_{1TT}^{\perp\perp} \, .
  \label{eq:GTT}
\end{align}
All functions depend on $(z,k_T^2)$, and the tensors $k_T^{\lambda\sigma}$, $k_T^{\lambda\sigma\rho\tau}$, the invariants $\epsilon_T^{ab}$, $S_{TT}^{kk}$, $S_{TT}^{k\tilde k}$ and the symmetrization $a^{\{\lambda}b^{\sigma\}}$ are defined in the Appendix~\ref{sec:conventions}.
The parametrization follows the one of  Ref.~\cite{Boer:2016xqr} for gluons and spin-1 hadrons, adapted to fragmentation by  $\{n,x\}\to\{\nbar,1/z\}$, up to normalization conventions.
The eighteen functions are collected in Tab.~\ref{tab:TMDFFs}.
In four dimensions, the identity $S_{TT\,\alpha}^{\{\lambda}k_T^{\sigma\}\alpha}=g_T^{\lambda\sigma}\,k_T^{\alpha\beta}S_{TT\,\alpha\beta}$ makes the structures of $D_{1TT}$ and $H_{1TT}^{\perp}$ coincide, so that they enter only through the combination $\Phi\equiv H_{1TT}^{\perp}-D_{1TT}$, and Tab.~\mbox{\ref{tab:TMDFFs}} has nineteen entries but only eighteen independent functions.
In the following, we therefore give results only for $\Phi$.

Note that the factor $1/3$ in Eq.~\eqref{eq:decomp} is chosen such that, summing over the three polarization states of $H$ (for which the vector and tensor polarizations average to zero), one recovers
\begin{equation}
    \sum_{S}\Delta^{\lambda\sigma}(z,k_T;S)=\Gamma^{\lambda\sigma}_U(z,k_T) \, .
\end{equation}

\begin{table}[t]
\centering
\renewcommand{\arraystretch}{1.25}
\begin{tabular}{c|ccc}
\toprule
 & \multicolumn{3}{c}{Gluon polarization}\\
Hadron pol. & Unpolarized ($-g_T$) & Circular ($i\epsilon_T$) & Linear (sym.\ traceless)\\
\midrule
$U$  & $D_1$            &                & $H_1^\perp$ \\
$L$  &                  & $G_{1L}$       & $H_{1L}^\perp$ \\
$T$  & $D_{1T}^\perp$   & $G_{1T}$       & $H_{1T},\ H_{1T}^\perp$ \\
$LL$ & $D_{1LL}$        &                & $H_{1LL}^\perp$ \\
$LT$ & $D_{1LT}$        & $G_{1LT}$      & $H_{1LT},\ H_{1LT}^\perp$ \\
$TT$ & $D_{1TT}$        & $G_{1TT}$      & $H_{1TT},\ H_{1TT}^{\perp},\ H_{1TT}^{\perp\perp}$ \\
\bottomrule
\end{tabular}
\caption{Leading-twist gluon TMD fragmentation functions into a spin-1 hadron, classified by the gluon polarization (columns) and by the polarization parameter of the hadron to which they are proportional (rows), cf.~Ref.~\cite{Boer:2016xqr}. 
The labels $L,T,LL,LT,TT$ refer to the spin parameters, not to helicity states of $H$. In four dimensions, $D_{1TT}$ and $H_{1TT}^{\perp}$ multiply the same tensor structure and enter only through the combination $\Phi\equiv H_{1TT}^{\perp}-D_{1TT}$. Since the structure of $H_{1LT}$ in Eq.~\mbox{\eqref{eq:GLT}} has a trace, the function associated with an unpolarized gluon in the $LT$ row is $D_{1LT}-H_{1LT}$, see Sec.~\mbox{\ref{sec:projectors}}.
Empty spaces mean functions forbidden by symmetries of the theory.}
\label{tab:TMDFFs}
\end{table}

\subsection{Projectors onto TMD distributions}
\label{sec:projectors}

Each TMDFF of Tab.~\ref{tab:TMDFFs}, namely $F$ in short, is extracted from the tensors of Eqs.~\eqref{eq:GU}--\eqref{eq:GTT} by contraction with a projector $\mathcal{P}[F]_{\lambda\sigma}$ satisfying
\begin{equation}
    \mathcal{P}[F_i]_{\lambda\sigma}\,\mathcal{B}_j^{\lambda\sigma}=\delta_{ij} \, ,
\end{equation}
where $\mathcal{B}_j$ are the tensor structures which appear after the tensorial decomposition of $\Gamma_X^{\lambda \sigma}$ for $X \in \{U,L,T,LL,LT,TT\}$.
The traditional way to construct them is to invert the Gram matrix of the structures. Here we use a property of the two-dimensional transverse plane that makes the inversion trivial and gives all projectors in a single and uniform form.

Note that every polarization tensor has the following generic form,
\begin{equation}
  \Gamma_X^{\lambda\sigma}
  = -g_T^{\lambda\sigma}\, a_X\, D_X
    + i\epsilon_T^{\lambda\sigma}\, b_X\, G_X
    + \mathcal{A}_X^{\lambda\sigma}\, H_X
    + \mathcal{B}_X^{\lambda\sigma}\, H'_X \, ,
  \label{eq:generic}
\end{equation}
where $a_X$ and $b_X$ are scalars, and $\mathcal{A}_X$, $\mathcal{B}_X$ are symmetric traceless transverse tensors. For $X=LT$ this form is reached after separating the trace of the $H_{1LT}$ structure in Eq.~\eqref{eq:GLT}, $g_{T\lambda\sigma}S_{LT}^{\{\lambda}k_T^{\sigma\}}=2\,k_T\cdot S_{LT}$: writing $S_{LT}^{\{\lambda}k_T^{\sigma\}}=\big[S_{LT}^{\{\lambda}k_T^{\sigma\}} 
-(k_T\cdot S_{LT})\,g_T^{\lambda\sigma}\big]+(k_T\cdot S_{LT})\,g_T^{\lambda\sigma}$, the coefficient of $-g_T^{\lambda\sigma}(k_T\cdot S_{LT})/M$ is $D_{1LT}-H_{1LT}$, which is the function associated with an unpolarized gluon. The three pieces are then mutually orthogonal, so the unpolarized and circularly polarized functions follow from a single contraction,
\begin{equation}
  D_X = -\frac{g_{T\lambda\sigma}\,\Gamma_X^{\lambda\sigma}}{2\,a_X} ,
  \qquad
  G_X = -\frac{i\,\epsilon_{T\lambda\sigma}\,\Gamma_X^{\lambda\sigma}}{2\,b_X} .
  \label{eq:proj_fg}
\end{equation}
The linearly polarized sector is two-dimensional.
To separate $H_X$ from $H'_X$ we introduce the dual of a symmetric traceless transverse tensor,
\begin{equation}
  \widetilde{\mathcal{X}}^{\lambda\sigma}
  \equiv \epsilon_T^{\lambda}{}_{\rho}\,\mathcal{X}^{\rho\sigma} ,
  \label{eq:dual}
\end{equation}
which generalizes the dual vector $\tilde k_T^\alpha=\epsilon_T^{\alpha\beta}k_{T\beta}$ of Appendix~\ref{sec:conventions}.
In two dimensions $\widetilde{\mathcal{X}}$ is again symmetric and traceless, and
\begin{equation}
  \widetilde{\widetilde{\mathcal{X}}} = -\mathcal{X} ,
  \qquad
  \widetilde{\mathcal{X}}\cdot\mathcal{Y} = -\mathcal{X}\cdot\widetilde{\mathcal{Y}} ,
  \qquad
  \mathcal{X}\cdot\widetilde{\mathcal{X}} = 0 ,
  \qquad
  \widetilde{\mathcal{X}}\cdot\widetilde{\mathcal{X}} = \mathcal{X}\cdot\mathcal{X} ,
  \label{eq:dual_props}
\end{equation}
with $\mathcal{X}\cdot\mathcal{Y}\equiv\mathcal{X}_{\lambda\sigma}\mathcal{Y}^{\lambda\sigma}$. 
Geometrically, $\mathcal{X}\to\widetilde{\mathcal{X}}$ rotates the plane of linear polarization by $\pi/4$, i.e., it multiplies the helicity-flip amplitudes $e^{\pm 2i\phi}$ by a phase $\pm i$. Since $\mathcal{B}_X\cdot\widetilde{\mathcal{B}}_X=0$, contracting Eq.~\eqref{eq:generic} with $\widetilde{\mathcal{B}}_X$ removes $H'_X$, and vice versa.
Defining
\begin{equation}
  C_X \equiv \mathcal{A}_X\cdot\widetilde{\mathcal{B}}_X
      = -\widetilde{\mathcal{A}}_X\cdot\mathcal{B}_X ,
  \label{eq:DX}
\end{equation}
one finds
\begin{equation}
  H_X = \frac{\widetilde{\mathcal{B}}_{X\lambda\sigma}\,\Gamma_X^{\lambda\sigma}}{C_X} ,
  \quad
  H'_X = -\frac{\widetilde{\mathcal{A}}_{X\lambda\sigma}\,\Gamma_X^{\lambda\sigma}}{C_X} ,
  \quad\text{and} \quad
  H_X = \frac{\mathcal{A}_{X\lambda\sigma}\,\Gamma_X^{\lambda\sigma}}{\mathcal{A}_X\cdot\mathcal{A}_X} \,\text{ if } \mathcal{B}_X=0 .
  \label{eq:proj_h}
\end{equation}
In words, the projector onto one helicity-flip structure is the dual of its partner, divided by a single normalization, and Eqs.~\eqref{eq:proj_fg} and~\eqref{eq:proj_h} replace the Gram-matrix inversion.
By the two-dimensional Lagrange identity, the Gram determinant of the helicity-flip block is $(\mathcal{A}_X\cdot\mathcal{A}_X)(\mathcal{B}_X\cdot\mathcal{B}_X)-(\mathcal{A}_X\cdot\mathcal{B}_X)^2=C_X^2$.
As a result, the spin- and angle-dependent invariants appear in the projectors only linearly, not quadratically. Equivalently, if one maps the helicity-flip part onto a complex amplitude with a circular basis of transverse vectors, $C_X$ is (up to a convention-dependent sign) the imaginary part of $\mathcal{A}_X^{(+2)}\mathcal{B}_X^{(+2)*}$.

Evaluating Eqs.~\eqref{eq:proj_fg} and~\eqref{eq:proj_h} with $a_X$, $b_X$ and $\mathcal A_X$, $\mathcal B_X$, and using the duals
\begin{gather}
  \widetilde{k}_T^{\lambda\sigma}
  = \tfrac12\tilde k_T^{\{\lambda}k_T^{\sigma\}}
  = -\tfrac12\epsilon_T^{k_T\{\lambda}k_T^{\sigma\}} ,
  \qquad
  \widetilde{\big(k_T^{\lambda\sigma\rho\tau}S_{TT\rho\tau}\big)}
  = -k_T^{\lambda\sigma\rho\tau}\widetilde{S}_{TT\rho\tau} ,
  \nonumber\\
  \widetilde{\big(\epsilon_T^{k_T\{\lambda}S^{\sigma\}}
  +\epsilon_T^{S\{\lambda}k_T^{\sigma\}}\big)}
  = 2\big(S^{\{\lambda}k_T^{\sigma\}}-(k_T\cdot S)\,g_T^{\lambda\sigma}\big) ,
  \label{eq:dual_ids}
\end{gather}
valid for any transverse vector $S$, we obtain the projectors in Tab.~\ref{tab:projectors}. We have checked explicitly that they satisfy $\mathcal{P}[F_i]\cdot\mathcal{B}_j=\delta_{ij}$ within each sector and that they agree with the Gram-matrix construction.

\begin{table}[t]
\centering
\renewcommand{\arraystretch}{2.4}
\begin{tabular}{c l l}
\hline
$X$ & Unpolarized / circular & Linear (helicity flip) \\
\hline
$U$ &
$\mathcal{P}[D_1] = -\dfrac{g_{T\lambda\sigma}}{2}$ &
$\mathcal{P}[H_1^\perp] = \dfrac{2M^2}{k_T^4}\,k_{T\lambda\sigma}$ \\
$L$ &
$\mathcal{P}[G_{1L}] = -\dfrac{i\,\epsilon_{T\lambda\sigma}}{2S_L}$ &
$\mathcal{P}[H_{1L}^\perp] = \dfrac{2M^2}{S_L\,k_T^4}\,\tilde k_{T\lambda\sigma}$ \\
$T$ &
$\mathcal{P}[D_{1T}^\perp] = -\dfrac{M\,g_{T\lambda\sigma}}{2\,\epsilon_T^{S_T k_T}}$ &
$\mathcal{P}[H_{1T}] = -\dfrac{M}{k_T^2}\Big[\dfrac{k_{T\lambda\sigma}}{\epsilon_T^{S_T k_T}}-\dfrac{\tilde k_{T\lambda\sigma}}{k_T\cdot S_T}\Big]$ \\
 &
$\mathcal{P}[G_{1T}] = -\dfrac{iM\,\epsilon_{T\lambda\sigma}}{2\,k_T\cdot S_T}$ &
$\mathcal{P}[H_{1T}^\perp] = -\dfrac{2M^3}{k_T^4}\Big[\dfrac{k_{T\lambda\sigma}}{\epsilon_T^{S_T k_T}}+\dfrac{\tilde k_{T\lambda\sigma}}{k_T\cdot S_T}\Big]$ \\
$LL$ &
$\mathcal{P}[D_{1LL}] = -\dfrac{g_{T\lambda\sigma}}{2S_{LL}}$ &
$\mathcal{P}[H_{1LL}^\perp] = \dfrac{2M^2}{S_{LL}\,k_T^4}\,k_{T\lambda\sigma}$ \\
$LT$ &
$\mathcal{P}[D_{1LT}] = -\dfrac{M\,g_{T\lambda\sigma}}{2\,k_T\cdot S_{LT}}+\mathcal{P}[H_{1LT}]$ &
$\mathcal{P}[H_{1LT}] = \dfrac{M}{2k_T^2}\Big[\dfrac{k_{T\lambda\sigma}}{k_T\cdot S_{LT}}+\dfrac{\tilde k_{T\lambda\sigma}}{\epsilon_T^{S_{LT} k_T}}\Big]$ \\
 &
$\mathcal{P}[G_{1LT}] = -\dfrac{iM\,\epsilon_{T\lambda\sigma}}{2\,\epsilon_T^{S_{LT} k_T}}$ &
$\mathcal{P}[H_{1LT}^\perp] = \dfrac{2M^3}{k_T^4}\Big[\dfrac{k_{T\lambda\sigma}}{k_T\cdot S_{LT}}-\dfrac{\tilde k_{T\lambda\sigma}}{\epsilon_T^{S_{LT} k_T}}\Big]$ \\
$TT$ &
$\mathcal{P}[\Phi] = \dfrac{M^2\,g_{T\lambda\sigma}}{2S_{TT}^{kk}}$ &
$\mathcal{P}[H_{1TT}] = -\dfrac{k_{T\lambda\sigma\rho\tau}\,\widetilde S_{TT}^{\rho\tau}}{S_{TT}^{kk}\,S_{TT}^{k\tilde k}}$ \\
 &
$\mathcal{P}[G_{1TT}] = -\dfrac{iM^2\,\epsilon_{T\lambda\sigma}}{2S_{TT}^{k\tilde k}}$ &
$\mathcal{P}[H_{1TT}^{\perp\perp}] = -\dfrac{M^4\,\widetilde S_{TT\lambda\sigma}}{S_{TT}^{kk}\,S_{TT}^{k\tilde k}}$ \\
\hline
\end{tabular}
\caption{Projectors onto the eighteen independent gluon TMDFFs ($F$ in short),
$F=\mathcal{P}[F]_{\lambda\sigma}\Gamma^{\lambda\sigma}_X$, obtained from Eqs.~(2.17) and (2.21).
Here $\tilde k_T^{\lambda\sigma}=\epsilon^{\lambda}_{T\rho}k_T^{\rho\sigma}=-\tfrac12\epsilon_T^{k_T\{\lambda}k_T^{\sigma\}}$,
$\widetilde S^{\lambda\sigma}_{TT}=\epsilon^{\lambda}_{T\rho}S^{\rho\sigma}_{TT}$, see Eq.~(2.18), and $k_T^4\equiv(k_T^2)^2$.
In two transverse dimensions $S^{\lambda}_{TT\,\rho}k_T^{\rho\sigma}+S^{\sigma}_{TT\,\rho}k_T^{\rho\lambda}=g_T^{\lambda\sigma}S^{kk}_{TT}$,
so the structures of $D_{1TT}$ and $H^\perp_{1TT}$ coincide and only $\Phi\equiv H^\perp_{1TT}-D_{1TT}$ can be projected.
Because the $H_{1LT}$ structure carries a trace, $\mathcal{P}[D_{1LT}]$ also contains helicity-flip terms.}
\label{tab:projectors}
\end{table}

\subsection{NRQCD factorization}
\label{sec:nrqcd}

In the NRQCD factorization formalism and in the kinematic regime $k_T\gg m_Qv$, with $m_Qv$ the typical relative momentum of the heavy-quark pair, regardless of the size of $k_T$ relative to $M$ (see Sec.~\mbox{\ref{sec:ope}}), the TMDFFs factorize into perturbative short-distance coefficients (SDCs) and NRQCD long-distance matrix elements (LDMEs).
Therefore, when the produced spin-1 hadron considered in the discussion of the previous sections is a quarkonium state, each of the functions in Tab.~\ref{tab:TMDFFs} is factorized as follows,
\begin{equation}
F^X_{g\to \mathcal Q}(z,b_T;\mu,\zeta)
= \sum_n d^{[F^X]}_{g\to Q\bar Q[n]}(z,b_T;\mu,\zeta ,\mu_\Lambda)\,
\big\langle\mathcal{O}^{\mathcal Q}(n)\big\rangle (\mu_\Lambda) 
\label{eq:NRQCDfact}
\end{equation}
where $n={}^{2S+1}L_J^{[c]}$ labels the color and angular-momentum state of the produced heavy-quark pair in $d^{[F^X]}_{g \to Q \bar{Q}[n]}$.
Here $\mu_\Lambda$ is the NRQCD factorization scale.
The SDCs have a residual dependence on $\mu_\Lambda$ inherited from the subtraction of infrared poles in the matching onto NRQCD, while the LDMEs depend on this scale through operator mixing.
The dependence on this scale cancels in the sum over $n$, so that the corresponding TMDFF is independent of it.
The SDCs are computed as a power series in $\alpha_s$, while the sum over $n$ is organized according to the scaling of the LDMEs with $v$.
In this work, we calculate the former at $\mathcal{O}(\alpha_s^2)$ meaning that the previous formula is written as
\begin{equation}
    F_{g \to \mathcal{Q}}^X(z,b_T;\mu,\zeta) = \sum_{n = {}^3S_1^{[8]}, {}^1S_0^{[8]}, {}^3P_J^{[8]}} d^{[F^X]}_{g\to Q\bar Q[n]}(z,b_T;\mu,\zeta,\mu_\Lambda)\,
\big\langle\mathcal{O}^{\mathcal Q}(n)\big\rangle (\mu_\Lambda)  + \mathcal{O}(\alpha_s^3) \, ,
\end{equation}
For gluon fragmentation into a quarkonium with quantum numbers $L=0$ and $J=1$, i.e., $J/\psi(1S)$, $\Upsilon(nS)$, $\psi(2S)$, the lowest order in $\alpha_s$ is reached through the color-octet ${}^3S_1^{[8]}$ channel, which starts at $\mathcal{O}(\alpha_s)$.
At $\mathcal{O}(\alpha_s^2)$, the ${}^1S_0^{[8]}$ and ${}^3P_J^{[8]}$ channels also contribute.
 These channels enter at the same order in the NRQCD velocity expansion as ${}^3S_1^{[8]}$, namely $\mathcal{O}(v^7)$. 

 Therefore, to obtain the complete  color-octet contribution through $\mathcal{O}(\alpha_s^2)$ at this order in $v$, we compute the next-to-leading-order (NLO) corrections to the ${}^3S_1^{[8]}$ channel together with the leading-order (LO) contributions of the ${}^1S_0^{[8]}$ and ${}^3P_J^{[8]}$ channels. With relativistic normalization of the hadron states, the corresponding LDMEs are
\begin{align}
\big\langle\mathcal{O}^{\mathcal{Q}}({}^3S_1^{[8]})\big\rangle
&= \frac{1}{2M}\sum_{S}\sum_X
\langle 0|\chi^\dagger\sigma^iT^a\psi|\mathcal{Q}(S),X\rangle
\langle \mathcal{Q}(S),X|\psi^\dagger\sigma^iT^a\chi|0\rangle ,
\label{eq:LDME3s18}\\
\big\langle\mathcal{O}^{\mathcal{Q}}({}^1S_0^{[8]})\big\rangle
&= \frac{1}{2M}\sum_{S}\sum_X
\langle 0|\chi^\dagger T^a\psi|\mathcal{Q}(S),X\rangle
\langle \mathcal{Q}(S),X|\psi^\dagger T^a\chi|0\rangle ,
\label{eq:LDME1s08}\\
\big\langle\mathcal{O}^{\mathcal{Q}}({}^3P_J^{[8]})\big\rangle
&= \frac{1}{2M}\sum_{S}\sum_X
\Big\langle 0\Big|\chi^\dagger\Big(-\tfrac{i}{2}\overleftrightarrow{D}^{l}\Big)\sigma^{j}T^a\psi\Big|\mathcal{Q}(S),X\Big\rangle
\,\mathcal{P}_J^{lj,ki}\,\nonumber\\
&\qquad\Big\langle \mathcal{Q}(S),X\Big|\psi^\dagger\Big(-\tfrac{i}{2}\overleftrightarrow{D}^{k}\Big)\sigma^{i}T^a\chi\Big|0\Big\rangle ,
\label{eq:LDME3PJ8}
\end{align}
where $\psi$ ($\chi$) annihilates (creates) a heavy quark (antiquark), $\overleftrightarrow D=\overrightarrow D-\overleftarrow D$, and the gauge completion of the operators with eikonal lines is discussed in  Ref.~\cite{Nayak:2005rw}. In Eq.~\eqref{eq:LDME3PJ8}, $k,l$ are orbital and $i,j$ spin indices, and
\begin{equation}
\begin{aligned}
&\mathcal{P}_0^{lj,ki}=\tfrac13\,\delta^{lj}\delta^{ki},\qquad
\mathcal{P}_1^{lj,ki}=\tfrac12\big(\delta^{lk}\delta^{ji}-\delta^{li}\delta^{jk}\big),\\
&\mathcal{P}_2^{lj,ki}=\tfrac12\big(\delta^{lk}\delta^{ji}+\delta^{li}\delta^{jk}\big)-\tfrac13\,\delta^{lj}\delta^{ki}
\end{aligned}
\label{eq:PJ}
\end{equation}
project onto total angular momentum $J$, with $\delta^{kl}\delta^{ij}\,\mathcal P_J^{lj,ki}=2J+1$ and $\sum_J\mathcal P_J^{lj,ki}=\delta^{lk}\delta^{ji}$.
They reproduce the standard operators of  Ref.~\cite{Bodwin:1994jh}.

Heavy-quark spin symmetry determines the matrix elements for a hadron in a definite polarization state $S$ in terms of the LDMEs, up to corrections of relative order $v^2$~\cite{Cho:1994gb,Braaten:1996jt}. For the ${}^3S_1^{[8]}$ channel,
\begin{equation}
\sum_X
\langle 0|\chi^\dagger\sigma^jT^a\psi|\mathcal{Q}(S),X\rangle
\langle \mathcal{Q}(S),X|\psi^\dagger\sigma^iT^a\chi|0\rangle =
\frac{2M}{3}\, \varepsilon_S^{*i}\varepsilon_S^{j}\, \big\langle\mathcal{O}^{\mathcal{Q}}({}^3S_1^{[8]})\big\rangle
+\mathcal O(v^2).
\label{eq:s3-ldme}
\end{equation}
The polarization of the pair is transferred to the hadron, and the factor $1/3$ counts the three spin states of the pair. Summing over $S$ and setting $j=i$ reproduces Eq.~\eqref{eq:LDME3s18}.
In the ${}^3P_J^{[8]}$ channel the spin of the pair is likewise transferred to the hadron, while the orbital angular momentum is summed over, giving
\begin{equation}
\begin{aligned}
\sum_X &\Big\langle 0\Big| \chi^\dagger\Big(-\tfrac{i}{2}\overleftrightarrow D^{l}\Big)\sigma^jT^a\psi \Big|\mathcal{Q}(S),X\Big\rangle
\Big\langle \mathcal{Q}(S),X\Big|\psi^\dagger \Big(-\tfrac{i}{2}\overleftrightarrow D^{k}\Big) \sigma^iT^a\chi \Big|0\Big\rangle\\
=& 2M\,\delta^{kl}\,\varepsilon_S^{*i}\varepsilon_S^{j}\,
\big\langle\mathcal O^{\mathcal{Q}}({}^3P_0^{[8]})\big\rangle +\mathcal O(v^2).
\label{eq:P-ldme}
\end{aligned}
\end{equation}
Contracting with $\mathcal P_J^{lj,ki}$ and summing over $S$ gives
$\langle\mathcal O^{\mathcal{Q}}({}^3P_J^{[8]})\rangle=(2J+1)\langle\mathcal O^{\mathcal{Q}}({}^3P_0^{[8]})\rangle$, so the whole multiplet is described by $\langle\mathcal O^{\mathcal{Q}}({}^3P_0^{[8]})\rangle$, and the sum over $J$ is performed by $\delta^{lk} \delta^{ji}$. 
Unlike Eq.~\eqref{eq:s3-ldme}, Eq.~\eqref{eq:P-ldme} has no factor $1/3$, because $\langle\mathcal O^{\mathcal{Q}}({}^3P_0^{[8]})\rangle$ normalizes a single state of the pair. 
Compared with the ${}^3S_1^{[8]}$ channel, this produces an extra factor $3$ in the $P$-wave short-distance coefficients.

The ${}^1S_0^{[8]}$ channel is different, since the pair has spin zero. Because the operator is a scalar under rotations, the three polarization states of the hadron contribute equally after the sum over $X$:
\begin{equation}
\sum_X
\langle0|\chi^\dagger T^a\psi|\mathcal{Q}(S),X\rangle
\langle \mathcal{Q}(S),X|\psi^\dagger T^a\chi|0\rangle =
\frac{2M}{3}\,
\big\langle\mathcal O^{\mathcal{Q}}({}^1S_0^{[8]})\big\rangle
+\mathcal O(v^2)
\label{eq:s1-ldme}
\end{equation}
for each $\lambda_H=0,\pm1,$ where $\lambda_H$ is the helicity of the quarkonium.
This channel therefore produces an unpolarized hadron and contributes only to $D_1$ and $H_1^\perp$.

For the spin-triplet channels, the spin density matrix is parametrized as~\cite{Bacchetta:2000jk}, with the opposite orientation of $S_{LT}$
\begin{equation}
\varepsilon^{i}_S\varepsilon^{*j}_S
= \frac13\delta^{ij} - \frac{i}{2}\epsilon^{ijk}S^k - T^{ij} ,
\qquad
T^{ij} = \frac12
\begin{pmatrix}
-\frac23 S_{LL}+S_{TT}^{xx} & S_{TT}^{xy} & -S_{LT}^{x}\\
S_{TT}^{xy} & -\frac23 S_{LL}-S_{TT}^{xx} & -S_{LT}^{y}\\
-S_{LT}^{x} & -S_{LT}^{y} & \frac43 S_{LL}
\end{pmatrix} ,
\label{eq:rho}
\end{equation}
with $S=(S_T^x,S_T^y,S_L)$ in the hadron rest frame.
Boosting to the frame in which the $\mathcal{Q}$ has momentum $P$ using the boost matrices $L^\mu{}_i$, and defining the longitudinal polarization vector
\begin{equation}
\varepsilon_L^\mu = \frac{1}{M}\Big(P^+n^\mu-\frac{M^2}{2P^+}\nbar^\mu\Big),
\qquad \varepsilon_L\cdot P=0 ,\quad \varepsilon_L^2=-1 ,
\label{eq:epsL}
\end{equation}
one obtains, with $\epsilon^{0123}=+1$,
\begin{align}
L^\mu{}_iL^\nu{}_j\,\varepsilon^{i}_S\varepsilon^{*j}_S
&= -\frac13\Big(g^{\mu\nu}-\frac{P^\mu P^\nu}{M^2}\Big)
+ \frac{i}{2M}\epsilon^{\mu\nu\alpha\beta}P_\alpha S_\beta
- T^{\mu\nu} ,
\label{eq:rho_cov}\\
S^\mu = S_L\,\varepsilon_L^\mu + S_T^\mu , &\qquad
T^{\mu\nu}
= \frac12\Big[S_{LL}\Big(\frac43\varepsilon_L^\mu\varepsilon_L^\nu
+ \frac23 g_T^{\mu\nu}\Big)
- \varepsilon_L^{\{\mu}S_{LT}^{\nu\}} + S_{TT}^{\mu\nu}\Big].
\label{eq:LT_cov}
\end{align}
The tensor $T^{\mu\nu}$ is symmetric, traceless and transverse to $P$, and Eq.~\eqref{eq:LT_cov} reproduces the rest-frame matrix~\eqref{eq:rho}.
Using $2P^+P^-=M^2$, the transverse projection of the vector term is
$\epsilon^{\lambda\sigma\alpha\beta}P_\alpha S_\beta=-M\,S_L\,\epsilon_T^{\lambda\sigma} + \epsilon^{\lambda \sigma \alpha \beta} P_\alpha S_{T\beta}$,
with $\epsilon_T^{\lambda\sigma}\equiv\epsilon^{\lambda\sigma\alpha\beta}n_\alpha \nbar_\beta$, while $S_T$
does not contribute. 
With this convention, we note that $S_L = \pm 1$ when $\lambda = \pm 1$.

After the non-relativistic expansion of the heavy-quark spinors (see discussion in App.~\ref{app:polarisationandexpansion}), the correlator for the production of a $Q\bar Q[{}^3S_1^{[8]}]$ pair takes the form
\begin{equation}
\Delta^{\lambda\sigma}_{g\to Q\bar Q[{}^3S_1^{[8]}]}(z,k_T)
= \mathcal{D}^{\lambda\sigma}_{1} {}_{\mu\nu}(z,k_T)\,L^\mu{}_iL^\nu{}_j\,
\xi^\dagger\sigma^iT^a\eta\;\eta^\dagger\sigma^jT^a\xi ,
\label{eq:Dtensor}
\end{equation}
and, matching onto Eq.~\eqref{eq:s3-ldme}, the SDC of any TMDFF is
\begin{equation}
d^{[F^X]}_{g\to Q\bar Q[{}^3S_1^{[8]}]}(z,k_T)
= \mathcal{P}[F^X]_{\lambda\sigma}\,
\mathcal{D}_{1}^{\lambda\sigma}{}_{\mu\nu}(z,k_T)\,
\Pi_X^{\mu\nu} .
\label{eq:SDC_general}
\end{equation}
Here $\Pi_X^{\mu\nu}$ is the part of Eq.~\eqref{eq:rho_cov} proportional to the polarization parameter $X$:
\begin{equation}
\begin{aligned}
\label{eq:PiX}
\Pi_U^{\mu \nu} & = -\frac{1}{3}\Big(g^{\mu \nu} - \frac{P^\mu P^\nu}{M^2}\Big) , \\
\Pi_L^{\mu \nu} & = S_L\,\frac{i}{2M}\, \epsilon^{\mu \nu \alpha \beta} P_\alpha \varepsilon_{L \beta} , \qquad
\Pi_T^{\mu \nu} = \frac{i}{2M}\, \epsilon^{\mu \nu \alpha \beta} P_\alpha S_{T \beta} , \\
\Pi^{\mu \nu}_{LL} & = - \frac{S_{LL}}{2} \Big( \frac{4}{3} \varepsilon^\mu_L \varepsilon^\nu_L + \frac{2}{3} g_T^{\mu \nu} \Big) , \quad \Pi^{\mu\nu}_{LT} =+ \frac{1}{2}\,\varepsilon_L^{\{\mu} S_{LT}^{\nu\}}
, \quad \Pi^{\mu \nu}_{TT} = - \frac{1}{2}\, S_{TT}^{\mu \nu} .
\end{aligned}
\end{equation}
For the ${}^3P_J^{[8]}$ channel, the P-wave projection introduces an additional orbital index through differentiation with respect to the relative momentum $q$ of the heavy-quark pair. 
The perturbative correlator therefore carries both spin and orbital indices,
\begin{equation}
\Delta^{\lambda\sigma}_{g\to Q\bar Q[{}^3P_J^{[8]}]}(z,k_T)
= \mathcal{D}_P^{\lambda\sigma}{}_{\mu\rho,\nu\kappa}(z,k_T)\,
q^\rho q^\kappa\,L^\mu{}_iL^\nu{}_j\,
\xi^\dagger\sigma^iT^a\eta\;\eta^\dagger\sigma^jT^a\xi ,
\label{eq:Dtensor_P}
\end{equation}
where $q^\rho=L^\rho{}_r\,q^r$, and $\mathcal{D}_P$ is normalized as $\mathcal{D}_{1}$ in Eq.~\eqref{eq:Dtensor} and is obtained from the first derivatives of the amplitudes with respect to $q$ at $q=0$.
Matching onto Eq.~\eqref{eq:P-ldme}, the SDC for the complete P-wave contribution is
\begin{equation}
d^{[F^X]}_{g\to Q\bar Q[{}^3P_J^{[8]}]}(z,k_T)
= 3\,\mathcal{P}[F^X]_{\lambda\sigma}\,
\mathcal{D}_P^{\lambda\sigma}{}_{\mu\rho,\nu\kappa}(z,k_T)\,
\Pi_X^{\mu\nu}\Pi^{\rho\kappa},
\qquad
\Pi^{\rho\kappa}
= -g^{\rho\kappa}+\frac{P^\rho P^\kappa}{M^2},
\label{eq:SDC_general_P}
\end{equation}
as coefficients of $\langle\mathcal{O}^{\mathcal{Q}}({}^3P_0^{[8]})\rangle$. The factor $3$ arises because Eq.~\eqref{eq:P-ldme}, unlike Eqs.~\eqref{eq:s3-ldme} and~\eqref{eq:s1-ldme}, has no factor $\tfrac13$ to be identified with the overall $\tfrac13$ of Eq.~\eqref{eq:decomp}. 
The sum over the orbital states is performed by $\Pi^{\rho\kappa}$, while $\Pi_X^{\mu\nu}$ describes the polarization of the produced hadron, exactly as in the ${}^3S_1^{[8]}$ channel.
For the ${}^1S_0^{[8]}$ channel, the heavy-quark pair has neither spin nor orbital polarization indices.
Its perturbative correlator takes the form
\begin{equation}
\Delta^{\lambda\sigma}_{g\to Q\bar Q[{}^1S_0^{[8]}]}(z,k_T)
= \mathcal{D}_0^{\lambda\sigma}(z,k_T)\,
\xi^\dagger T^a\eta\;
\eta^\dagger T^a\xi .
\label{eq:Dtensor_1S0}
\end{equation}
Matching onto Eq.~\eqref{eq:s1-ldme}, the short-distance coefficients are obtained directly from the gluon TMD projectors, without any contraction with the hadron spin-density matrix, i.e.,
\begin{equation}
d^{[F]}_{g\to Q\bar Q[{}^1S_0^{[8]}]}(z,k_T)
= \mathcal{P}[F]_{\lambda\sigma}\,
\mathcal{D}_0^{\lambda\sigma}(z,k_T) .
\label{eq:SDC_1S0}
\end{equation}
Since this channel produces an unpolarized hadron at the considered order in $v$, it contributes only to $D_1$ and $H_1^\perp$. All TMDFFs associated with vector or tensor polarization of the final hadron vanish for this channel at this accuracy.

For the $\Sone$ channel, at $\mathcal{O}(\alpha_s)$, $k_T=0$ and $\mathcal{D}_1$ just projects Eq.~\eqref{eq:rho_cov} onto the transverse plane.
Taking into account the ordering $\varepsilon^{\lambda}\varepsilon^{*\sigma}$,
\begin{equation}
\Delta^{\lambda\sigma}_{g\to H}\Big|_{\rm LO}
\propto
-\frac13\big(1+S_{LL}\big)\,g_T^{\lambda\sigma}
-\frac{i}{2}S_L\,\epsilon_T^{\lambda\sigma}
- \frac12 S_{TT}^{\lambda\sigma} .
\label{eq:LOpattern}
\end{equation}
Comparing with Eq.~\eqref{eq:decomp}, only four functions are nonzero at LO, all proportional to $\delta(1-z)$.
This agrees with  Ref.~\cite{Copeland:2023wbu}.
With our conventions, a positive-helicity gluon contributes $\varepsilon_{+}^{\lambda}\varepsilon_{+}^{*\sigma} \sim -\tfrac{i}{2}\epsilon_T^{\lambda\sigma}$ to $\Delta^{\lambda\sigma}$, so that the negative LO value of $G_{1L}$ implied by Eq.~\mbox{\eqref{eq:LOpattern}} corresponds to the transfer of helicity from the gluon to the hadron.
Eq.~\eqref{eq:LOpattern} also makes clear why no $T$- or $LT$-type function arises at this order: $S_T$ drops out of the transverse projection of the vector term, and $S_{LT}$ always comes with the longitudinal vector $\varepsilon_L$.

At $\mathcal O(\alpha_s^2)$, the calculation includes the NLO corrections to the ${}^3S_1^{[8]}$ channel, together with the LO contributions from ${}^1S_0^{[8]}$ and ${}^3P_J^{[8]}$. 
The real emission introduces a non-vanishing $k_T$. 
The polarization parameters $S_T$ and $S_{LT}$, which do not contribute at LO, can now contribute through their correlations with $k_T$.
This allows the spin-triplet channels to generate the different unpolarized, vector-polarized and tensor-polarized TMDFFs in Tab.~\ref{tab:TMDFFs}.
The projectors of Tab.~\ref{tab:projectors} extract each SDC, providing the complete set of gluon TMDFFs through $\mathcal O(\alpha_s^2)$ for all gluon and hadron polarization configurations at the considered order in $v$.

We should note that the color-singlet ${}^3S_1^{[1]}$ channel does not contribute to the gluon TMDFF at the order considered in this work.
A single gluon cannot produce a color-singlet pair, and the process $g \to Q\bar Q[{}^3S_1^{[1]}]+g$ is forbidden by color conservation together with charge conjugation: two gluons in a color-singlet state have $C=+1$, whereas a ${}^3S_1$ pair has $C=-1$.
This channel first enters at $\mathcal{O}(\alpha_s^3)$, through $g\to Q\bar Q[{}^3S_1^{[1]}]+gg$.
However, in the combined counting in $\alpha_s$ and $v$, the ratio between this contribution, $\mathcal{O}(\alpha_s^3v^3)$, and the terms computed here, $\mathcal{O}(\alpha_s^2v^7)$, is $\alpha_s/v^4$.
For charmonium, with $\alpha_s\simeq0.25$ and $v^2\simeq0.3$, this ratio is of order 3, so the color-singlet channel is not suppressed by the power counting. 
However, the corresponding collinear color-singlet fragmentation function~\mbox{\cite{Braaten:1993rw}} is numerically small, and gluon fragmentation into $J/\psi$ and $\psi(2S)$ at large transverse momentum is expected to be dominated by the color-octet ${}^3S_1^{[8]}$ channel~\mbox{\cite{Braaten:1994vv}}.
Although an analysis of the interplay between the two channels is therefore needed, the TMD version of the color-singlet contribution requires the phase space of two real gluons at fixed $k_T$, a calculation of next-to-next-to-leading-order complexity that is beyond the scope of this work. 
We then leave this calculation for future work.


\section{Results}
\label{sec:results}
In this section we present the gluon TMDFFs at $\mathcal O(\alpha_s^2)$ for the ${}^3S_1^{[8]}$, ${}^1S_0^{[8]}$ and ${}^3P_J^{[8]}$ channels. We first outline the calculation in Sec.~\ref{sec:outline}. 
We then give the unsubtracted TMDFFs in $k_T$ space (Sec.~\ref{sec:kt_unsub}) and the subtracted ones after soft-factor subtraction and renormalization (Sec.~\ref{sec:kt_sub}), followed by their $b_T$-space expressions (Sec.~\ref{sec:bt_results}). 
We derive the collinear fragmentation functions (Sec.~\ref{sec:collinear}) and their relation to the small-$b_T$ limit of the TMDFFs through the operator product expansion (Sec.~\ref{sec:ope}).

\subsection{Outline of the calculation}
\label{sec:outline}

\begin{figure}[tbp]
    \centering
    \includegraphics[width=0.4\linewidth]{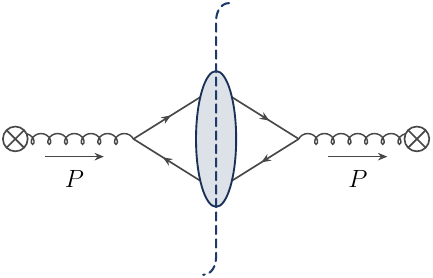}
    \caption{Tree-level $\mathcal O(\alpha_s)$ diagram for the fragmentation of a gluon into a $Q\bar Q$ pair in the ${}^3S_1^{[8]}$ state. }
    \label{fig:tree_level}
\end{figure}

\begin{figure}[tbp]
    \centering
    \includegraphics[width=\linewidth]{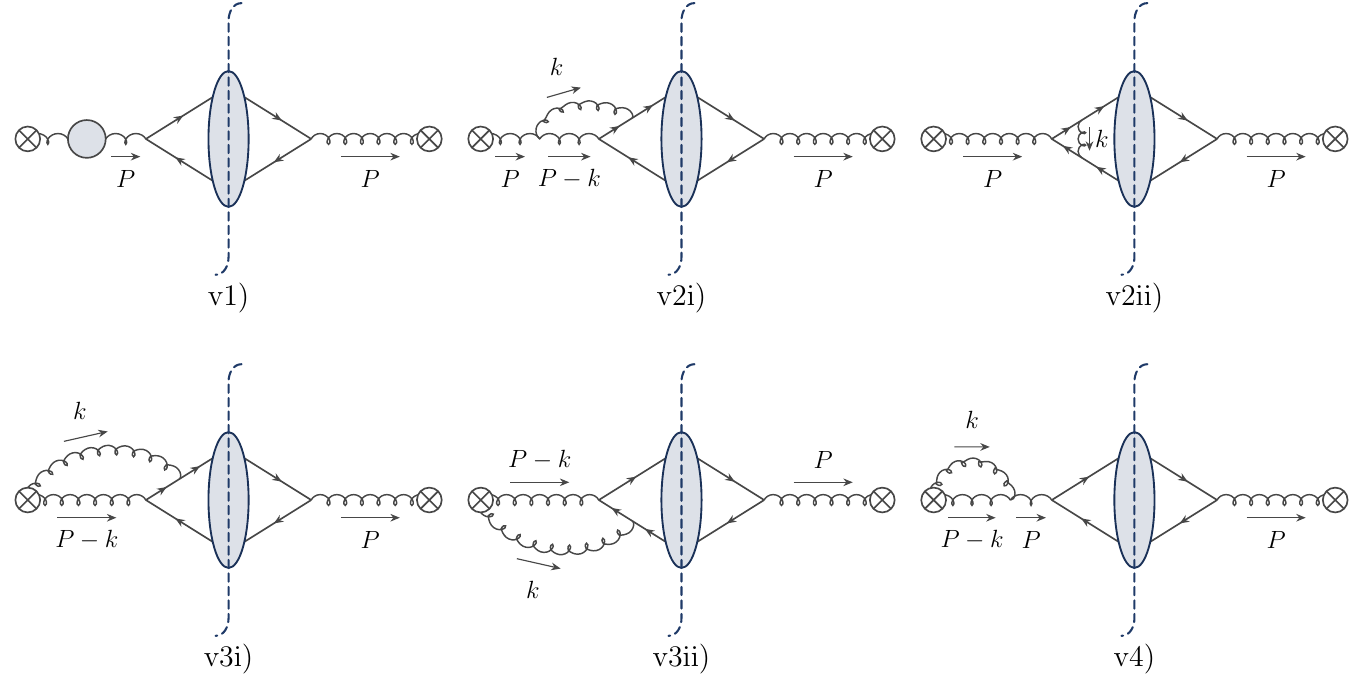}
    \caption{One-loop virtual diagrams for the fragmentation of a gluon into a $Q\bar Q$ pair in the ${}^3S_1^{[8]}$ state at $\mathcal O(\alpha_s^2)$. Only the diagrams on the left of the cut are shown; their complex conjugates are understood.}
    \label{fig:virtual_diagrams}
\end{figure}

\begin{figure}[tbp]
    \centering
    \includegraphics[width=\linewidth]{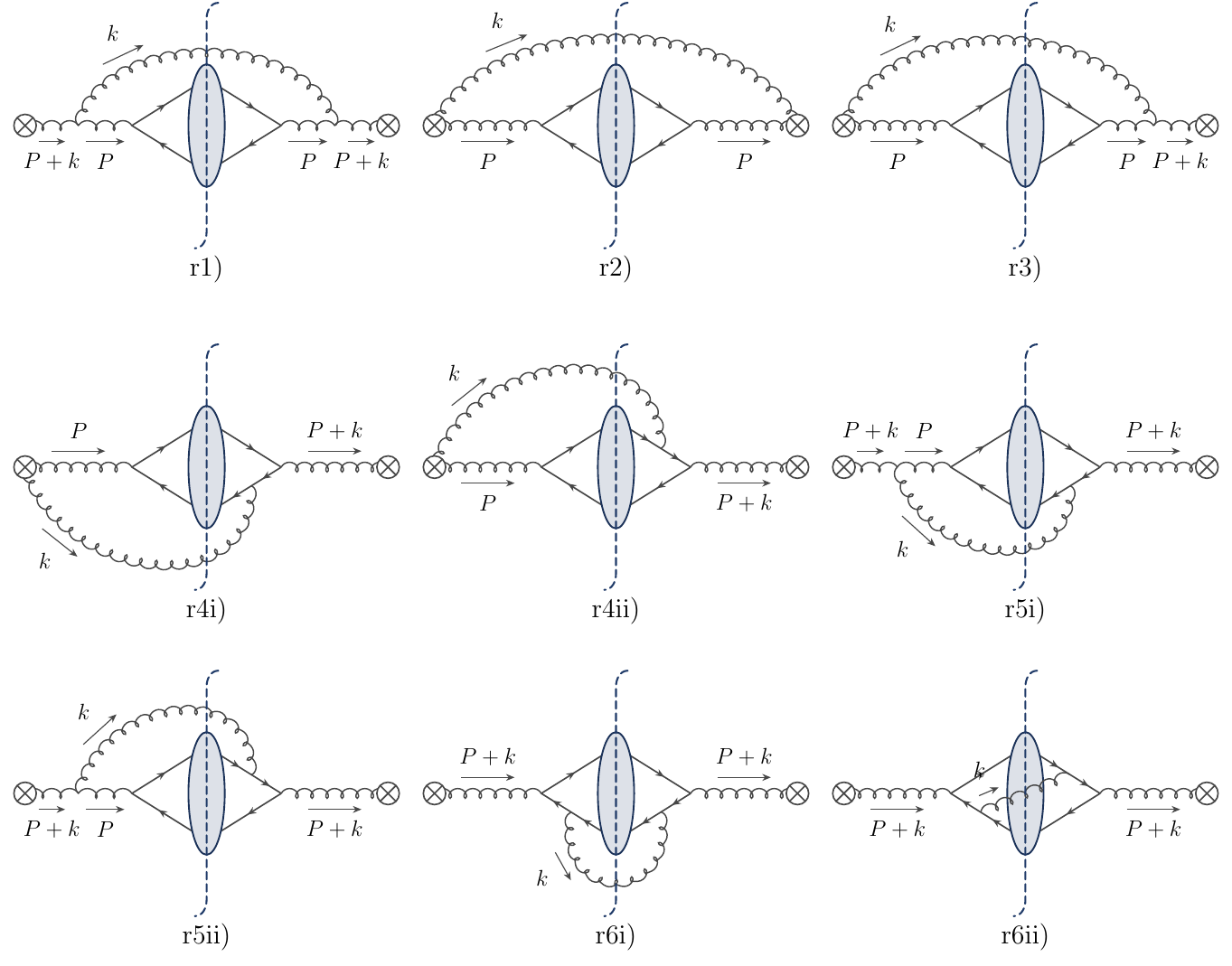}
    \caption{Real-emission diagrams for $g\to Q\bar Q+g$ at $\mathcal O(\alpha_s^2)$. They give the NLO correction to the ${}^3S_1^{[8]}$ channel and the LO contribution to the ${}^1S_0^{[8]}$ and ${}^3P_J^{[8]}$ channels.}
    \label{fig:real_diagrams}
\end{figure}

All eighteen SDCs of Tab.~\ref{tab:TMDFFs} are obtained from a single set of Feynman diagrams, which are shown in Figs.~\ref{fig:tree_level},~\ref{fig:virtual_diagrams} and \ref{fig:real_diagrams}.
The polarizations enter only through two insertions, which are independent of each other and of the diagram topology: the hadron spin projector $\Pi_X^{\mu\nu}$ of Sec.~\ref{sec:nrqcd}, inserted on the heavy-quark side of the diagram, and the gluon projector $\mathcal{P}[F]_{\lambda\sigma}$ of Tab.~\ref{tab:projectors}, contracted with the open indices of the operator $\mathcal{B}_{n\perp}^{\lambda}\mathcal{B}_{n\perp}^{\sigma}$.
The computation therefore factorizes into a polarization-independent part, the short-distance tensor $\mathcal{D}^{\lambda\sigma}{}_{\mu\nu}$, and two contractions, as in Eq.~\eqref{eq:SDC_general}.
The Lorentz and Dirac algebra is carried out with the Mathematica package \textsc{FeynCalc}~\cite{Mertig:1990an,Shtabovenko:2023idz}. 

We decompose four-vectors as $u^\mu = u^+ n^\mu + u^-\bar n^\mu + u_T^\mu$, with $u^+\equiv\bar n\cdot u$, $u^-\equiv n\cdot u$ and $u\cdot v = u^+v^- + u^-v^+ + u_T\cdot v_T$, so that $u_T^2=-\boldsymbol u_\perp^2$. In the hadron frame of Eq.~\eqref{eq:hadron_frame} the heavy-quark pair carries momentum $P$, with $P^2=M^2$ and $P_T=0$. 
The heavy quark and antiquark carry $P/2$ each, with $m_Q=M/2$, since the relative momentum is set to zero at the order in $v$ in which we work. 
At NLO the fragmenting gluon carries momentum $P+k$, where $k$ is the momentum of the emitted on-shell gluon, and
\begin{equation}
  k^2=0 ,\qquad
  k^+ = P^+\,\frac{1-z}{z} ,\, 
  \label{eq:kin_NLO}
\end{equation}
where $k_T$ is defined as the transverse momentum of the fragmenting gluon.
Rapidity divergences are regularized with the $\delta$-regulator in all Wilson lines~\cite{Echevarria:2011epo,Echevarria:2015byo,Echevarria:2016scs}.
The dimensionless combination $\delta\equiv\delta^+/P^+$ appears in the results. 
Ultraviolet and infrared divergences are regularized dimensionally, with $d=4-2\epsilon$.
 
After the nonrelativistic expansion of the heavy-quark spinors at zero relative momentum (App.~\ref{app:polarisationandexpansion}), every heavy-quark bilinear in the spin-triplet state reduces to the boosted spin factor $L^\mu{}_iL^\nu{}_j\,\xi^\dagger\sigma^iT^a\eta\,\eta^\dagger\sigma^jT^a\xi$, while spin-singlet states reduce to $\xi^\dagger T^a\eta\;\eta^\dagger T^a\xi$.
Through Eqs.~\eqref{eq:s3-ldme},~\eqref{eq:P-ldme},~\eqref{eq:s1-ldme} and~\eqref{eq:rho_cov}, these factors are traded for the projector $\Pi_X^{\mu\nu}$ of the hadron polarization $X$, with the overall normalization absorbed into the LDME~\cite{Echevarria:2023dme,Copeland:2023wbu}.
For the $\Sone$ channel, several diagrams appear.
For those in which the gluon couples directly to the $Q\bar Q$ pair, this amounts to
\begin{equation}
  \big[\bar u(p_1)\gamma^\alpha v(p_2)\big]\big[\bar v(p_2)\gamma^\nu u(p_1)\big]
  \;\longrightarrow\; M^2\,\Pi_X^{\alpha\nu} .
  \label{eq:spin_rule_1}
\end{equation}
When a gluon is emitted from a heavy-quark line, the three-gamma structures reduce to
\begin{align}
  \big[\bar u\,\gamma^\alpha(\slashed{l}+m_Q)\gamma^\rho v\big]
  \big[\bar v\,\gamma^\nu u\big]
  \;\longrightarrow\;
  &\;4m_Q^2\big(l^\alpha\,\Pi_X^{\rho\nu}+l^\rho\,\Pi_X^{\alpha\nu}
   -g^{\alpha\rho}\,l_\beta\,\Pi_X^{\beta\nu}\big)
  \nonumber\\
  &+2m_Q^2\big(P^\alpha\,\Pi_X^{\rho\nu}-P^\rho\,\Pi_X^{\alpha\nu}\big) ,
  \qquad l=-\tfrac{P}{2}-k ,
  \label{eq:spin_rule_3}
\end{align}
and analogously when both sides of the cut contain an emission.
For the ${}^1S_0^{[8]}$ and ${}^3P_J^{[8]}$ channels only this last case contributes, i.e., 
\begin{align}
{}^3S_1^{[8]}:\quad
  &\big[\bar u\,\gamma^\alpha(\slashed l+m_Q)\gamma^\rho v\big]\big[\bar v\,\gamma^\sigma(\slashed l+m_Q)\gamma^\nu u\big]
  \;\longrightarrow\;
  m_Q^2\,\Pi_{X\,\mu\lambda}\;
  W^{\alpha\rho\mu}\,W^{\nu\sigma\lambda},\\[4pt]
  {}^1S_0^{[8]}:\quad
  &\big[\bar u\,\gamma^\alpha(\slashed l+m_Q)\gamma^\rho v\big]\big[\bar v\,\gamma^\sigma(\slashed l+m_Q)\gamma^\nu u\big]
  \;\longrightarrow\;
  \epsilon^{\alpha\rho P l}\,\epsilon^{\nu\sigma P l},
  \label{eq:spin_rule_1S0}\\[4pt]
  {}^3P_J^{[8]}:\quad
  &\big[\bar u\,\gamma^\alpha(\slashed l+m_Q)\gamma^\rho v\big]\big[\bar v\,\gamma^\sigma(\slashed l+m_Q)\gamma^\nu u\big]
  \;\longrightarrow\;
  \frac{3}{4m_Q^2}\,\Pi_{\beta\gamma}\,\Pi_{X\,\mu\lambda}\;
  V^{\alpha\rho\mu\beta}\,V^{\nu\sigma\lambda\gamma},
  \label{eq:spin_rule_3PJ}
\end{align}
where $\epsilon^{\alpha\rho Pl}\equiv\epsilon^{\alpha\rho\mu\lambda}P_\mu l_\lambda$, 
\begin{equation}
    W^{\alpha\rho\mu}=2\big(l^\alpha g^{\rho\mu}+l^\rho g^{\alpha\mu}-g^{\alpha\rho}\,l^\mu\big)+P^\alpha g^{\rho\mu}-P^\rho g^{\alpha\mu},
\end{equation}
$\Pi_{\beta\gamma}=-g_{\beta\gamma}+P_\beta P_\gamma/M^2$ is the sum over the orbital polarization, and $V^{\alpha\rho\mu\beta}$ is the derivative of the ${}^3P_J^{[8]}$ spinorial structure with respect to the relative momentum $q_\beta$ at $q=0$, defined in App.~\ref{app:polarisationandexpansion}.  
A ${}^1S_0^{[8]}$ pair produces an unpolarized hadron, so Eq.~\eqref{eq:spin_rule_1S0} carries no hadron projector. 
In Eq.~\eqref{eq:spin_rule_3PJ} the spin indices are contracted with the same projectors $\Pi_X$ as for ${}^3S_1^{[8]}$, and the prefactor includes the factor 3 of the normalization to $\langle\mathcal O({}^3P_0^{[8]})\rangle$.
Since Eqs.~\eqref{eq:spin_rule_1} and~\eqref{eq:spin_rule_3} are linear in $\Pi_X$, each diagram is computed once as a function of an arbitrary tensor $\chi^{\mu\nu}$. 
The six cases of Eq.~\eqref{eq:PiX} are inserted only at the end. In this step we use the four-dimensional decomposition~\eqref{eq:rho} of the spin density matrix.

The virtual corrections are displayed in Fig.~\ref{fig:virtual_diagrams}, and we compute them in Feynman gauge.
Diagram $v1$ is the self-energy of the gluon line of virtuality $P^2=M^2$ that connects the operator to the $Q\bar Q$ vertex, which is an internal line and therefore receives no LSZ factor; $v2i$ and $v2ii$ are the non-abelian and abelian corrections to the $gQ\bar Q$ vertex, and $v3i$, $v3ii$ and $v4$ connect the Wilson line of the operator $\mathcal{B}^\mu_{n\perp}$ to the heavy quark, the heavy antiquark and the gluon line, respectively.
The amplitude is obtained from the amputated Green function of $\mathcal{B}^\mu_{n\perp}$ with an on-shell heavy quark and antiquark, $p_{1,2}^2=m_Q^2$, through the Lehmann–Symanzik–Zimmermann (LSZ) reduction formula.
Self-energy insertions on the external heavy-quark legs are therefore not included as diagrams; instead, each external heavy-quark leg contributes a factor $\sqrt{Z_2^{\rm OS}}$, where $Z_2^{\rm OS}=1+\delta Z_2^{\rm OS}$ is the residue of the heavy-quark propagator at its pole, with
\begin{equation}
\delta Z_2^{\rm OS}=-\frac{\alpha_s C_F}{4\pi}\Big[\frac{1}{\epsilon_{\rm UV}}
+\frac{2}{\epsilon_{\rm IR}}+3\ln\frac{\mu^2}{m_Q^2}+4\Big] \, .
\label{eq:Z2OS}
\end{equation}
At one loop the virtual amplitude thus receives $\delta Z_2^{\rm OS}$ times the LO amplitude (one half per leg), i.e.\ $2\,\delta Z_2^{\rm OS}$ times the LO coefficient in the squared amplitude.
The heavy-quark mass is the pole mass, $M=2m_Q$, where the identification of the hadron mass with $2m_Q$ holds at leading order in $v$.
Since the LO amplitude contains no internal heavy-quark propagator, no mass counterterm enters at this order.
The coupling is renormalized in the $\overline{\rm MS}$ scheme with $n_f$ light flavors, $\alpha_s^{\rm bare}=\mu^{2\epsilon} (e^{\gamma_E}/4\pi)^{\epsilon}Z_\alpha\,\alpha_s(\mu)$ with $Z_\alpha=1-\beta_0\alpha_s/(4\pi\epsilon)$.
The heavy-quark loop is not included in the gluon self-energy v1; its size is discussed below Eq.~\mbox{\eqref{eq:cV}}. 
Moreover, the abelian vertex correction v2ii contains the Coulomb singularity $\propto (C_F-C_A/2)\,\pi^2m_Q/|\boldsymbol{q}|$, which is reproduced by the one-loop NRQCD matrix element of the free $Q\bar Q$ pair and cancels in the matching onto NRQCD.
At $\boldsymbol{q}=0$ in dimensional regularization the potential region is scaleless, so it does not contribute to the short-distance coefficient.
The infrared poles of the virtual diagrams, including those of $\delta Z_2^{\rm OS}$, cancel against those of the real emission. At $\boldsymbol q=0$ the heavy quark and antiquark have the same velocity, and the soft poles of $\delta Z_2^{\rm OS}$, proportional to $2C_F$, and of $v2ii$, proportional to $C_A-2C_F$, add up to the color charge of the octet pair, $(\mathbf T_Q+\mathbf T_{\bar Q})^2=C_A$, where $\mathbf T_Q$ and $\mathbf T_{\bar Q}$ are the color generators acting on the heavy quark and antiquark; the ultraviolet poles of $v1$, $v2i$, $v2ii$ and $\delta Z_2^{\rm OS}$, where the $C_F$ terms cancel between the vertex corrections and $\delta Z_2^{\rm OS}$, are removed, together with those of the Wilson-line diagrams, by $Z_\alpha$ in the LO coefficient and by $Z_g$.
Finally, since $P^2=M^2>0$, the virtual diagrams have imaginary parts, which cancel against those of the complex-conjugate diagrams, while the timelike phases $e^{i\pi\epsilon}$ that multiply double poles generate real $\pi^2$ terms.
 
The real corrections are the cut diagrams of Fig.~\ref{fig:real_diagrams}.
We group them into six topologies according to how the emitted gluon is attached on each side of the cut. In diagram $r1$ it is attached to the triple-gluon vertex on both sides, and in diagram $r2$ to the Wilson line $\mathcal{B}^{(1)}$ on both sides.
Diagram $r3$ connects the triple-gluon vertex to the Wilson line.
Diagrams $r4i$--$r4ii$ connect the heavy-quark line to the Wilson line, and diagrams $r5i$--$r5ii$ connect the former to the triple-gluon vertex.
In diagrams $r6i$--$r6ii$ the gluon is attached to the heavy-quark line on both sides.
Each contribution has the form
\begin{equation}
\begin{aligned}
  \Delta^{\lambda\sigma}_{(i)}(z,k_T;X)
  & = \mathcal{N}_d\,c_i\int dk^+dk^-\,\delta(k^2)\,\theta(k^+)\,
    \delta\Big(k^+-P^+\frac{1-z}{z}\Big)\,
    \mathcal{D}^{\lambda\sigma}_{(i)\mu\nu}(P,k;\delta^+)\,\Pi_X^{\mu\nu} \, ,\\
  \mathcal{N}_d & = \frac{P^+\,g_s^4}{(N_c^2-1)(2\pi)^{d-1}} ,
  \label{eq:real_generic}
\end{aligned}
\end{equation}
where $(N_c^2-1)$ comes from the normalization of Eq.~\eqref{eq:corr_b}, and $(2\pi)^{-(d-1)}$ from the phase space of the emitted gluon, $\mathcal{D}=\mathcal{D}_{1,0,P}$ depending on the channel.
The color factors are $c_{r1}=c_{r2}=c_{r3}=C_A$ for diagrams $r1$, $r2$, $r3$. For diagrams $r4$ and $r5$, $c_{r4}=c_{r5}=2\times C_A/2$, where the factor $2$ accounts for the two equivalent attachments to the quark and antiquark lines. 
For diagram $r6$, the two orderings give the  color factors $C_F-1/(2N_c)$ and $C_F-C_A/2-1/(2N_c)$, where $-1/(2N_c)$ removes the  color-singlet component. Charge conjugation fixes their relative sign. 
For the C-odd ${}^3S_1^{[8]}$ state they enter with opposite signs, and the singlet terms cancel, i.e., $c_{r6}^{\,{}^3S_1}=\frac{C_A}{2}$, which is the $f^{abc}$ structure. For the C-even ${}^1S_0^{[8]}$ and ${}^3P_J^{[8]}$ states they enter with the same sign, so only the $d^{abc}$ structure survives. Hence, we have $c_{r6}^{\,{}^1S_0,\,{}^3P_J}=\frac{N_c^2-4}{2N_c}=2B_F$. 
Diagrams $r4$--$r6$ are accompanied by their mirror images.

Regarding the light-cone integrations, the $k^-$ integral is fixed by the on-shell condition, $\int dk^-\,\delta(2k^+k^-+k_T^2)=1/(2k^+)$, and the $k^+$ integral by the momentum fraction~\eqref{eq:kin_NLO}.
This leaves a function of $z$, $k_T^2$ and $\delta$ in which $P^+$ cancels against $\mathcal{N}_d$.
Two denominators appear,
\begin{equation}
  \boldsymbol k_\perp^2+\Lambda_1^2 ,\qquad
  \boldsymbol k_\perp^2+\Lambda_2^2 ,\qquad
  \Lambda_1^2=\frac{M^2(1-z)}{z^2} ,\quad
  \Lambda_2^2=\frac{M^2(1-z)^2}{z^2} ,
  \label{eq:Lambdas}
\end{equation}
which come from the gluon and heavy-quark propagators, respectively.
The $\delta$-regulator is relevant only for topologies $r3$ and $r4$. These develop the rapidity-divergent factor $(1-z)/[(1-z)^2+\delta^2z^2]$, and we keep $\delta$ finite in them.
In topologies $r1$, $r2$, $r5$, $r6$ we take $\delta\to0$. Topology $r2$ does not depend on $k_T$ and therefore does not contribute for $b_T\neq0$.

\subsection{Unsubtracted TMDFFs in transverse-momentum space}
\label{sec:kt_unsub}

In this section we collect the complete NLO SDCs as distributions in $z$ and in the transverse momentum $\bm{k}_\perp$ of the fragmenting gluon.
In the following, we separately write the contribution of the different states.

\subsubsection{\texorpdfstring{${}^{3}S_1^{[8]}$}{3S1[8]} state}
We use $a_s \equiv \alpha_s C_A/(2 \pi)$ and the LO coefficient $d_{\rm LO} = \pi \alpha_s /(3M^3)$.
We also use 
\begin{equation}
    \mathcal K_2 \equiv z^2 \bm{k}_\perp^2 + M^2 (1-z)^2 = z^2 (\bm{k}_\perp^2 + \Lambda_2^2) \, \quad \mathcal N \equiv \frac{3 a_s}{\pi} d_{\rm LO} \, ,
\end{equation}
and the unpolarized real-emission kernel $\mathcal{P}_{gg}(z)=2\big[\tfrac{z}{(1-z)_+}+\tfrac{1-z}{z}+z(1-z)\big]$.
The constants of the virtual corrections are collected in\footnote{This corrects Ref.~\cite{Echevarria:2023dme}, where the $\pi^2$ term in $c_V$ was $-4\pi^2/3$ instead of $-\pi^2/3$, because of a mismatch in the timelike phases $e^{i\pi\varepsilon}$ of the master integrals $I_{ABC}$ and $I_{BCD}$. The present value agrees with Ref.~\mbox{\cite{Ma:2013yla}}, and with Ref.~\mbox{\cite{Braaten:2000pc}} up to the $4\ln^22$ term discussed in Sec.~\mbox{\ref{sec:collinear}}.}
\begin{equation}
  c_V\equiv\frac{16}{3}\ln2-\frac{\pi^2}{3}+\frac{59}{9}-\frac{10\,n_f}{9C_A} ,
  \qquad \beta_0=\frac{11}{3}C_A-\frac23n_f.
  \label{eq:cV}
\end{equation}
Here $n_f$ counts only the light flavors. If the heavy-quark loop in diagram v1 were included with a zero-momentum (decoupling) subtraction, so that $\alpha_s$ remains the $n_f$-flavor coupling, its finite remainder at the threshold $P^2=4m_Q^2$ would add $-32T_F/(9C_A)$ to $c_V$.
The singular behavior at $\boldsymbol k_\perp\to0$ is described by the standard plus distributions (see, e.g.,  Ref.~\cite{Ebert:2016gcn})
\begin{equation}
  \mathcal{L}_n(\boldsymbol k_\perp,\mu)
  \equiv\frac{1}{\pi}\left[\frac{\ln^n(\boldsymbol k_\perp^2/\mu^2)}{\boldsymbol k_\perp^2}\right]_+^{\mu} ,
  \qquad \int_{|\boldsymbol k_\perp|\leq\mu}\!d^2\boldsymbol k_\perp\,\mathcal{L}_n(\boldsymbol k_\perp,\mu)=0 .
  \label{eq:Ln}
\end{equation}
We also need two $z$-dependent distributions in $\boldsymbol k_\perp$:
\begin{align}
  \mathcal{R}_{\Lambda_2}(\boldsymbol k_\perp)
  &\equiv \frac{1}{\pi(\boldsymbol k_\perp^2+\Lambda_2^2)}-\mathcal{L}_0(\boldsymbol k_\perp,\Lambda_2) ,\\
  &\int\! d^2\boldsymbol k_\perp\,e^{i\boldsymbol k_\perp\cdot\boldsymbol b_T}\,\mathcal{R}_{\Lambda_2}
   = 2\Big[K_0(b_T\Lambda_2)+\ln\frac{b_T\Lambda_2 e^{\gamma_E}}{2}\Big] ,
  \label{eq:RLambda}\\
  \mathcal{Q}_{\Lambda_2}(\boldsymbol k_\perp)
  &\equiv \frac{\Lambda_2^2}{\pi(\boldsymbol k_\perp^2+\Lambda_2^2)^2}-\delta^{(2)}(\boldsymbol k_\perp) ,\\
  &\int\! d^2\boldsymbol k_\perp\,e^{i\boldsymbol k_\perp\cdot\boldsymbol b_T}\,\mathcal{Q}_{\Lambda_2}
   = b_T\Lambda_2\,K_1(b_T\Lambda_2)-1 \, .
  \label{eq:QLambda}
\end{align}
Both distributions vanish for $\Lambda_2\to0$, as their Fourier transforms show.
This guarantees that their products with $1/(1-z)$ are well defined at $z=1$.

In terms of these definitions, we now write our results for the SDCs up to $\mathcal{O}(a_s)$.
We start with the TMDFF of an unpolarized fragmenting gluon into an unpolarized quarkonium:
\begin{align}
  d^{[D_1]}_{\Sone}&(z,\boldsymbol k_\perp;\delta)
  = d_{\rm LO}\,\delta(1-z)\,\delta^{(2)}(\boldsymbol k_\perp)
   \nonumber\\
  &+a_s\,d_{\rm LO}\bigg\{\delta(1-z)\bigg[
     \delta^{(2)}(\boldsymbol k_\perp)\Big(\frac{1}{\epsilon_{\rm UV}}\Big(\frac{\beta_0}{C_A}+2\ln\delta\Big)
     +\frac{\beta_0}{C_A}\ln\frac{\mu^2}{M^2}+c_V\Big)
   \nonumber\\
  &\hspace{3.3cm}
     -2\ln\delta\;\mathcal{L}_0(\boldsymbol k_\perp,\mu)\bigg]
     +\mathcal{R}^{\rm real}(z,\boldsymbol k_\perp)\bigg\} ,
  \label{eq:D1_unsub_full}
\end{align}
where the regular part of the real emission is
\begin{equation}
    \begin{aligned}
  \mathcal{R}^{\rm real}(z,\boldsymbol k_\perp)
  &=\mathcal{P}_{gg}(z)\Big[\mathcal{L}_0(\boldsymbol k_\perp,\mu)+\mathcal{R}_{\Lambda_2}(\boldsymbol k_\perp)
     +\ln\frac{z^2\mu^2}{M^2}\,\delta^{(2)}(\boldsymbol k_\perp)\Big]\\
  &\quad-\delta^{(2)}(\boldsymbol k_\perp)\left[\frac{2z(2-2z+z^2)}{(1-z)_+}
   +\frac{4(1-z+z^2)^2}{z}\left(\frac{\ln(1-z)}{1-z}\right)_+\right] \\
   &\quad-\frac{2z(2-2z+z^2)}{1-z}\,\mathcal{Q}_{\Lambda_2}(\boldsymbol k_\perp).
  \label{eq:Rreal}
\end{aligned}
\end{equation} 
The IR poles of the real and virtual contributions cancel in the sum,
and so do the $\ln^2\delta$ terms. What remains is the UV pole and the
rapidity divergence, $\ln\delta$.

For $F\in\{D_{1LL},G_{1L},H_{1TT}\}$, the virtual corrections and the soft region of the real emission are proportional to the LO tensor, Eq.~\eqref{eq:LOpattern}. In this region a soft gluon is emitted eikonally and does not alter the spin structure of the heavy-quark pair.
The complete coefficients can therefore be written as
\begin{equation}
  d^{[F]}_{\Sone}(z,\boldsymbol k_\perp;\delta)
  = c_F\,d^{[D_1]}_{\Sone}(z,\boldsymbol k_\perp;\delta)+\mathcal{N}\,\Delta T_F(z,\boldsymbol k_\perp) ,
  \label{eq:pol_full}
\end{equation}
with $c_{D_{1LL}}=1$, $c_{G_{1L}}=-
\tfrac32$ and $c_{H_{1TT}}=-\tfrac32$, and
\begin{align}
  \Delta T_{D_{1LL}} &= -\frac{2M^2z(1-z)^3}{\mathcal{K}_2^2} \, ,
  \label{eq:dT_LL}\\
  \Delta T_{G_{1L}} &= \frac{z(1-z)^3\big(M^2+z^2\boldsymbol k_\perp^2\big)}{\mathcal{K}_2^2} \, ,
  \label{eq:dT_G1L}\\
  \Delta T_{H_{1TT}} &= \frac{z(1-z)\big[M^2(1-z)^2+z^2(1+z^2)\,\boldsymbol k_\perp^2\big]}{\mathcal{K}_2^2} \, .
  \label{eq:dT_H1TT}
\end{align}
The remainders $\Delta T_F$ are ordinary functions. 
They vanish at least as $(1-z)$ for $z\to1$, they are integrable around $(z,\boldsymbol k_\perp)=(1, \boldsymbol 0)$, and they carry no IR pole, no rapidity divergence and no $\delta^{(2)}(\boldsymbol k_\perp)$ term. 
Consequently, all poles, all $\ln\delta$ terms and the full $\delta(1-z)\,\delta^{(2)}(\boldsymbol k_\perp)$ coefficient of a polarized function are $c_F$ times those of $D_1$.

The rest of the functions in Tab.~\ref{tab:TMDFFs} have no LO term and no virtual correction since they start one order higher in $\alpha_s$ as they need $g \to Q \bar{Q} + g$.
They contain no $\delta^{(2)}(\boldsymbol k_\perp)$ term, and they are free of rapidity divergences (at this perturbative order).
Their complete NLO coefficients are
\begin{equation}
    \begin{aligned}
  d^{[H_1^\perp]}_{\Sone} = d^{[H_{1LL}^\perp]}
  &= \frac{4}{3}\,\mathcal{N}\,\frac{M^2z^5(1-z)}{\mathcal{K}_2^2} ,
  &
  d^{[G_{1T}]}_{\Sone} &= - \mathcal{N}\,\frac{M^2z^3(1-z)(2-z)}{\mathcal{K}_2^2} ,\\
  d^{[D_{1LT}]}_{\Sone} &= -2\,\mathcal{N}\,\frac{M^2z^2(1-z)^2}{\mathcal{K}_2^2},
    &
d^{[H_{1LT}]}_{\Sone} &= \mathcal{N}\,\frac{M^2z^4(1-z)}{\mathcal{K}_2^2} ,\\
d^{[\Phi]}_{\Sone} &=\mathcal{N}\,\frac{M^2z^3(1-z)}{\mathcal{K}_2^2} .
  \label{eq:kT_noLO}
\end{aligned}
\end{equation}

The remaining eight functions vanish at this order, in particular
\begin{equation}
\begin{aligned}
    d^{[D_{1T}^\perp]}_{\Sone}&=d^{[H_{1T}]} _{\Sone}=d^{[G_{1LT}]}_{\Sone}=\mathcal{O}(\delta) \, ,
  \qquad\\
  d^{[H_{1L}^\perp]}_{\Sone}&=d^{[H_{1T}^\perp]}_{\Sone}=d^{[H_{1LT}^\perp]}_{\Sone}=d^{[G_{1TT}]}_{\Sone}=d^{[H_{1TT}^{\perp\perp}]}_{\Sone}=0 \, .
\end{aligned}
  \label{eq:kT_zero}
\end{equation}
The first three receive contributions only from the imaginary part of the eikonal propagator. These are odd in $\delta$ and vanish when the regulator is removed.
The rest cancels out when combining the contributions from the real-emission diagrams.

\subsubsection{\texorpdfstring{${}^{1}S_0^{[8]}$}{1S0[8]} state}
\label{sec:1S08-kt}

The heavy-quark pair, with momenta $p_{1,2}=P/2\pm q$ and $m_Q=M/2$, is projected onto the spin singlet with the covariant projector
\begin{equation}
\Pi_0\propto\Big(\tfrac{\slashed P}{2}-\slashed q-m_Q\Big)\gamma_5\Big(\tfrac{\slashed P}{2}+\slashed q+m_Q\Big),
\end{equation}
evaluated at $q=0$.
Since the ${}^1S_0^{[8]}$ operator is a scalar under spatial rotations, rotational invariance implies that its contribution is independent of the polarization of the produced hadron. 
Consequently, at the considered accuracy, only the unpolarized-hadron TMDFFs $D_1$ and $H_1^\perp$ receive contributions from this channel. 
As discussed in Sec.~\ref{sec:outline}, only topology $r6$ survives, i.e., the diagram with the gluon emission coming from the heavy-quark lines. 

The scales appearing are both $\Lambda_1$ and $\Lambda_2$ defined in Eq.~\eqref{eq:Lambdas}, with\footnote{We note that $2 P \cdot k = \tfrac{\mathcal{K}_2}{z(1-z)}$ and that $(P + k )^2 = \tfrac{\mathcal{K}_1}{z(1-z)}$.
Therefore, $\mathcal{K}_2$ is actually the heavy-quark propagator denominator and $\mathcal{K}_1$ is the virtuality of the fragmenting gluon.
} 
\begin{equation}
\mathcal{K}_1 \equiv z^2 \boldsymbol k_\perp ^2 +M^2 
(1-z)=z^2\big(\boldsymbol k_\perp^2+{\Lambda}_1^2\big).
\label{eq:sp_K1}
\end{equation}
\begin{equation}
    \mathcal K_2 \equiv z^2 \bm{k}_\perp^2 + M^2 (1-z)^2 = z^2 (\bm{k}_\perp^2 + \Lambda_2^2) \, ,
    \label{eq:sp_K2}
    \end{equation}
The overall factor is 
\begin{equation}
\mathcal{N}_B=\frac{B_F\,\alpha_s^2}{\pi M^3}, 
\label{eq:sp_NB}
\end{equation}
with $B_F=(N_c^2 − 4)/(4N_c),$ $B_F = 5/12 $ for $N_c = 3$,
and the two nonzero coefficients are
\begin{equation}
\begin{aligned}
d^{[D_1]}_{{}^1S_0^{[8]}}&=\mathcal{N}_BM^2z^3(1-z)\\
&\,\frac{\big[z^4\big(z^2+(1-z)^2\big)\boldsymbol k_\perp^4+2M^2z^2(1-z)^3 \boldsymbol k_\perp^2+M^4(1-z)^4\big]}{\mathcal{K}_1^2\,\mathcal{K}_2^2},\\
d^{[H_1^\perp]}_{{}^1S_0^{[8]}}&=-4\,\mathcal{N}_B\,\frac{M^4z^6(1-z)^2}{\mathcal{K}_1\,\mathcal{K}_2^2},
\end{aligned}
\label{eq:sp_S0}
\end{equation}
These coefficients are infrared finite and they have no rapidity divergence, since the Wilson-line diagrams vanish and there is no LO term to be multiplied by the soft factor.
Hence, none of the distributions $\mathcal{L}_n$, $\mathcal{R}_\Lambda$ and $\mathcal{Q}_\Lambda$ is needed in this channel.

\subsubsection{\texorpdfstring{${}^{3}P_J^{[8]}$}{3PJ[8]} state}
\label{sec:3PJ8-kt}
Here, as in the previous case, only diagrams with topology $r6$ contribute.
Because of $C$ parity, the  color factor is also the same as for ${}^{1}S_0^{[8]}$, i.e., $2B_F$.
The pair is projected onto the spin triplet with
\begin{equation}
\Pi_1^\alpha\propto\Big(\tfrac{\slashed P}{2}-\slashed q-m_Q\Big)\gamma^\alpha\Big(\tfrac{\slashed P}{2}+\slashed q+m_Q\Big) \, ,
\end{equation}
and we take the derivative of the amplitude with respect to $q^\beta$ at $q=0$.
With the definitions of Eqs.~\eqref{eq:sp_K1},~\eqref{eq:sp_K2} and~\eqref{eq:sp_NB}, twelve functions are nonzero:
\begin{equation}
\begin{aligned}
d^{[D_1]}_{{}^3P_0^{[8]}}&=\frac{4\mathcal{N}_B\,z^3(1-z)}{\mathcal{K}_1^2\mathcal{K}_2^3}\Big[3z^6\big[z^2+(1-z)^2\big]\kperp^6\\
&\qquad-M^2z^4(1-z)\,(14z^3-26z^2+17z-9)\,\kperp^4\\
&\qquad-M^4z^2(1-z)^2(14z^3-29z^2+16z-9)\,\kperp^2\\
&\qquad+M^6(1-z)^4(7z^2-2z+3)\Big],\\
d^{[H_1^\perp]}_{{}^3P_0^{[8]}}&=-16\,\mathcal{N}_B\,\frac{M^2z^6(1-z)^2\big[3z^2\kperp^2-M^2(1-z)(7z-3)\big]}{\mathcal{K}_1\mathcal{K}_2^3},\\
d^{[G_{1L}]}_{{}^3P_0^{[8]}}&= - 24\,\mathcal{N}_B\,M^2z^4(1-z)^2\\
&\qquad\frac{\big[z^4\kperp^4+2M^2z^2(1-z)(1+z-z^2)\kperp^2+M^4(1-z)^3(1+z)\big]}{\mathcal{K}_1^2\mathcal{K}_2^3},\\
d^{[G_{1T}]}_{{}^3P_0^{[8]}}&=24\,\mathcal{N}_B\,\frac{M^2z^5(1-z)^2\big[z^4\kperp^4+M^2z^2(1-z)(2+z)\kperp^2+M^4(1-z)^3\big]}{\mathcal{K}_1^2\mathcal{K}_2^3},\\
d^{[D_{1LL}]}_{{}^3P_0^{[8]}}&=-16\,\mathcal{N}_B\,\frac{M^2z^4(1-z)^2}{\mathcal{K}_1^2\mathcal{K}_2^3}\Big[z^4(5-8z+2z^2)\kperp^4\\
&\hspace{6.2em}-2M^2z^2(1-z)(1+z-z^2)\kperp^2-M^4(1-z)^3(1+z)\Big],\\
d^{[H_{1LL}^\perp]}_{{}^3P_0^{[8]}}&=64\,\mathcal{N}_B\,\frac{M^4z^7(1-z)^3\big[M^2(1-z)-z(3-z)\kperp^2\big]}{\mathcal{K}_1^2\mathcal{K}_2^3},\\
d^{[D_{1LT}]}_{{}^3P_0^{[8]}}&=-24\,\mathcal{N}_B\,\frac{M^2z^7(1-z)^3\,\kperp^2\big[z^2\kperp^2-M^2(1-z)\big]}{\mathcal{K}_1^2\mathcal{K}_2^3},\\
d^{[H_{1LT}^\perp]}_{{}^3P_0^{[8]}}&=-96\,\mathcal{N}_B\,\frac{M^4z^7(1-z)^3\big(z^2\kperp^2-M^2(1-z)\big)}{\mathcal{K}_1^2\mathcal{K}_2^3},\\
d^{[H_{1LT}]}_{{}^3P_0^{[8]}}&=-24\,\mathcal{N}_B\,\frac{M^2z^5(1-z)^2\big[z^5\kperp^4+3M^2z^2(1-z)\kperp^2+M^4(1-z)^3\big]}{\mathcal{K}_1^2\mathcal{K}_2^3},\\
d^{[\Phi]}_{{}^3P_0^{[8]}}&=-48\,\mathcal{N}_B\,\frac{M^4z^7(1-z)^4\big(z\kperp^2-M^2\big)}{\mathcal{K}_1^2\mathcal{K}_2^3},\\
d^{[H_{1TT}]}_{{}^3P_0^{[8]}}&=-24\,\mathcal{N}_B\,M^2z^4(1-z)^2 \\
&\qquad\frac{\big[z^2(2-z)\kperp^2+M^2(1-z)^2\big]\big[z^3\kperp^2+M^2(1-z)(1+z)\big]}{\mathcal{K}_1^2\mathcal{K}_2^3},\\
d^{[H_{1TT}^{\perp\perp}]}_{{}^3P_0^{[8]}}&=96\,\mathcal{N}_B\,\frac{M^6z^8(1-z)^4}{\mathcal{K}_1^2\mathcal{K}_2^3}.
\end{aligned}
\label{eq:sp_PJ}
\end{equation}
Here we recall that $\langle \mathcal{O}^{\mathcal{Q}} ({}^3P_J^{[8]})\rangle=(2J+1)\langle\mathcal \mathcal{O}^{\mathcal{Q}}({}^3P_0^{[8]})\rangle$. The remaining six functions, $H_{1L}^\perp$, $D_{1T}^\perp$, $H_{1T}$, $H_{1T}^\perp$, $G_{1LT}$ and $G_{1TT}$, vanish.
Compared with ${}^{3}S_1^{[8]}$, the functions $H_{1LT}^\perp$ and $H_{1TT}^{\perp\perp}$ are nonzero, and no naive T-odd function appears.
As for ${}^{1}S_0^{[8]}$, there are no rapidity and UV divergences, hence $\mathcal{L}_n$ and $\mathcal{R}_\Lambda$ do not appear.
However, Eq.~\eqref{eq:sp_PJ} holds for $ \boldsymbol k_\perp \neq\boldsymbol 0$.
The four functions $D_1$, $G_{1L}$, $D_{1LL}$ and $H_{1TT}$, which are the only ones with a non-vanishing LO coefficient in the ${}^3S_1^{[8]}$, are singular at the point $(z,  \boldsymbol k_\perp) = (1, \boldsymbol 0)$.
In this region, for $\boldsymbol k_\perp \sim \Lambda_2$, the four coefficients reduce to 
\begin{equation}
    \left. d^{[F]}_{{}^3P_J^{[8]}} (z,\boldsymbol k_\perp)  \right|_{z\to1,\boldsymbol k_\perp \sim \Lambda_2} \simeq c_F \frac{32 \pi \mathcal{N}_B}{M^2} \, \frac{1}{1-z}
    \,\frac{\Lambda_2^2}{\pi (\boldsymbol k_\perp^2 + \Lambda_2^2)^2} \, ,
\label{eq:corner_divergence}
\end{equation}
which is not integrable at $z = 1$ since it produces an infrared pole, whose origin and treatment are discussed below.
The remaining eight non-vanishing functions are free of this singularity.

\subsection{Subtracted TMDFFs in transverse-momentum space}
\label{sec:kt_sub}

The subtraction procedure described in this section applies only to the $^3S_1^{[8]}$ channel, since it is the only channel starting at $\mathcal{O}(\as)$.
Its NLO real and virtual corrections therefore contain UV and rapidity divergences, which are removed by renormalization and by the soft function.
The $^1S_0^{[8]}$ and $^3P_J^{[8]}$ channels start at $\mathcal{O}(\alpha_s^2)$ and their contributions fall off as $1/\boldsymbol k_\perp^4$ and have no UV or rapidity divergences, so they need no soft function subtraction.
The infrared pole of the ${}^3P_J^{[8]}$ channel is instead removed by the renormalization of the ${}^3S_1^{[8]}$ LDME.
In the following, we will therefore discuss only the $^3S_1^{[8]}$ channel. 

Rapidity divergences are removed by the factor $R_g=S^{-1/2}$ of Eq.~\eqref{eq:renorm}, and the UV divergences by $Z_g$.
At one loop the adjoint soft function, e.g., the sum of its virtual and real parts in  Ref.~\cite{Echevarria:2015byo}, reads in momentum space
\begin{equation}
\begin{aligned}
  S(\boldsymbol k_\perp;\delta,\zeta)
  & = \delta^{(2)}(\boldsymbol k_\perp)
   \\
   & +a_s\bigg[\delta^{(2)}(\boldsymbol k_\perp)\Big(-2\, \veuv^{-2}
     +2 \, \veuv^{-1} \, \ln\frac{\delta^2\zeta}{\mu^2}+\frac{\pi^2}{6}\Big)
     \\
     &\qquad+2\mathcal{L}_1(\boldsymbol k_\perp,\mu)
     -2\ln\frac{\delta^2\zeta}{\mu^2}\,\mathcal{L}_0(\boldsymbol k_\perp,\mu)\bigg] .
  \label{eq:soft_full}
\end{aligned}
\end{equation}
At this order the convolution with $S^{-1/2}$ therefore amounts to subtracting $\tfrac12S^{(1)}$ times the LO coefficient.
Moreover, the UV counterterm is
\begin{equation}
    Z_g=1-a_s\left[\veuv^{-2}+\veuv^{-1}\left(\frac{\beta_0}{2C_A}+\ln\frac{\mu^2}{\zeta}\right)\right] \, ,
\end{equation}
which together with the renormalization of $\alpha_s$ in $d_{\rm LO}$ removes all poles.
The  $\ln\delta$ terms of Eq.~\eqref{eq:D1_unsub_full} cancel exactly against those of the soft factor.
The renormalized unpolarized TMDFF is then
\begin{align}
  d^{[D_1]}_{\Sone}(z,\boldsymbol k_\perp;\mu,\zeta)
  &= d_{\rm LO}\,\delta(1-z)\,\delta^{(2)}(\boldsymbol k_\perp)
   +a_s\,d_{\rm LO}\bigg\{\delta(1-z)\bigg[
     -\mathcal{L}_1(\boldsymbol k_\perp,\mu)
   \nonumber\\
  &\quad-\ln\frac{\mu^2}{\zeta}\,\mathcal{L}_0(\boldsymbol k_\perp,\mu)
     +\delta^{(2)}(\boldsymbol k_\perp)\Big(\frac{\beta_0}{C_A}\ln\frac{\mu^2}{M^2}
     +c_V-\frac{\pi^2}{12}\Big)\bigg]\\
     &\quad
     +\mathcal{R}^{\rm real}(z,\boldsymbol k_\perp)\bigg\} ,
  \label{eq:D1_sub_full}
\end{align}
with $\mathcal{R}^{\rm real}$ given in Eq.~\eqref{eq:Rreal}.

The subtracted polarized functions $F \in \{ D_{1LL}, G_{1L}, H_{1TT} \}$ follow from Eq.~\eqref{eq:pol_full}:
\begin{equation}
  d^{[F]}_{\Sone}(z,\boldsymbol k_\perp;\mu,\zeta)
  = c_F\,d^{[D_1]}_{\Sone}(z,\boldsymbol k_\perp;\mu,\zeta)
   +\mathcal{N}\,\Delta T_F(z,\boldsymbol k_\perp) ,
  \label{eq:pol_sub_full}
\end{equation}
with $F\in\{D_{1LL},G_{1L},H_{1TT}\}$, $\Delta T_F$ given in Eqs.~\eqref{eq:dT_LL}--\eqref{eq:dT_H1TT}.
Finally, the coefficients of Eqs.~\eqref{eq:kT_noLO} and~\eqref{eq:kT_zero} are unchanged by the subtraction and by the renormalization at this order. 

Note that Eq.~\eqref{eq:D1_sub_full} depends on $\zeta$ only through the
$\delta(1-z)\ln(\mu^2/\zeta)\mathcal{L}_0$ term.
Hence, for any TMDFF, the $\zeta$-evolution is given by
\begin{equation}
  \frac{\partial\,d^{[F]}}{\partial\ln\zeta}
  = c_F\,a_s\,d_{\rm LO}\,\delta(1-z)\,\mathcal{L}_0(\boldsymbol k_\perp,\mu)
  = -\big[\widetilde{\mathcal{D}}_g\otimes d^{[F]}_{\rm LO}\big](z,\boldsymbol k_\perp) ,
  \label{eq:zeta_kT}
\end{equation}
where $\widetilde{\mathcal{D}}_g=-a_s\mathcal{L}_0(\boldsymbol k_\perp,\mu)$ is the momentum-space form of the one-loop rapidity anomalous dimension $\mathcal{D}_g=a_sL_T$. This shows explicitly that all polarized TMDFFs share the rapidity evolution of the unpolarized one, as expected.

\subsubsection{Infrared pole of the \texorpdfstring{${}^{3}P_J^{[8]}$}{3PJ[8]} channel}

The singular term identified after Eq.~\eqref{eq:sp_PJ} arises from the region in which the gluon emitted from the heavy-quark lines becomes soft, i.e., as already mentioned, when $z \to 1$ and $ \boldsymbol k_\perp \to  \boldsymbol 0$.
The soft gluon with a wavelength much larger than the size of the pair couples to it, at leading order in the multipole expansion, through its total  color charge and through its color dipole.
The former cannot change the state of the pair. In the ${}^3S_1^{[8]}$ channel its soft singularities cancel between real and virtual corrections, while in the ${}^3P_J^{[8]}$ channel it does not contribute at this order, since there is no Born amplitude $g^*\to Q\bar Q[{}^3P_J^{[8]}]$ and hence no virtual correction.
When a $L=1$ state ($P$-wave) is considered, the term linear in the relative momentum survives and it scales as the inverse gluon energy, i.e., the known chromoelectric dipole coupling in NRQCD.
This transition has no energy threshold, and the soft-gluon spectrum extends down to a gluon with energy tending to zero, producing the logarithmic divergence.

To isolate the divergence, we split the coefficient as
\begin{equation}
\begin{aligned}
    d^{[F]}_{{}^3P_0^{[8]}}(z, \boldsymbol k_\perp) & =  d^{[F]}_{\rm sub}(z,  \boldsymbol k_\perp) + c_F \frac{32 \pi \mathcal{N}_B}{M^2} \, \frac{1}{1-z} \, \frac{\Lambda_2^2}{\pi( \boldsymbol k_\perp^2 + \Lambda_2^2)^2} \; ,
\end{aligned}
\label{eq:sub}
\end{equation}
where the second term is the soft limit of Eq.~\eqref{eq:corner_divergence}, and the first term
is integrable over the full range of $z$ and $ \boldsymbol k_\perp$, defined by Eq.~\eqref{eq:sub}.
We observe that the second term can be written in terms of
\begin{equation}
     \frac{\Lambda_2^2}{\pi( \boldsymbol k_\perp^2 + \Lambda_2^2)^2} = \mathcal{Q}_{\Lambda_2}( \boldsymbol k_\perp) + \delta^{(2)}( \boldsymbol k_\perp) \, ,
\end{equation}
so that the term proportional to $\mathcal{Q}_{\Lambda_2}( \boldsymbol k_\perp)$ is a well-defined distribution, while the piece proportional to $\delta^{(2)}( \boldsymbol k_\perp)$ is not integrable at $z = 1$.
Regularizing this divergence by dimensional regularization, we obtain the following
\begin{equation}
\begin{aligned}
    d^{[F]}_{{}^3P_0^{[8]}}(z, \boldsymbol k_\perp) & =  d^{[F]}_{\rm sub}(z,  \boldsymbol k_\perp) + c_F \frac{32 \pi \mathcal{N}_B}{M^2}  \frac{\mathcal{Q}_{\Lambda_2}( \boldsymbol k_\perp)}{1-z} \\
    & + c_F \frac{32 \pi \mathcal{N}_B}{M^2} \delta^{(2)}( \boldsymbol k_\perp) \left[ \frac{1}{(1-z)_+} - \frac{\delta(1-z)}{2} \left( \frac{1}{\ve} + \ln \frac{\mu^2}{M^2} \right) \right] \, ,
\end{aligned}
\label{eq:sub_2}
\end{equation}
where here $\mu$ is the scale of dimensional regularization, and we used the $\overline{\text{MS}}$ scheme.
The residue is evaluated in four dimensions.
Keeping the $d$-dimensional orbital average: $3(d-2)/(2(d-1)) = 1 - \epsilon/3$ would add $\tfrac{16}{3} c_F\,\delta(1-z)\,\delta^{(2)}(\boldsymbol k_\perp)\,B_F\alpha_s^2/M^5$, which accounts for the difference in  Ref.~\cite{Ma:2014eja}.
The two conventions differ by a finite renormalization of the ${}^3S_1^{[8]}$ LDME.
Since the total TMDFF does not depend on the convention, $\langle\mathcal O({}^3S_1^{[8]})\rangle=\langle\mathcal O({}^3S_1^{[8]})\rangle_d+16B_F\alpha_s\langle\mathcal O({}^3P_0^{[8]})\rangle/(\pi M^2)$, where the subscript $d$ denotes the convention with the $d$-dimensional orbital average. This relation has to be used when our ${}^3P_J^{[8]}$ coefficients are combined with LDMEs extracted in that convention.

Moreover, producing a ${}^3P_J^{[8]}$ pair together with a gluon softer than the binding energy (which is part of the hadronization process) is indistinguishable from producing a ${}^3S_1^{[8]}$ pair that radiates this gluon during the hadronization (information encoded in the ${}^3S_1^{[8]}$ LDME by definition in NRQCD).
Accordingly, the pole is removed by the renormalization of the ${}^3S_1^{[8]}$ operator, which mixes with the ${}^3P_{J}^{[8]}$ operators~\cite{Bodwin:1994jh,Petrelli:1997ge}.
In dimensional regularization and in the $\overline{\text{MS}}$ scheme, the renormalized matrix element obeys
\begin{equation}
    \langle \mathcal{O}({}^3S_1^{[8]}) \rangle_{\rm bare} = \langle \mathcal{O}({}^3S_1^{[8]}) \rangle (\mu_\Lambda) + \frac{48 B_F \alpha_s}{\pi M^2} \left( \frac{1}{\ve} + \ln \frac{\mu^2}{\mu_\Lambda^2} \right) \langle \mathcal{O}({}^3P_0^{[8]}) \rangle \, ,
\end{equation}
where $\mu_\Lambda$ is the NRQCD factorization scale introduced above, which separates gluons attributed to the matrix element from those attributed to the SDC.
Inserting this equation into the LO term of the ${}^3S_1^{[8]}$ channel generates a contribution proportional to $\langle \mathcal{O}({}^3P_0^{[8]}) \rangle$ which is exactly the residue of the pole in Eq.~\eqref{eq:sub_2}.
In dimensional regularization, both poles are the same one, the scale $\mu$ drops out, and $\ln(\mu^2/M^2)$ is replaced by $\ln(\mu_\Lambda^2/M^2)$.
Absorbing this contribution into the ${}^3P_J^{[8]}$ coefficient, the renormalized coefficient reads
\begin{equation}
\begin{aligned}
    d^{[F]}_{{}^3P_0^{[8]}}(z, \boldsymbol k_\perp;\mu_\Lambda) & =  d^{[F]}_{\rm sub}(z,  \boldsymbol k_\perp) + c_F \frac{32 \pi \mathcal{N}_B}{M^2}  \frac{\mathcal{Q}_{\Lambda_2}( \boldsymbol k_\perp)}{1-z} \\
    & + c_F \frac{32 \pi \mathcal{N}_B}{M^2} \delta^{(2)}( \boldsymbol k_\perp) \left[ \frac{1}{(1-z)_+} - \frac{\delta(1-z)}{2} \,  \ln \frac{\mu_\Lambda^2}{M^2}  \right] \, .
\end{aligned}
\label{eq:sub_final}
\end{equation}
We remind that this discussion is for $F \in \{ D_1, G_{1L}, D_{1LL}, H_{1TT} \}$ since the remaining functions are free of this singularity so require no renormalization. 
We also note that we directly developed this discussion for ${}^3P_0^{[8]}$ but it is direct to generalize to ${}^3P_J^{[8]}$ by using heavy-quark spin symmetry.

\subsection{Subtracted TMDFFs in impact-parameter space}
\label{sec:bt_results}

We now transform the renormalized coefficients of Sec.~\ref{sec:kt_sub} for the $^3S_1^{[8]}$ channel and those of Secs.~\ref{sec:1S08-kt} and~\ref{sec:3PJ8-kt} to impact-parameter space with Eq.~\eqref{eq:ft}.
The transform depends on the rank $n$ of the irreducible tensor in $k_T$ that multiplies each function.
Let $\mathcal{T}_n(k_T)$ be a symmetric traceless tensor of rank $n$ built from $k_T$. Then~\cite{Boer:2011xd}
\begin{equation}
  \int d^2\boldsymbol k_\perp\,e^{i\boldsymbol k_\perp\cdot\boldsymbol b_T}\,
  \frac{\mathcal{T}_n(k_T)}{M^n}\,F(z,\boldsymbol k_\perp^2)
  = \frac{(iM)^n}{n!}\,\mathcal{T}_n(b_T)\,\widetilde F^{(n)}(z,b_T) ,
  \label{eq:FT_rank_n}
\end{equation}
with $\widetilde F^{(n)}$ the Bessel-weighted transform.
For $n=0$ this is the ordinary Fourier transform, which we denote simply by $\widetilde F$.
The non-vanishing structures of Eqs.~\eqref{eq:GU}--\eqref{eq:GTT} become
\begin{equation}
  \begin{aligned}
       \widetilde\Gamma^{\lambda\sigma}_U
  &= -g_T^{\lambda\sigma}\widetilde D_1-\frac{M^2}{2}\,b_T^{\lambda\sigma}\,\widetilde H_1^{\perp(2)} \, ,\\
  \widetilde\Gamma^{\lambda\sigma}_L 
  & = S_L\,i\epsilon_T^{\lambda\sigma}\,\widetilde G_{1L}  - \frac{M^2 \epsilon_{T \alpha}^{\{\lambda } b_T^{\sigma \} \alpha}S_L }{4} \tilde{H}_{1L}^{\perp (2)} \, ,\\
  \widetilde\Gamma^{\lambda\sigma}_T & = -i M g_T^{\lambda \sigma} \epsilon_T^{S_T b_T} \tilde{D}_{1T}^{\perp (1)} -M\,\epsilon_T^{\lambda\sigma}\,(b_T\cdot S_T)\,\widetilde G_{1T}^{(1)} \\
  &- \frac{iM ( \epsilon_T^{b_T \{ \lambda} S_T^{\sigma \}} +  \epsilon_T^{S_T \{ \lambda} b_T^{\sigma \}})}{4} \tilde{H}_{1T}^{(1)} + \frac{i M^3 \epsilon_{T \alpha}^{\{ \lambda} b_T^{\sigma \} \alpha S_T} }{12} \tilde{H}_{1T}^{\perp (3)},
   \\
  \widetilde\Gamma^{\lambda\sigma}_{LL}
  &= S_{LL}\Big[-g_T^{\lambda\sigma}\widetilde D_{1LL}-\frac{M^2}{2}\,b_T^{\lambda\sigma}\,\widetilde H_{1LL}^{\perp(2)}\Big] ,
   \\
  \widetilde\Gamma^{\lambda\sigma}_{LT}
  &=  -iM\,g_T^{\lambda\sigma}(b_T\cdot S_{LT})\,\widetilde D_{1LT}^{(1)}
  - M \epsilon_T^{\lambda \sigma} \epsilon_T^{S_{LT} b_T} \tilde{G}_{1LT}^{(1)}
     \\
     & +iM S_{LT}^{\{\lambda}b_T^{\sigma\}} \widetilde H_{1LT}^{(1)} - \frac{i M^3 b_{T}^{\lambda \sigma \alpha} S_{LT \alpha}}{6} \widetilde{H}_{1 LT}^{\perp (3)},
   \\
  \widetilde\Gamma^{\lambda\sigma}_{TT}
  &= -\frac{M^2 g_T^{\lambda \sigma} b_T^{\alpha \beta} S_{TT \alpha \beta}}{2}\, \widetilde{\Phi}^{(2)}
  - \frac{i M^2 \epsilon_T^{\lambda \sigma} \epsilon_{T \gamma}^\beta b_T^{\gamma \alpha} S_{TT \alpha \beta} }{2} \widetilde{G}_{1TT}^{(2)}
     \\
     & 
     +S_{TT}^{\lambda\sigma}\,\widetilde H_{1TT} + \frac{M^4 b_T^{\lambda \sigma \alpha \beta} S_{TT \alpha \beta} }{24} \widetilde{H}_{1TT}^{\perp \perp (4)},
  \label{eq:Gamma_b}
  \end{aligned}  
\end{equation}
where $b_T^{\lambda\sigma}=b_T^\lambda b_T^\sigma-\tfrac12b_T^2g_T^{\lambda\sigma}$ and $S_{TT}^{bb}=b_{T\alpha}S_{TT}^{\alpha\beta}b_{T\beta}$, and as in momentum space, we write $\widetilde F_n=\widetilde d^{[F]}\,\langle\mathcal{O}^\mathcal{Q}(n)\rangle$.

Using the dictionary of Sec.~\ref{sec:kt_unsub} and the transforms in Eqs.~\eqref{eq:RLambda} and~\eqref{eq:QLambda}, Eq.~\eqref{eq:D1_sub_full} becomes
\begin{align}
  \widetilde d^{[D_1]}_{\Sone}(z,b_T;\mu,\zeta)
  & = d_{\rm LO}\,\delta(1-z)
   \nonumber\\
   & +a_s\,d_{\rm LO}\bigg\{\delta(1-z)\Big[-\frac{L_T^2}{2}+L_T\ln\frac{\mu^2}{\zeta}
     +\frac{\beta_0}{C_A}\ln\frac{\mu^2}{M^2}+c_V-\frac{\pi^2}{12}\Big]
   \nonumber\\
  &\quad-\mathcal{P}_{gg}(z)\Big[L_T-\ln\frac{\mu^2}{M^2}-2A(z,b_T)\Big] \label{eq:D1_b}
     \\ &\quad -2(2-2z+z^2)\,b_TM\,B(z,b_T)
   \nonumber\\
  &\quad-\frac{2z(2-2z+z^2)}{(1-z)_+}
     -\frac{4(1-z+z^2)^2}{z}\left(\frac{\ln(1-z)}{1-z}\right)_+\bigg\} \nonumber,
\end{align}
with
\begin{equation}
  A(z,b_T)=K_0(b_T\Lambda_2)+\ln\frac{b_T\Lambda_2 e^{\gamma_E}}{2}+\ln z ,
  \qquad
  B(z,b_T)=K_1(b_T\Lambda_2)-\frac{1}{b_T\Lambda_2} .
  \label{eq:AB}
\end{equation}
We note that both functions are regular at $z=1$, and for $b_TM\to0$, they behave as $A(z,b_T)\to\ln z$ and $b_TM B(z,b_T)\to0$.

From Eq.~\eqref{eq:pol_sub_full}, the Fourier transform of the functions $D_{1LL}$, $G_{1L}$ and $H_{1TT}$ are
\begin{equation}
  \widetilde d^{[F]}_{\Sone}(z,b_T;\mu,\zeta)
  = c_F\,\widetilde d^{[D_1]}_{\Sone}(z,b_T;\mu,\zeta)+\Delta\widetilde d^{[F]}_{\Sone}(z,b_T) \, .
  \label{eq:pol_b}
\end{equation}
where the remainders are the Fourier transforms of $\mathcal{N}\Delta T_F$, Eqs.~\eqref{eq:dT_LL}--\eqref{eq:dT_H1TT}:
\begin{align}
  \Delta\widetilde d^{[D_{1LL}]}_{\Sone}
  &= -6a_s\,d_{\rm LO}\,\frac{(1-z)^2}{z^2}\;b_TM\,K_1(b_T\Lambda_2) ,
  \label{eq:dd_LL}\\
  \Delta\widetilde d^{[G_{1L}]}_{\Sone}
  &= 3a_s\,d_{\rm LO}\,\frac{(1-z)^2}{z}\Big[(2-z)\,b_TM\,K_1(b_T\Lambda_2)+2(1-z)\,K_0(b_T\Lambda_2)\Big] ,
  \label{eq:dd_G1L}\\
  \Delta\widetilde d^{[H_{1TT}]}_{\Sone}
  &= 3a_s\,d_{\rm LO}\,(1-z)\Big[\frac{2(1+z^2)}{z}\,K_0(b_T\Lambda_2)-(1-z)\,b_TM\,K_1(b_T\Lambda_2)\Big] .
  \label{eq:dd_H1TT}
\end{align}
To obtain them we used
\begin{equation}
  \int d^2\boldsymbol k_\perp\,\frac{e^{i\boldsymbol k_\perp\cdot\boldsymbol b_T}}{\mathcal{K}_2}=\frac{2\pi}{z^2}K_0(b_T\Lambda_2) ,
  \qquad
  \int d^2\boldsymbol k_\perp\,\frac{e^{i\boldsymbol k_\perp\cdot\boldsymbol b_T}}{\mathcal{K}^2_2}=\frac{\pi b_T}{z^4\Lambda_2}K_1(b_T\Lambda_2) \, .
  \label{eq:FT_master}
\end{equation}
We notice that the remainders are finite, vanish for $z\to1$, and carry no dependence on $\mu$ or $\zeta$.
The dependence on the renormalization and rapidity scales of all four functions is therefore that of $D_1$, multiplied by $c_F$. Note also that only the scale $\Lambda_2=M(1-z)/z$ appears.
The second scale $M\sqrt{1-z}/z$ of the individual diagrams, Eq.~\eqref{eq:Lambdas}, cancels in the sum of topologies, for the polarized functions as well as for $D_1$.

For the functions in Eq.~\eqref{eq:kT_noLO}, by using
\begin{equation}
  \int_0^\infty dk\,\frac{k^{n+1}J_n(b_Tk)}{(k^2+\Lambda_2^2)^2}=\frac12\,b_T\Lambda_2^{n-1}K_{n-1}(b_T\Lambda_2) ,
  \qquad n=1,2,
  \label{eq:bessel_master}
\end{equation}
we find
\begin{align}
  \widetilde d^{[H_1^\perp](2)}_{\Sone} = \widetilde d^{[H_{1LL}^\perp](2)}_{\Sone}
  &= 8\,a_s\,d_{\rm LO}\,\frac{(1-z)^2\,K_1(b_T\Lambda_2)}{b_TM} ,
  \nonumber\\
  \widetilde d^{[\Phi](2)}_{\Sone}
  &= 6\,a_s\,d_{\rm LO}\,\frac{(1-z)^2\,K_1(b_T\Lambda_2)}{z^2\,b_TM} ,
  \nonumber\\
  \widetilde d^{[G_{1T}](1)}_{\Sone}
  &= -3\,a_s\,d_{\rm LO}\,\frac{(1-z)(2-z)}{z}\,K_0(b_T\Lambda_2) ,\label{eq:b_noLO}
  \\
  \widetilde d^{[D_{1LT}](1)}_{\Sone}
  &= -6\,a_s\,d_{\rm LO}\,\frac{(1-z)^2}{z^2}\,K_0(b_T\Lambda_2),
  \nonumber\\
  \widetilde d^{[H_{1LT}](1)}_{\Sone}
  &= 3\,a_s\,d_{\rm LO}\,(1-z)\,K_0(b_T\Lambda_2) \, \nonumber.
\end{align}
We find that the coefficients in Eq.~\eqref{eq:b_noLO} are also finite.
They are independent of $\mu$ and $\zeta$ and contain neither $\delta(1-z)$ nor plus distributions, and they decay exponentially for $b_T\Lambda_2\gg1$.
For $b_TM\ll1$, the rank-1 functions grow logarithmically and the rank-2 functions as $1/(b_TM)^2$.
The rank-1 functions multiply structures proportional to $b_TM$, so their contributions are power suppressed and have no twist-2 collinear counterpart. The $1/(b_TM)^2$ growth of the rank-2 functions instead compensates the factor $M^2b_T^{\lambda\sigma}$ of their structures, so that $H_1^\perp$, $H_{1LL}^\perp$ and $\Phi$ contribute at leading power and match onto the twist-2 collinear FFs, see Sec.~\mbox{\ref{sec:ope}}.
For the $^1S_0^{[8]}$ and $^3P_J^{[8]}$ channels the TMDs in $b_T$-space follow directly from Eq.~\eqref{eq:FT_rank_n}, applying the same procedure as for $^3S_1^{[8]}$.

For the spin singlet case, the only non-zero TMDs are those in the correlator $\widetilde\Gamma^{\lambda\sigma}_U $ in Eq.~\eqref{eq:Gamma_b}, i.e.,
\begin{equation}
\begin{aligned}
    \widetilde d^{[D_1]}_{\Szero}(z,b_T)=& \,\frac{B_F\,\alpha_s^2}{M^3}\Big[z\,b_T\Lambda_1K_1(b_T\Lambda_1)
+2z(1-z)\,b_T\Lambda_2\,K_1(b_T\Lambda_2) \\
&+4(1-z)\big(K_0(b_T\Lambda_1)-K_0(b_T\Lambda_2)\big)\Big],
\end{aligned}
\end{equation}
\begin{equation}
    \begin{aligned}
        \widetilde d^{[H_1^\perp](2)}_{\Szero}(z,b_T)=& \,\frac{8\,B_F\,\alpha_s^2}{b_T^2 M^5}(1-z)
\Big[2(1-z)\,K_2(b_T\Lambda_2)- z\,b_T\Lambda_2\,K_1(b_T\Lambda_2)\\
&-2\,K_2(b_T\Lambda_1)\Big].
    \end{aligned}
\end{equation}

After the renormalization of $\langle {}^3S_1^{[8]} \rangle$, the ${}^3P_J^{[8]}$ contributions to $D_1$, $G_{1L}$, $D_{1LL}$ and $H_{1TT}$ share the following structure 
\begin{equation}
\tilde d^{[F]}_{\PJ}(z,b_T;\mu_\Lambda)=\frac{B_F\,\alpha_s^2}{M^5}\left\{32\,c_F\left[\frac{1}{(1-z)_+}-\frac{1}{2}\,\delta(1-z)\ln\frac{\mu_\Lambda^2}{M^2}\right]+\tilde R_F(z,b_T)\right\},
\label{eq:PJ_bspace}
\end{equation}
where
\begin{align}
\tilde R_{D_1}={}&16(13-7z)\big[K_0(b_T\Lambda_1)-K_0(b_T\Lambda_2)\big]-8z^2\,b_T^2\Lambda_2^2\,K_0(b_T\Lambda_2) \nonumber\\
&+\frac{8z(5z^2-12z+11)}{1-z}\big[b_T\Lambda_2K_1(b_T\Lambda_2)-1\big]\nonumber\\
&+28z\big[b_T\Lambda_1K_1(b_T\Lambda_1)-1\big]-4(10z^2-21z+8),\\[4pt]
\tilde R_{G_{1L}}=-\Bigg[ {}&\frac{48(2z^2-5z+6)}{z}\big[K_0(b_T\Lambda_1)-K_0(b_T\Lambda_2)\big]-12z\,b_T^2\Lambda_2^2\,K_0(b_T\Lambda_2)\nonumber\\
&+24(2-z)\big[b_T\Lambda_1K_1(b_T\Lambda_1)-1\big]+96(1-z)\\
&+\frac{24(3z^2-5z+4)}{1-z}\big[b_T\Lambda_2K_1(b_T\Lambda_2)-1\big]\nonumber\Bigg],\\[4pt]
\tilde R_{D_{1LL}}={}&\frac{32(18-30z+17z^2-2z^3)}{z^2}\big[K_0(b_T\Lambda_1)-K_0(b_T\Lambda_2)\big]\nonumber\\
&-8(z^2-3z+3)\,b_T^2\Lambda_2^2\,K_0(b_T\Lambda_2)+\frac{16(1-z)(z^2-5z+18)}{z}\\
&+\frac{16(z^4-6z^3+22z^2-27z+12)}{z(1-z)}\big[b_T\Lambda_2K_1(b_T\Lambda_2)-1\big]\nonumber\\
&+\frac{16(z^2-6z+6)}{z}\big[b_T\Lambda_1K_1(b_T\Lambda_1)-1\big]\nonumber,\\[4pt]
\tilde R_{H_{1TT}}={}&-\frac{48(2z^2-2z+3)}{z^2}\big[K_0(b_T\Lambda_1)-K_0(b_T\Lambda_2)\big]
\nonumber\\
&-\frac{12(z^4-z^3+5z^2-5z+4)}{z(1-z)}\big[b_T\Lambda_2K_1(b_T\Lambda_2)-1\big] \\
&-\frac{24}{z}\big[b_T\Lambda_1K_1(b_T\Lambda_1)-1\big]+\frac{12(z^3+5z-6)}{z}\nonumber\\
&+6(1+z^2)\,b_T^2\Lambda_2^2\,K_0(b_T\Lambda_2)\nonumber.
\end{align}
The singular part at $z\to1$ contains the plus distribution and the $\delta(1-z)$ term with the IR pole. It is independent of $b_T$, hence the same renormalization of $\langle\mathcal O({}^3S_1^{[8]})\rangle$ removes the pole at every $b_T$.
The functions $\tilde R_F$ are the regular ones, which contain the $b_T$ dependence.

The remaining ${}^3P_J^{[8]}$ functions read
\begin{align}
\widetilde d^{[H_1^\perp](2)}_{\Pzero}&=\frac{B_F\,\alpha_s^2}{M^5} \frac{32(1-z)}{(b_T M)^3}\Big\{b_TM(1-z)(b_T^2M^2+14)K_0(b_T\Lambda_2)\\
&\qquad+7\big[4z-b_T^2M^2(1-z)\big]K_1(b_T\Lambda_2) -14b_TM\,K_2(b_T\Lambda_1)\Big\}\nonumber,\\[4pt]
\widetilde d^{[G_{1T}](1)}_{\Pzero}&= - \frac{B_F\,\alpha_s^2}{M^5}\frac{12}{b_TM z}\Big\{(1-z)\big[b_T^2M^2(1-z)+4(6-z)\big]K_1(b_T\Lambda_2)
\nonumber\\
&\qquad-4(6-z)\sqrt{1-z}\,K_1(b_T\Lambda_1)-2b_TM(1-z)\big[2K_0(b_T\Lambda_1)\\
&\qquad+(4-z)K_0(b_T\Lambda_2)\big]\Big\},\nonumber\\[4pt]
\widetilde d^{[H_{1LL}^\perp](2)}_{\Pzero}&=\frac{B_F\,\alpha_s^2}{M^5}\frac{32}{(b_T M)^3 z^2}\Big\{b_TM(1-z)^2\big[b_T^2M^2(z^2-3z+3)\nonumber\\
&\qquad+8(z^2-3z+9)\big]K_0(b_T\Lambda_2)\nonumber\\
&\qquad-8b_TM(1-z)(z^2-3z+9)K_0(b_T\Lambda_1)
\nonumber\\
&\qquad+4(1-z)\big[4z(z^2-3z+9)\\
&\qquad-b_T^2M^2(1-z)(z^2-3z+6)\big]K_1(b_T\Lambda_2)\nonumber\\
&\qquad-4\sqrt{1-z}\big[3b_T^2M^2(1-z)\nonumber\\
&\qquad+4z(z^2-3z+9)\big]K_1(b_T\Lambda_1)\Big\}\nonumber\\
\widetilde d^{[D_{1LT}](1)}_{\Pzero}&=\frac{B_F\,\alpha_s^2}{M^5}\frac{6(1-z)^{3/2}}{b_T M z^2}\Big\{(2-z)\big[24K_1(b_T\Lambda_1)\nonumber\\
&\qquad-\sqrt{1-z}\,\big(b_T^2M^2(1-z)+24\big)K_1(b_T\Lambda_2)\big]\\
&\qquad+4b_TM\sqrt{1-z}\big[2K_0(b_T\Lambda_1)+(4-3z)K_0(b_T\Lambda_2)\big]\Big\}\nonumber\\
\widetilde d^{[H_{1LT}](1)}_{\Pzero}&=\frac{B_F\,\alpha_s^2}{M^5} \frac{6}{b_TM z^2}\Big\{(1-z)\big[b_T^2M^2(1-z)(z^2-z+2)\nonumber\\
&\qquad+8(2z^2-3z+6)\big]K_1(b_T\Lambda_2)\\
&\qquad-8(2z^2-3z+6)\sqrt{1-z}\,K_1(b_T\Lambda_1)
\nonumber\\
&\qquad-4b_TM(1-z)\big[2K_0(b_T\Lambda_1)+(2z^2-3z+4)K_0(b_T\Lambda_2)\big]\Big\}\nonumber,\\
\widetilde d^{[H_{1LT}^\perp](3)}_{\Pzero}&=\frac{B_F\,\alpha_s^2}{M^5}\frac{144}{(b_T M)^5z^2}\Big\{(1-z)\big[b_T^4M^4(1-z)^2(2-z)\nonumber\\
&\qquad+8b_T^2M^2(1-z)(2z^2-11z+6)+64z^2(6-z)\big]K_1(b_T\Lambda_2)\nonumber\\
&\qquad-4b_TM(1-z)^2\big[b_T^2M^2(1-z)(4-z)-8z(6-z)\big]K_0(b_T\Lambda_2)\\
&\qquad-8b_TM(1-z)\big[b_T^2M^2(1-z)+4z(6-z)\big]K_0(b_T\Lambda_1)\nonumber\\
&\qquad-8\sqrt{1-z}\big[b_T^2M^2(1-z)(6+z)+8z^2(6-z)\big]K_1(b_T\Lambda_1)\Big\}\nonumber,\\
\widetilde d^{[\Phi](2)}_{\Pzero}&=\frac{B_F\,\alpha_s^2}{M^5} \frac{24}{(b_T M)^3 z^2}\Big\{b_TM(1-z)^2\big[b_T^2M^2(z^2-z+1) \nonumber\\
&\qquad+8(z^2-z+3)\big]K_0(b_T\Lambda_2) \nonumber\\
&\qquad-8b_TM(1-z)(z^2-z+3)K_0(b_T\Lambda_1)\nonumber\\
&\qquad+4(1-z)\big[4z(z^2-z+3)\\
&\qquad-b_T^2M^2(1-z)(z^2-z+2)\big]K_1(b_T\Lambda_2)\nonumber\\
&\qquad-4\sqrt{1-z}\big[b_T^2M^2(1-z)+4z(z^2-z+3)\big]K_1(b_T\Lambda_1)\Big\}\nonumber,\\
\widetilde d^{[H_{1TT}^{\perp\perp}](4)}_{\Pzero}&=\frac{B_F\,\alpha_s^2}{M^5}\frac{576}{(b_T M)^7z^2}\Big\{b_TM(1-z)^2\big[b_T^4M^4(1-z)^2
\nonumber\\
&\qquad -8b_T^2M^2(1-z)(7z-3)+576z^2\big]K_0(b_T\Lambda_2)\nonumber\\
&\qquad+2(1-z)\big[b_T^4M^4(1-z)^2(5z-4)\\
&\qquad-32b_T^2M^2z(1-z)(4z-3)+576z^3\big]K_1(b_T\Lambda_2)\nonumber\\
&\qquad-8b_TM(1-z)\big[b_T^2M^2(1-z)(2z+3)+72z^2\big]K_0(b_T\Lambda_1)\nonumber\\
&\qquad-4\sqrt{1-z}\big[b_T^4M^4(1-z)^2+8b_T^2M^2z(1-z)(z+6)\nonumber\\
&\qquad+288z^3\big]K_1(b_T\Lambda_1)\Big\}\nonumber 
\end{align}
They are free of infrared poles and need no plus distributions. Since these are $P$-wave contributions, the coefficients scale as $M^{-5}$. The rest of the TMDFFs are zero, i.e., the T-odd TMDs
\begin{equation}
\widetilde d^{[H_{1L}^\perp](2)}_{\Pzero}=\widetilde d^{[D_{1T}^\perp](1)}_{\Pzero}=\widetilde d^{[H_{1T}](1)}_{\Pzero}=\widetilde d^{[H_{1T}^\perp](3)}
_{\Pzero}=\widetilde d^{[G_{1LT}](1)}_{\Pzero}=\widetilde d^{[G_{1TT}](2)}_{\Pzero}=0 .
\end{equation}
\subsection{Twist-2 collinear functions}
\label{sec:collinear}

Integrating the gluon TMD correlator over the transverse momentum of the gluon removes every structure that depends on the azimuth of $k_T$, so in Eqs.~\eqref{eq:GU}--\eqref{eq:GTT}, only the structures $-g_T^{\lambda\sigma}$, $i\epsilon_T^{\lambda\sigma}S_L$, $-g_T^{\lambda\sigma}S_{LL}$ and $S_{TT}^{\lambda\sigma}$ survive.
Therefore, at leading twist (or twist-2), exactly four gluon FFs into a spin-1 hadron exist~\cite{Ji:1993vw,Bacchetta:2000jk,Boer:2016xqr}: $D_1(z)$, $G_{1L}(z)$, $D_{1LL}(z)$ and $H_{1TT}(z)$.
The remaining fourteen TMDFFs have no twist-2 collinear counterpart, since their $k_T$-weighted moments are related to collinear functions of higher twist.

We define the bare collinear correlator as
\begin{align}
  \Delta^{\lambda\sigma}_{g\to H}&(z;S)
  = z^{2-2\epsilon}\int d^{d-2}\boldsymbol k_\perp\,
    \Delta^{\lambda\sigma}_{g\to H}(z,\boldsymbol k_\perp;S)
  \nonumber\\
  &= \frac13\Big[-g_T^{\lambda\sigma}D_1(z)+i\epsilon_T^{\lambda\sigma}S_L\,G_{1L}(z)
     -g_T^{\lambda\sigma}S_{LL}\,D_{1LL}(z)+S_{TT}^{\lambda\sigma}H_{1TT}(z)\Big] \, ,
  \label{eq:coll_corr}
\end{align}
which is the matrix element of Eq.~\eqref{eq:corr_b} at $b_T=0$, times the factor $z^{2-2\epsilon}$.

By Eq.~\eqref{eq:pol_full}, the bare collinear coefficients satisfy
\begin{equation}
  d^{[F]}_{\Sone}(z)=c_F\,d^{[D_1]}_{\Sone}(z)
  +z^{2-2\epsilon}\!\int d^{d-2}\boldsymbol k_\perp\,\mathcal{N}\,
   \big[\Delta T_F(z,\boldsymbol k_\perp)+\epsilon\,\delta T_F(z,\boldsymbol k_\perp)+\mathcal{O}(\epsilon^2)\big] ,
  \label{eq:coll_method}
\end{equation}
where $\Delta T_F$ are the four-dimensional remainders of Eqs.~\eqref{eq:dT_LL}--\eqref{eq:dT_H1TT}, and $\epsilon\,\delta T_F$ is their $\mathcal{O}(\epsilon)$ part, averaged over the directions of $\boldsymbol k_\perp$ in $d-2$ dimensions. Since the remainder has no IR-sensitive region, $\delta T_F$ contributes only through its UV tail, $\delta T_F\to b_F(z)/\boldsymbol k_\perp^2$ for $\boldsymbol k_\perp^2\to\infty$, which yields the finite term $\pi b_F(z)$. We obtained $b_F$ by evaluating the projected real-emission diagrams of Sec.~\ref{sec:outline} in integer dimensions, with the polarization structures confined to the physical subspace.
The result is most transparently expressed through the $d$-dimensional real-emission kernels, defined by the large-$\boldsymbol k_\perp$ limit $\Delta^{[F]}(z,\boldsymbol k_\perp)\to(\alpha_s/2\pi^2\boldsymbol k_\perp^2)\,c_F\,d_{\rm LO}\,C_A\,\widehat P_F^{(d)}(z)$:
\begin{equation}
\begin{aligned}
  \widehat P^{(d)}_{D_1}(z) &= \mathcal{P}_{gg}(z) ,
  &\qquad
  \widehat P^{(d)}_{G_{1L}}(z) &= 2\Big[\frac{1}{1-z}+1-2z+2\epsilon(1-z)\Big] ,
  \\
  \widehat P^{(d)}_{D_{1LL}}(z) &= \mathcal{P}_{gg}(z) 
  ,
  &\qquad
  \widehat P^{(d)}_{H_{1TT}}(z) &= \frac{2z}{1-z} ,
\end{aligned}
\label{eq:kernels_d}
\end{equation}
for $z<1$.
The $\mathcal{O}(\epsilon)$ term of the helicity kernel coincides with the known result for $\Delta P_{gg}$ in the HVBM scheme~\cite{tHooft:1972fi,Breitenlohner:1977hr,Vogelsang:1996im,Vogelsang:1995vh}, which is a nontrivial check of the procedure.
The other kernels have no $\mathcal{O}(\epsilon)$ term.

After the cancellation of the IR poles and of the $\ln\delta$ terms between real and virtual contributions, we remove the remaining UV poles in the $\overline{\rm MS}$ scheme, including the renormalization of $\alpha_s$ in $d_{\rm LO}$.
All four renormalized coefficients can then be written in the common form
\begin{equation}
    \begin{aligned}
  \frac{1}{z^2}\,d^{[F]}_{\Sone}&(z;\mu)
  = c_F\,d_{\rm LO}\bigg\{\delta(1-z)
   +a_s\bigg[\Big(\widehat P_F(z)+\frac{\beta_0}{C_A}\,\delta(1-z)\Big)\ln\frac{\mu^2}{M^2}\\
  &\quad
    +c_V\,\delta(1-z)-\frac{2z(2-2z+z^2)}{(1-z)_+}
   -2\,p_F(z)\left(\frac{\ln(1-z)}{1-z}\right)_+
   +\hat r_F(z)\bigg]\bigg\} ,
  \label{eq:coll_result}
\end{aligned}
\end{equation}
with $c_V$ given in Eq.~\eqref{eq:cV}. 
Here $\widehat P_F(z)=p_F(z)/(1-z)_+$ is the four-dimensional kernel.
The functions $p_F$ and $\hat r_F$ are collected in Tab.~\ref{tab:coll}.
Explicitly,
\begin{align}
  \widehat P_{D_1}=\widehat P_{D_{1LL}}&=\mathcal{P}_{gg}=2\Big[\frac{z}{(1-z)_+}+\frac{1-z}{z}+z(1-z)\Big] ,
  \nonumber\\
  \widehat P_{G_{1L}}&=2\Big[\frac{1}{(1-z)_+}+1-2z\Big] ,
  \qquad
  \widehat P_{H_{1TT}}=\frac{2z}{(1-z)_+} .
  \label{eq:kernels_4}
\end{align}

\begin{table}[t]
\centering
\renewcommand{\arraystretch}{1.9}
\begin{tabular}{c c c c c}
\toprule
$F$ & $c_F$ & $p_F(z)$ & $\hat r_F(z)$ & $r_F(z)$ \\
\midrule
$D_1$ & $1$ & $\dfrac{2(1-z+z^2)^2}{z}$ & $0$ & $0$ \\
$D_{1LL}$ & $1$ & $\dfrac{2(1-z+z^2)^2}{z}$ & $-\dfrac{6(1-z)}{z}$ & 0 \\
$G_{1L}$ & $-\tfrac32$ & $2(2-3z+2z^2)$ & $2z(1-z)$ & $-4(1-z)$ \\
$H_{1TT}$ & $-\tfrac32$ & $2z$ & $2z(1-z)$ & $0$ \\
\bottomrule
\end{tabular}
\caption{Functions entering the NLO collinear coefficients,
Eq.~\eqref{eq:coll_result}.
The kernels are written as
$\widehat P_F=p_F/(1-z)_+$, while the finite terms $\hat r_F$ and $r_F$ of the polarized functions refer to the $\overline{\rm MS}$ scheme, with the HVBM prescription for $\epsilon_T$ in the case of $G_{1L}$.
In particular, $r_F$ equals minus the $\mathcal{O}(\epsilon)$ coefficient of $\widehat P^{(d)}_F$ in Eq.~\eqref{eq:kernels_d}.
}
\label{tab:coll}
\end{table}
As a cross-check of these results, we consider the $\mu$-dependence of Eq.~\eqref{eq:coll_result}, which is given by the DGLAP evolution equation:
\begin{equation}
  \frac{d}{d\ln\mu^2}\,\frac{d^{[F]}_{\Sone}(z;\mu)}{z^2}
  = c_F\,d_{\rm LO}\,a_s\Big[\widehat P_F(z)+\frac{\beta_0}{C_A}\delta(1-z)\Big] \, ,
  \label{eq:DGLAP_check}
\end{equation}
where half of the $\beta_0$ term compensates the running of $\alpha_s$ in $d_{\rm LO}$, the derivative being taken at fixed $\alpha_s$.
The other half, together with $\widehat P_F$, reconstructs the LO time-like $g\to g$ kernels acting on the LO coefficient.
Hence each collinear function obeys the DGLAP equation appropriate to its gluon polarization, and $D_{1LL}$ evolves like $D_1$.

In the spin-singlet channel ${}^1S_0^{[8]}$, $D_1$ is the only function that survives the $ \boldsymbol k_\perp$ integration, since $H_1^\perp$ is a rank-2 function.
We find
\begin{equation}
    \frac{1}{z^2}d^{[D_1]}_{\Szero}(z)
=\frac{B_F \,\alpha_s^2}{M^3}\Big[\,3z-2z^2+2(1-z)\ln(1-z)\Big],
\end{equation}
which is finite as $z\to1$. 

After absorbing the infrared pole of the ${}^3P_J^{[8]}$ channel into the renormalized LDME $\langle\mathcal O({}^3S_1^{[8]})\rangle$, we find
\begin{equation} 
\frac{1}{z^2}\,d^{[F]}_{\Pzero}(z;\mu_\Lambda)=\frac{B_F \,\alpha_s^2}{M^5}\bigg\{32\,c_F\Big[\frac{z^2}{(1-z)_+}
-\frac12\,\delta(1-z)\ln\frac{\mu_\Lambda^2}{M^2}\Big]+R_F(z)\bigg\},
\label{eq:collPJ}
\end{equation}
with $c_{D_1}=c_{D_{1LL}}=1$, $c_{G_{1L}}=-\tfrac32$, $c_{H_{1TT}}=-\tfrac32$, and
\begin{equation}
\begin{aligned}
R_{D_1}(z)&=8(13-7z)\ln(1-z)+4z(29-10z),\\[2pt]
R_{G_{1L}}(z)&= -\Bigg[ \frac{24(2z^2-5z+6)}{z}\ln(1-z)+48(3-z)\Bigg],\\[2pt]
R_{D_{1LL}}(z)&=\frac{16(18-30z+17z^2-2z^3)}{z^2}\ln(1-z)+\frac{16(2-z)(3-z)^2}{z},\\[2pt]
R_{H_{1TT}}(z)&=-\frac{24(2z^2-2z+3)}{z^2}\ln(1-z)-\frac{12(6-z+4z^2-z^3)}{z}.
\end{aligned}
\label{eq:RPJ}
\end{equation}
We compare our collinear functions with  Ref.~\cite{Ma:2013yla}. Our $D_1$ agrees with  Ref.~\cite{Ma:2013yla} for ${}^1S_0^{[8]}$ and ${}^3S_1^{[8]}$.
For ${}^3P_J^{[8]}$, summed over $J$, our result differs by $\frac{B_F \,\as^2}{M^5}\frac{16}{3}\,\delta(1-z)$ from Ref.~\cite{Ma:2014eja}, due to the renormalization scheme of $\langle\mathcal O({}^3S_1^{[8]})\rangle$ (see discussion below Eq.~\eqref{eq:sub_2}).
For ${}^3S_1^{[8]}$, we compare our results with those in Ref.~\cite{Braaten:2000pc}. In particular, we relate our $D_1$ with the sum over the $J/\psi$ spin states, $2d_T+d_L$, where $d_T=D_{(\pm1)}$ and $d_L=D_{(0)}$.
On the other hand, all the polarized results are new and, to our knowledge, the only result that we can compare regards $D_{1LL}$. For an unpolarized gluon and a $J/\psi$ with helicity $\lambda$ we have
\begin{equation}
  D_{(\lambda)}(z)=\frac13\Big[D_1(z)+S_{LL}(\lambda)\,D_{1LL}(z)\Big] \, ,
  \label{eq:D_lambda}
\end{equation}
where $S_{LL}=1/2$ for $\lambda=\pm1$ and $S_{LL}=-1$ for $\lambda=0$. 
Hence, given $d_L$ and $d_T$ as in  Ref.~\cite{Braaten:2000pc}, the following relation holds for $D_{1LL}$
\begin{equation}
  D_{1LL}=2\,(d_T-d_L) \, .
\end{equation}
We find that our results for $D_1$ and $D_{1LL}$ agree with Ref.~\cite{Braaten:2000pc}, up to a term $4\ln^22$ in $d_T$, which originates from the integral $I_{ACD}$ of  Ref.~\cite{Braaten:2000pc}, as also noted in  Ref.~\cite{Ma:2013yla}.

Ref.~\cite{Ma:2015yka} gives the fragmentation functions of an unpolarized gluon into a longitudinally polarized $J/\psi$.
They correspond to $d_L=(D_1-D_{1LL})/3$, and agree with our results for both ${}^3S_1^{[8]}$ and ${}^3P_J^{[8]}$. Together with the comparison of $D_1$, this checks our $D_{1LL}$.
For ${}^1S_0^{[8]}$, our $D_1$ agrees with the  color-singlet $g\to\eta_c$ result of  Ref.~\cite{Braaten:1993rw} after replacing the  color factor.
For ${}^3P_J^{[8]}$, summing the  color-singlet results of  Ref.~\cite{Braaten:1994kd} over $J$ and replacing the  color factor reproduces our $D_1$.
Only the constant of the $\delta(1-z)$ term differs because of the different regularization scheme.
The longitudinal function $d_L=(D_1-D_{1LL})/3$ agrees with  Ref.~\cite{Beneke:1995yb} for both ${}^3S_1^{[8]}$ and ${}^3P_J^{[8]}$. For ${}^1S_0^{[8]}$, the result has the same $z$ dependence as ours but a different normalization, while our normalization agrees with  Refs.~\cite{Braaten:1993rw,Ma:2013yla}.
For ${}^1S_0^{[8]}$, our $D_1$ also agrees with  Ref.~\cite{Feng:2018ulg}.
Our $d_L$ and the plus-distribution terms of $d_T$, including their $\mu$ dependence, agree with Ref.~\cite{Feng:2026owq}.
The $\delta(1-z)$ term of $d_T$ of Ref.~\cite{Feng:2026owq} differs from ours and from  Refs.~\cite{Ma:2013yla,Zhang:2020atv}.

\subsection{Operator product expansion}
\label{sec:ope}
For $b_T M \ll 1$, every TMDFF admits an operator product expansion (OPE) in terms of collinear FFs.
For a function multiplying a structure of rank $n$, a leading-power matching onto twist-2 collinear FFs requires $\tilde{F}^{(n)} \sim (b_T M)^{-n}$.
In this section we derive the OPE of all eighteen TMDFFs at $\mathcal{O}(\alpha_s^2)$.

The TMDFF can be related to the NRQCD LDMEs in two ways. For $1/b_T\gg m_Qv$, the soft radiation from the heavy-quark pair changes the transverse momentum only by power-suppressed amounts, and the TMDFF is matched directly onto the LDMEs, Eq.~\eqref{eq:NRQCDfact}, with the SDCs of Secs.~\ref{sec:kt_unsub}--\ref{sec:bt_results}, which are exact in $b_TM$. For $1/b_T\gg M$, the TMDFF can alternatively be expanded onto collinear FFs through the OPE derived in this section, and the collinear FFs are in turn matched onto the same LDMEs, with the collinear SDCs of Sec.~\ref{sec:collinear}. For $1/b_T\gg M$ both routes apply and must coincide: the small-$b_T$ limit of the TMD SDCs has to reproduce the OPE coefficients convoluted with the collinear SDCs, up to corrections of $\mathcal{O}(b_TM)$,
\begin{equation}
\widetilde d^{[F]}_{g\to Q\bar Q[n]}(z,b_T;\mu,\zeta)
=\Big[C_{[F]}(b_T;\mu,\zeta)\otimes\frac{d^{[F]}_{g\to Q\bar Q[n]}(\mu)}{z^2}\Big](z)+\mathcal{O}(b_TM)\,,
\label{eq:two_routes}
\end{equation}
where $C_{[F]}$ are the OPE coefficients given in Eq.~\eqref{eq:C_matching} below, $\otimes$ denotes the Mellin convolution, and $d^{[F]}_{g\to Q\bar Q[n]}(z;\mu)$ are the collinear SDCs of Sec.~\ref{sec:collinear}. This relation holds for the functions with a twist-2 collinear counterpart, and analogously for the off-diagonal terms of Eq.~\eqref{eq:ope2}.
We verify this below for the three channels. The direct matching keeps all powers of $b_TM$, while the OPE allows the resummation of the logarithms of $b_TM$ through DGLAP evolution. These statements concern the TMDFF itself and do not involve the hard scale $Q$, which only restricts the use of TMD factorization in the cross section to $q_T\ll Q$. For charmonium, $m_c\simeq1.5$~GeV and $v^2\simeq0.3$ give $m_cv\simeq0.8$~GeV, while for bottomonium $m_b\simeq4.7$~GeV and $v^2\simeq0.1$ give $m_bv\simeq1.5$~GeV. The window between $m_Qv$ and $M$ is therefore narrow for charmonium and wider for bottomonium.

The large-$\bm{k}_\perp$ behavior of the correlator is the momentum-space image of the small-$b_T$ limit.
In this region the fragmenting gluon, with polarization indices $\{\lambda, \sigma\}$, splits collinearly into an unobserved gluon and a gluon of momentum fraction $z$, which produces the $Q \bar{Q}$ pair.
The full NLO correlator then factorizes as
\begin{equation}
  \Gamma^{ij}(z,\boldsymbol k_\perp)\simeq\frac{2a_s}{\pi\boldsymbol k_\perp^2}\;
  \mathcal{T}^{ij;ab}(z,\hat{\boldsymbol k}_\perp)\;\Gamma^{ab}_{\rm LO} ,
  \qquad \boldsymbol k_\perp^2\gg M^2 ,
  \label{eq:split_fact}
\end{equation}
where $\Gamma_{\rm LO}^{ab}$ is the LO tensor~\eqref{eq:LOpattern}. Here
$i,j$ ($a,b$) are transverse indices of the parent (daughter) gluon, in
Euclidean notation. The splitting tensor is
\begin{equation}
  \mathcal{T}^{ij;ab}=z(1-z)\sum_{c}\mathcal{A}^{iac}\mathcal{A}^{jbc} ,
  \qquad
  \mathcal{A}^{iac}=\frac{\delta^{ia}\hat k^c}{1-z}+\frac{\delta^{ic}\hat k^a}{z}-\delta^{ac}\hat k^i ,
  \label{eq:split_tensor}
\end{equation}
which is the squared tree-level $g\to gg$ collinear vertex summed over the polarizations of the emitted gluon. It is the spin-correlated $g\to gg$ kernel of  Ref.~\cite{Catani:1999ss}, generalized to a polarized daughter.
Projecting Eq.~\eqref{eq:split_fact} onto the structures of Eqs.~\eqref{eq:GU}--\eqref{eq:GTT}, we find that every leading-power large-$\boldsymbol k_\perp$ coefficient obtained in Sec.~\ref{sec:kt_unsub} is reproduced exactly.
With
$\hat k^{ij}=\hat k^i\hat k^j-\tfrac12\delta^{ij}$:
\begin{align}
  \mathcal{T}^{ij;ab}\,\delta^{ab}
  &= \frac{\mathcal{P}_{gg}(z)}{2}\,\delta^{ij}+2z(1-z)\,\hat k^{ij} ,
  \label{eq:T_U}\\
  \tfrac12\epsilon^{ij}\,\mathcal{T}^{ij;ab}\,\epsilon^{ab}
  &= \frac{1}{1-z}+1-2z ,
  \qquad
  \mathcal{T}^{(ij);ab}\,\epsilon^{ab}=0 ,
  \label{eq:T_L}\\
  \mathcal{T}^{ij;ab}\,S_{TT}^{ab}
  &= \frac{z}{1-z}\,S_{TT}^{ij}+\frac{1-z}{z}\,S_{TT}^{\hat k\hat k}\,\delta^{ij} ,
  \label{eq:T_TT}
\end{align}
for $z<1$. 
The traces give the diagonal kernels of Sec.~\ref{sec:collinear}.
The term $2z(1-z)\hat k^{ij}$ in Eq.~\eqref{eq:T_U} is the linear-polarization projection of the splitting with respect to the parent gluon: an unpolarized daughter, i.e., an unpolarized or $S_{LL}$-polarized hadron, comes from a linearly polarized fragmenting gluon.\footnote{It generates $H_1^\perp$ and $H_{1LL}^\perp$, and reproduces $d^{[H_1^\perp]}\simeq(4a_s/\pi)\,z(1-z)\,M^2\boldsymbol k_\perp^{-4}\,d_{\rm LO}$ of Eq.~\eqref{eq:kT_noLO}.
In fact, this is the azimuthal term $-2(1-\epsilon)z(1-z)k_\perp^\mu k_\perp^\nu/k_\perp^2$ of the spin-correlated kernel of  Ref.~\cite{Catani:1999ss} at $\epsilon=0$.}
It is the timelike counterpart of the coefficient that relates the linearly polarized gluon TMDPDF to the unpolarized gluon PDF~\mbox{\cite{Sun:2011iw,Echevarria:2015uaa,Gutierrez-Reyes:2019rug}}.
The term $\tfrac{1-z}{z}S_{TT}^{\hat k\hat k}\delta^{ij}$ in Eq.~\eqref{eq:T_TT} is the linear-polarization projection with respect to the daughter gluon in the splitting: an unpolarized fragmenting gluon produces a linearly polarized daughter, which transfers its polarization to the $S_{TT}$ tensor of the spin-1 hadron, and it generates $\Phi$.

In impact-parameter space, the four TMDFFs with a twist-2 collinear counterpart are given by the convolution of a perturbative coefficient with the collinear FFs of Sec.~\ref{sec:collinear}:
\begin{equation}
  \widetilde F(z,b_T;\mu,\zeta)
  = \Big[C_{[F]}(b_T;\mu,\zeta)\otimes\frac{F(\mu)}{z^{2}}\Big](z)+\mathcal{O}(b_TM) ,
  \label{eq:OPE}
\end{equation}
where $F\in\{D_1,D_{1LL},G_{1L},H_{1TT}\}$, and $\otimes$ denotes the Mellin convolution.
At one loop,
\begin{equation}
  \begin{aligned}
      C_{[F]}(z,b_T;\mu,\zeta)=&\delta(1-z)+a_s\Big[\delta(1-z)\Big(-\frac{L_T^2}{2}+L_T\ln\frac{\mu^2}{\zeta}-\frac{\pi^2}{12}\Big)\\
   &-\widehat P_F(z)\,\big(L_T-2\ln z\big)+r_F(z)\Big] ,
  \end{aligned}
  \label{eq:C_matching}
\end{equation}
with the four-dimensional kernels $\widehat P_F$ of Eq.~\eqref{eq:kernels_4}
and the finite terms
\begin{equation}
  r_{D_1}=r_{H_{1TT}}=0 ,\qquad
  r_{G_{1L}}=-4(1-z) ,\qquad
  r_{D_{1LL}}= 0 .
  \label{eq:rF}
\end{equation}
We have checked that the small-$b_T$ expansion of Eqs.~\eqref{eq:D1_b} and~\eqref{eq:pol_b} coincides with Eq.~\eqref{eq:OPE} at $\mathcal{O}(\alpha_s^2)$ for all four functions, with the collinear FFs of Eq.~\eqref{eq:coll_result}. This is the agreement between the two routes of Eq.~\eqref{eq:two_routes} in the $\Sone$ channel; for the $\Szero$ and $\PJ$ channels it is shown in Eqs.~\eqref{eq:1S0_D1_OPE}--\eqref{eq:PJ_OPE}.
The logarithmic terms are governed by the kernel of the corresponding gluon polarization, as expected for a coefficient that does not depend on the hadronic state. The finite terms $r_F$ are scheme dependent and equal minus the $\mathcal{O}(\epsilon)$ coefficient of the $d$-dimensional kernels~\eqref{eq:kernels_d}. For the helicity function, $r_{G_{1L}}=-4(1-z)$ is the familiar finite term of polarized gluon coefficient functions in the HVBM scheme.\footnote{The scheme dependence cancels
in the product $C_{[F]}\otimes F$, so the TMDFFs are scheme independent.}

In general, the full gluon correlators can be written at leading power of the OPE as follows:
\begin{equation}
\begin{aligned}
  \widetilde\Gamma^{\lambda\sigma}_U
  &\simeq -g_T^{\lambda\sigma}\Big[C_{[D_1]}\otimes\frac{D_1}{z^2}\Big]
   -4a_s\,\frac{b_T^{\lambda\sigma}}{\boldsymbol b_T^2}\Big[z(1-z)\otimes\frac{D_1}{z^2}\Big] ,\\
  \widetilde\Gamma^{\lambda\sigma}_L
  &\simeq S_L\,i\epsilon_T^{\lambda\sigma}\Big[C_{[G_{1L}]}\otimes\frac{G_{1L}}{z^2}\Big] ,\\
  \widetilde\Gamma^{\lambda\sigma}_{LL}
  &\simeq S_{LL}\Big\{-g_T^{\lambda\sigma}\Big[C_{[D_{1LL}]}\otimes\frac{D_{1LL}}{z^2}\Big]
   -4a_s\,\frac{b_T^{\lambda\sigma}}{\boldsymbol b_T^2}\Big[z(1-z)\otimes\frac{D_{1LL}}{z^2}\Big]\Big\} ,\\
  \widetilde\Gamma^{\lambda\sigma}_{TT}
  &\simeq S_{TT}^{\lambda\sigma}\Big[C_{[H_{1TT}]}\otimes\frac{H_{1TT}}{z^2}\Big]
   +2a_s\,g_T^{\lambda\sigma}\,\frac{S_{TT}^{bb}}{\boldsymbol b_T^2}\Big[\frac{1-z}{z}\otimes\frac{H_{1TT}}{z^2}\Big] ,
  \\
  \widetilde\Gamma^{\lambda\sigma}_{T}
  &\simeq\widetilde\Gamma^{\lambda\sigma}_{LT}\simeq0 ,
  \label{eq:OPE_full}
\end{aligned}
\end{equation}
up to corrections of $\mathcal{O}(b_TM)$ and of higher order in $\alpha_s$.
In terms of the Bessel-weighted functions, the off-diagonal terms read
\begin{equation}
\begin{aligned}
  \widetilde H_{1}^{\perp(2)}(z,b_T)
  &=\frac{8a_s}{\boldsymbol b_T^2M^2}\Big[z(1-z)\otimes\frac{D_1}{z^2}\Big](z) ,\\
  \widetilde H_{1LL}^{\perp(2)}(z,b_T)
  &=\frac{8a_s}{\boldsymbol b_T^2M^2}\Big[z(1-z)\otimes\frac{D_{1LL}}{z^2}\Big](z) ,
  \\
  \widetilde \Phi^{(2)}(z,b_T)
  &=-\frac{4a_s}{\boldsymbol b_T^2M^2}\Big[\frac{1-z}{z}\otimes\frac{H_{1TT}}{z^2}\Big](z) .
\end{aligned}
\label{eq:ope2}
\end{equation}
We have checked Eqs.~\eqref{eq:OPE_full} and~\eqref{eq:ope2} against the small-$b_T$ limit of the exact results of Sec.~\ref{sec:bt_results}, inserting the LO collinear FFs $D_1=D_{1LL}=d_{\rm LO}\,\delta(1-z)$ and $H_{1TT}=-\tfrac32d_{\rm LO}\,\delta(1-z)$ (in units of the LDME).
For instance,
\begin{equation}
\begin{aligned}
    \widetilde d^{[H_1^\perp](2)}_{\Sone} & = 8a_sd_{\rm LO}(1-z)^2 \frac{K_1(b_T\Lambda_2)}{(b_TM)} \to8a_sd_{\rm LO}\,\frac{z(1-z)}{(b_TM)^2} \, , \\
    \widetilde d^{[\Phi](2)}_{\Sone}& =6a_sd_{\rm LO}(1-z)^2 \frac{K_1(b_T\Lambda_2)}{(z^2b_TM)}\to  6a_sd_{\rm LO} \frac{(1-z)}{z(b_TM)^2}
\end{aligned}
\end{equation}
as required by Eq.~\eqref{eq:ope2}.
We note that the same tensor~\eqref{eq:split_tensor} would determine the OPE of the corresponding gluon TMD PDFs of a spin-1 target, with the roles of parent and daughter interchanged.

The rank-1 functions $G_{1T}$, $D_{1LT}$ and $H_{1LT}$ multiply structures proportional to $b_TM$, while Eq.~\eqref{eq:b_noLO} gives
\begin{equation}
    \widetilde F^{(1)}\propto K_0(b_T\Lambda_2)\sim\ln(b_TM) \, .
\end{equation}
Their contribution to $\widetilde\Gamma_X$ is therefore of $\mathcal{O}\big(b_TM\ln(b_TM)\big)$, so they have no leading-power OPE.
By contrast, they match onto collinear correlators of twist-3.
Correspondingly, their contributions to the momentum-space correlator fall as $M/|\boldsymbol k_\perp|^3$, i.e., they are suppressed by $M/|\boldsymbol k_\perp|$ with respect to the leading $1/\boldsymbol k_\perp^2$ behavior.
Finally, contributions in which the gluon TMDFF matches onto quark FFs, through $C_{g\leftarrow q}$, are of higher order here, since the quark FF into a ${}^3S_1^{[8]}$ pair starts at $\mathcal{O}(\alpha_s^2)$.

For the ${}^1S_0^{[8]}$ channel, the correlator contains only the two unpolarized structures, where their $b_T \to 0$ limit is given by
\begin{align}
  \widetilde d^{[D_1]}_{{}^1S_0^{[8]}}(z,b_T)
  &=\frac{B_F \,\alpha_s^2}{M^3}\,\big[3z-2z^2+2(1-z)\ln(1-z)\big],
  \label{eq:1S0_D1_OPE}\\[4pt]
  \widetilde d^{[H_1^\perp](2)}_{{}^1S_0^{[8]}}(z,b_T)
  &=\frac{B_F \,\alpha_s^2}{M^3} (1-z)^2\nonumber\\
  &\Bigg[ L_T-\ln\frac{\mu^2}{M^2}+\frac{1}{2z^2}\Big(4z^2\ln\frac{1-z}{z}-z^2-2z-2\ln(1-z)\Big)\Bigg],
  \label{eq:1S0_H1_OPE}
\end{align}
while all the polarized structures vanish. Since the channel has no leading-order term, the $\mathcal O(\alpha_s^2)$ result is free of divergences, and there is no infrared pole at $z\to1$, because both functions remain finite in this limit. The results are thus finite, require no renormalization, and are independent of $\mu$. For the latter, we recall that $L_T=\ln[{b_T^2\mu^2e^{2\gamma_E}}/{4}]$.

In the correlator, the linearly polarized structure enters as $-\tfrac{M^2}{2}\,b_T^{\lambda\sigma}\,\widetilde d^{[H_1^\perp](2)}$. Since $\widetilde d^{[H_1^\perp](2)}$ is finite as $b_T\to0$, up to a logarithm, this contribution is of order $\boldsymbol b_T^2M^2$ and has no leading-power matching onto collinear fragmentation functions. In the ${}^3S_1^{[8]}$ channel, instead, $\widetilde d^{[H_1^\perp](2)}\sim1/\boldsymbol b_T^2$, and the linearly polarized structure is matched onto the unpolarized collinear function through $z(1-z)\otimes D_1/z^2$. At leading power in $b_T$, the operator product expansion of the ${}^1S_0^{[8]}$ correlator therefore takes the form
\begin{equation}
  \widetilde\Gamma^{\lambda\sigma}_{U,{}^1S_0^{[8]}}
  \simeq -g_T^{\lambda\sigma}\Big[C_{[D_1]}\otimes\frac{D_1}{z^2}\Big],
  \qquad C_{[D_1]}(z)=\delta(1-z)+\mathcal O(\alpha_s).
\end{equation}
Since the channel starts at $\mathcal O(\alpha_s^2)$, the matching coefficient is needed only at leading order, and the small-$b_T$ limit of $\widetilde d^{[D_1]}$ coincides with the collinear fragmentation function and reproduces the $z$ dependence of the  color-singlet result of  Ref.~\cite{Braaten:1993rw}. The normalization differs by the octet-to-singlet  color factor.

We now turn to the ${}^3P_J^{[8]}$ channel. As for ${}^1S_0^{[8]}$, the channel has no leading-order term, and it first contributes at $\mathcal O(\alpha_s^2)$ through the emission from the heavy-quark lines. 
Its small-$b_T$ behavior is as follows.

 At leading power in $b_T$, only the four rank-0 functions of the ${}^3P_J^{[8]}$ channel survive, and the correlators match onto the collinear fragmentation functions,
\begin{equation}
\begin{aligned}
  \widetilde\Gamma^{\lambda\sigma}_{U}&\simeq -g_T^{\lambda\sigma}\Big[C_{[D_1]}\otimes\frac{D_1}{z^2}\Big],
  &\qquad
  \widetilde\Gamma^{\lambda\sigma}_{L}&\simeq i\epsilon_T^{\lambda\sigma}S_L\Big[C_{[G_{1L}]}\otimes\frac{G_{1L}}{z^2}\Big],\\
  \widetilde\Gamma^{\lambda\sigma}_{LL}&\simeq -g_T^{\lambda\sigma}S_{LL}\Big[C_{[D_{1LL}]}\otimes\frac{D_{1LL}}{z^2}\Big],
  &\qquad
  \widetilde\Gamma^{\lambda\sigma}_{TT}&\simeq S_{TT}^{\lambda\sigma}\Big[C_{[H_{1TT}]}\otimes\frac{H_{1TT}}{z^2}\Big],
\end{aligned}
\end{equation}
with $C_{[F]}(z)=\delta(1-z)+\mathcal O(\alpha_s)$, since the channel starts at $\mathcal O(\alpha_s^2)$, while $\widetilde\Gamma_T$ and $\widetilde\Gamma_{LT}$ vanish at this power.
At small $b_T$, Eq.~(\ref{eq:PJ_bspace}) becomes
\begin{equation}
\begin{aligned}
    \widetilde d^{[F]}_{\Pzero}(z,b_T;\mu_\Lambda)&=\frac{B_F \,\alpha_s^2}{M^5}\left\{32\,c_F\left[\frac{1}{(1-z)_+}-\frac12\,\delta(1-z)\ln\frac{\mu_\Lambda^2}{M^2}\right]+\tilde R_F(z)\right\}\\
    &\qquad+\mathcal O\big(b_T^2M^2\ln(b_TM)\big),
\end{aligned}
\label{eq:PJ_OPE}
\end{equation}
with
\begin{equation}
\begin{aligned}
\tilde R_{D_1}(z)&=8(13-7z)\ln(1-z)-4(10z^2-21z+8),\\
\tilde R_{G_{1L}}(z)&=- \frac{24\,(2z^2-5z+6)}{z}\ln(1-z)-96(1-z),\\
\tilde R_{D_{1LL}}(z)&=\frac{16\,(18-30z+17z^2-2z^3)}{z^2}\ln(1-z)+\frac{16\,(1-z)(z^2-5z+18)}{z},\\
\tilde R_{H_{1TT}}(z)&=-\frac{24\,(2z^2-2z+3)}{z^2}\ln(1-z)+\frac{12\,(z^3+5z-6)}{z}.
\end{aligned}
\end{equation}
The logarithms of $b_T$ cancel between $K_0(b_T\Lambda_1)$ and $K_0(b_T\Lambda_2)$.
Using $z^2/(1-z)_+=1/(1-z)_+-(1+z)$, Eq.~\eqref{eq:PJ_OPE} coincides with the collinear functions $d^{[F]}(z)/z^2$ of Eq.~\eqref{eq:collPJ}, as required by $C_{[F]}=\delta(1-z)$.

The functions of nonzero rank do not contribute at leading power. Their structures in the correlator carry $(M\,b_T)^n b_T^{\lambda_1\dots\lambda_n}/\boldsymbol b_T^{\,n}$, and since the channel has no leading-order term, none of the rank-$n$ functions develop the $1/\boldsymbol b_T^{\,n}$ singularity needed to compensate it. Their contributions are therefore suppressed by $(\boldsymbol b_TM)^n$. The rank-1 functions $G_{1T}$, $D_{1LT}$ and $H_{1LT}$, and the rank-2 functions $H_{1LL}^\perp$ and $\Phi=H_{1TT}^\perp-D_{1TT}$, approach finite values as $b_T\to0$. The rank-2 $H_1^\perp$, rank-3 $H_{1LT}^\perp$ and rank-4 $H_{1TT}^{\perp\perp}$ instead grow logarithmically, as $\ln(b_T\Lambda_2)$, because the corresponding $k_\perp$-weighted integrals diverge logarithmically. In none of these cases is there a matching onto collinear fragmentation functions at leading power. This differs from the ${}^3S_1^{[8]}$ channel, where the leading-order term and the spin-correlated $g\to gg$ kernel generate a leading-power contribution of the linearly polarized structures $H_1^\perp$ and $H_{1LL}^\perp$. The time-reversal-odd functions vanish at this order.


\section{Conclusions}
\label{sec:conclusions}
We have presented the complete set of gluon TMDFFs into quarkonium through NLO in perturbative QCD. More generally, the same set applies to any spin-1 hadron.
We have accounted for all polarization configurations of both the fragmenting gluon and the produced $J^{\rm PC} = 1^{--}$ quarkonium.
Within the NRQCD framework, we have included all relevant angular-momentum channels at the considered orders in $\as$ and in the heavy-quark relative velocity $v$.
These are the  color-octet ${}^3S_1^{[8]}$, ${}^1S_0^{[8]}$ and ${}^3P_J^{[8]}$ states.

Several structures in the calculation are not specific to quarkonium.
They follow from the spin-1 nature of the final-state hadron.
These include the decomposition of the hadron density matrix into vector and tensor polarizations and the corresponding projectors.
The quarkonium-specific input enters only through the NRQCD matching coefficients and the LDMEs.

For ${}^3S_1^{[8]}$, we have computed the real and virtual corrections at NLO.
We have also obtained the matching of the polarized TMDFFs onto their collinear counterparts.
For ${}^1S_0^{[8]}$ and ${}^3P_J^{[8]}$, we have obtained all nonvanishing functions at order $\as^2$.
In the ${}^3P_J^{[8]}$ channel, the four functions with a twist-2 counterpart are infrared divergent at $z \to 1$.
This divergence originates from the soft dipole transition of NRQCD and is absorbed into the ${}^3S_1^{[8]}$ LDME through operator mixing, so that the renormalized coefficients depend on the NRQCD factorization scale $\mu_\Lambda$, while the full TMDFFs do not.

Our results agree with the existing literature wherever a comparison is possible.
At the TMD level, the unpolarized functions agree with Ref.~\cite{Echevarria:2023dme}, except for the $\pi^2$ term of the constant $c_V$ of the virtual correction, which was incorrect there and is corrected here, and at LO the polarized TMDFFs agree with Ref.~\cite{Copeland:2023wbu}.
At the collinear level, $D_1$ agrees with Refs.~\cite{Braaten:1993rw,Ma:2013yla,Feng:2018ulg} for ${}^1S_0^{[8]}$ and with Refs.~\cite{Braaten:2000pc,Ma:2013yla} for ${}^3S_1^{[8]}$, apart from the $4\ln^2 2$ term in Ref.~\cite{Braaten:2000pc} already noted in Ref.~\cite{Ma:2013yla}. For ${}^3P_J^{[8]}$, it reproduces the sum over $J$ of Ref.~\cite{Braaten:1994kd}, and differs from Ref.~\cite{Ma:2014eja} only by a $\delta(1-z)$ term due to the renormalization scheme of $\langle\mathcal O({}^3S_1^{[8]})\rangle$, related by the finite renormalization given below Eq.~\mbox{\eqref{eq:sub_2}}.
Among the polarized functions, only $D_{1LL}$ can be compared with previous results, through the transverse and longitudinal fragmentation functions $d_T$ and $d_L$. These agree with Refs.~\cite{Braaten:2000pc,Ma:2015yka,Beneke:1995yb}, except for the normalization of the ${}^1S_0^{[8]}$ result in Ref.~\cite{Beneke:1995yb}, and with Ref.~\cite{Feng:2026owq} except for the $\delta(1-z)$ term of $d_T$, where our result agrees with Refs.~\cite{Ma:2013yla,Zhang:2020atv}.
All the other polarized TMDFFs are new.

Most of these results appear here for the first time. They include the polarized TMDFFs beyond LO and the full $k_\perp$ dependence in the ${}^1S_0^{[8]}$ and ${}^3P_J^{[8]}$ channels.
Together, they complete the perturbative description of how a gluon transfers its spin and transverse momentum to a quarkonium at this order in $\alpha_s$ and $v$. 
This opens the way to polarization-sensitive studies of quarkonium in jets, in SIDIS at the EIC and in $e^+e^-$ annihilation.
Moreover, since the ${}^1S_0^{[8]}$ channel produces an unpolarized hadron, the TMDFFs of a polarized hadron, i.e., those in the rows $L$, $T$, $LL$, $LT$ and $TT$ of Tab.~\mbox{\ref{tab:TMDFFs}}, do not depend on $\langle\mathcal O({}^1S_0^{[8]})\rangle$ at this order, and thus give direct access to the ${}^3S_1^{[8]}$ and ${}^3P_J^{[8]}$ LDMEs.

Our results are limited to the considered orders in $\alpha_s$ and $v$. The color-singlet ${}^3S_1^{[1]}$ channel first contributes at $\mathcal O(\alpha_s^3)$, through the emission of two real gluons, and its TMD version remains to be computed. Corrections of relative order $v^2$ are not included. The matching onto LDMEs holds for $q_T\gg m_Qv$, while at smaller transverse momenta TMD shape functions or soft transition functions are needed. Finally, the heavy-quark loop in the gluon self-energy is not included, and its effect is a constant shift of $c_V$, given below Eq.~\mbox{\eqref{eq:cV}}.

The Mathematica notebooks and the results for all polarized gluon TMDFFs presented in this work are available at
\begin{center}
\url{https://github.com/alessiabongallino/PolarizedGluonTMDFFs}.
\end{center}
\section*{Acknowledgments}

We thank Marston Copeland for helpful discussions.
SFR acknowledges financial support from the fellowship PIF22/219 of the EHU.
This work was supported by the Spanish Ministerio de Ciencia, Innovación y Universidades (MCIN/AEI/10.13039/501100011033) through the grant PID2025-174203NB-I00, by the Basque Government through the grant IT1977-26, as well as by the European Union's Horizon Europe project HADRON-2030 (No. 101292522) and the COST action SHARP (No. 24159).


\appendix

\section{Conventions}
\label{sec:conventions}
We work in the light-cone basis defined by $n$ and $\bar n$, with $n^2=\bar n^2=0$ and $n\cdot\bar n=1$.
The transverse metric and Levi-Civita tensor are
\begin{equation}
  \gT^{\mu\nu}=g^{\mu\nu}-\bar n^{\mu}n^{\nu}-n^{\mu}\bar n^{\nu},
  \qquad
  \eT^{\mu\nu}=\epsilon^{\mu \nu n \bar n}.
\end{equation}
With these conventions,
\begin{equation}
  \gT^{\mu\nu}\gT{}_{\mu\nu}=2,\qquad
  \eT^{\mu\nu}\eT{}_{\mu\nu}=2,\qquad
  \eT^{\alpha\beta}\eT^{\gamma\delta}=\gT^{\alpha\gamma}\gT^{\beta\delta}-\gT^{\alpha\delta}\gT^{\beta\gamma}.
\end{equation}
The transverse momentum of the gluon satisfies $\kTsq=-|\boldsymbol k_\perp|^2$. We define the dual vector and the shorthands
\begin{equation}
  \kt^{\alpha}\equiv\eT^{\alpha\beta}k_{T\beta},\qquad
  \eT^{k\lambda}\equiv\eT^{\rho\lambda}k_{T\rho},\qquad
  \eT^{ab}\equiv\eT^{\rho\sigma}a_{\rho}b_{\sigma}.
\end{equation}
Note that $\kt^2=\kTsq$ and $\kt\cdot\kT=0$.
Symmetrization without trace subtraction is written as
\begin{equation}
  a^{\{\lambda}b^{\sigma\}}\equiv a^{\lambda}b^{\sigma}+a^{\sigma}b^{\lambda}.
\end{equation}
The symmetric traceless transverse tensors built from $\kT$ are
\begin{align}
  k_T^{\lambda\sigma}&=k_T^{\lambda}k_T^{\sigma}-\tfrac12\kTsq\,\gT^{\lambda\sigma},\\
  k_T^{\lambda \sigma \rho}&= k_T^\lambda k_T^\sigma k_T^\rho 
  - \frac{k_T^2 }{4 } (g_T^{\lambda \sigma} k_T^\rho + g_T^{\lambda \rho} k_T^\sigma + g_T^{\sigma \rho} k_T^\lambda)\, ,\\
  k_T^{\lambda\sigma\rho\tau}&=k_T^{\lambda}k_T^{\sigma}k_T^{\rho}k_T^{\tau}
  -\frac{\kTsq}{6}\Big(\gT^{\lambda\sigma}k_T^{\rho\tau}+\gT^{\lambda\rho}k_T^{\sigma\tau}+\gT^{\lambda\tau}k_T^{\sigma\rho}
  +\gT^{\sigma\rho}k_T^{\lambda\tau}+\gT^{\sigma\tau}k_T^{\lambda\rho}+\gT^{\rho\tau}k_T^{\lambda\sigma}\Big)\nonumber\\
  &\quad-\frac{k_T^4}{8}\Big(\gT^{\lambda\sigma}\gT^{\rho\tau}+\gT^{\lambda\rho}\gT^{\sigma\tau}+\gT^{\lambda\tau}\gT^{\sigma\rho}\Big).
\end{align}
The hadron polarization is described by $S_L$, $S_T^{\mu}$, $S_{LL}$, $\SLT^{\mu}$ and the symmetric traceless tensor $\STT^{\mu\nu}$.
All of these are transverse to $n$, $\bar n$ and the hadron momentum $P$.
For the vector polarizations we use
\begin{equation}
  (\eT^{k_TS_T})^2=\kTsq S_T^2-(\kT\cdot S_T)^2,
\end{equation}
and the analogous relation for $\SLT$.
For the tensor polarization we define three invariants:
\begin{equation}
  \SSq\equiv \STT^{\alpha\beta}S_{TT\,\alpha\beta},\qquad
  \Skk\equiv k_{T\alpha}\STT^{\alpha\beta}k_{T\beta},\qquad
  \Skt\equiv \kt{}_{\alpha}\STT^{\alpha\beta}k_{T\beta}
  =\eT{}_{\beta\gamma}\,k_T^{\gamma\alpha}\,S_{TT\,\alpha}{}^{\beta}.
\end{equation}
In two transverse dimensions these invariants are related by
\begin{equation}
  \big(\Skk\big)^2+\big(\Skt\big)^2=\tfrac12\,\SSq\,k_T^4 .
  \label{eq:2Did}
\end{equation}

Expanding the gauge-invariant field~\eqref{eq:Bfield} in powers of $g_s$
gives the one- and two-gluon vertices~\cite{Echevarria:2016scs}
\begin{align}
  \mathcal{B}^{(0)\mu\nu}(p)
  &= g_T^{\mu\nu} - \frac{p_T^\mu\,\bar n^\nu}{p^+ - i\delta^+} ,
  \label{eq:B0}\\
  \mathcal{B}^{(1)\mu\alpha\beta}(p,q)
  &= \frac{g_T^{\mu\beta}\,\bar n^\alpha}{p^+-i\delta^+}
   - \frac{g_T^{\mu\alpha}\,\bar n^\beta}{q^+-i\delta^+}
   + \Big(\frac{p_T^\mu}{q^+-i\delta^+}-\frac{q_T^\mu}{p^+-i\delta^+}\Big)
     \frac{\bar n^\alpha\bar n^\beta}{p^++q^+-2i\delta^+} ,
  \label{eq:B1}
\end{align}
where the color structure $g_sf^{abc}$ of $\mathcal{B}^{(1)}$ has been
factored out.
These are supplemented by the standard QCD triple-gluon vertex,
$V^{\alpha_1\alpha_2\alpha_3}(p_1,p_2,p_3)=g^{\alpha_1\alpha_2}(p_1-p_2)^{\alpha_3}+g^{\alpha_2\alpha_3}(p_2-p_3)^{\alpha_1}+g^{\alpha_3\alpha_1}(p_3-p_1)^{\alpha_2}$,
and by the heavy-quark propagators. At LO only $\mathcal{B}^{(0)}(P)$ appears.
Since $P_T=0$, it reduces to $g_T^{\mu\nu}$.
 

\section{Quarkonium polarization and spinor expansions}
\label{app:polarisationandexpansion}
 
This appendix collects the heavy-quark spinor structures that enter the calculation of the gluon TMDFFs at $\mathcal O(\alpha_s)$ and $\mathcal O(\alpha_s^2)$ in the $^3S_1^{[8]}$, $^1S_0^{[8]}$ and $^3P_J^{[8]}$ channels, together with the covariant spin projectors and the polarization projectors of the hadron.
We follow Refs.~\cite{Braaten:1996jt, Copeland:2023qed, Boer:2016xqr, Echevarria:2023dme}.
The covariant spin projectors are those of
Refs.~\cite{Petrelli:1997ge, Bodwin:2002cfe}. All the identities below were verified numerically with explicit $4\times4$ Dirac matrices and explicit Pauli spinors.

\subsection{Kinematics, spinors and conventions}
\label{app:B_kin}
 
The pair has momentum $P^\mu=P^+ n^\mu+\frac{M^2}{2P^+}\nbar^\mu$ with $P^2=M^2$, and the heavy quark and antiquark carry
\begin{equation}
  p_1^\mu=\tfrac12P^\mu+q^\mu,\qquad p_2^\mu=\tfrac12P^\mu-q^\mu,\qquad
  q^\mu\equiv L^\mu{}_i\,q^i,\qquad P\cdot q=0,\quad q^2=-|\boldsymbol q|^2,
  \label{eq:B_mom}
\end{equation}
where $q^i$ are the components of the relative momentum in the rest frame of the pair, and $L^\mu{}_\nu$ is the boost from that frame to the frame in which the pair has momentum $P$. On shell, $P^2=4(m_Q^2+|\boldsymbol q|^2)$, and we set $M=2m_Q$.
In the rest frame, in the Dirac representation,
\begin{equation}
  u(p_1)=\sqrt{E+m_Q}\begin{pmatrix}\xi\\[2pt]\dfrac{\boldsymbol\sigma\!\cdot\!\boldsymbol q}{E+m_Q}\,\xi\end{pmatrix},
  \qquad
  v(p_2)=\sqrt{E+m_Q}\begin{pmatrix}-\dfrac{\boldsymbol\sigma\!\cdot\!\boldsymbol q}{E+m_Q}\,\eta\\[2pt]\eta\end{pmatrix},
  \label{eq:B_spinors}
\end{equation}
with $E=\sqrt{m_Q^2+|\boldsymbol q|^2}$, $\xi^\dagger\xi=\eta^\dagger\eta=1$ and $\bar uu=2m_Q$ at $q=0$.
To the order needed,
$u\simeq\sqrt{2m_Q}\,(\xi;\,\boldsymbol\sigma\!\cdot\!\boldsymbol q\,\xi/2m_Q)$ and
$v\simeq\sqrt{2m_Q}\,(-\boldsymbol\sigma\!\cdot\!\boldsymbol q\,\eta/2m_Q;\,\eta)$.
We use $\gamma_5=i\gamma^0\gamma^1\gamma^2\gamma^3$ and
$\mathrm{Tr}[\gamma^\mu\gamma^\nu\gamma^\rho\gamma^\sigma\gamma_5]
=-4i\,\epsilon^{\mu\nu\rho\sigma}$, hence $\epsilon^{0123}=+1$.
A different convention changes the sign of the terms with a single $\epsilon$.
This sign drops out of $|\mathcal M|^2$, and of the products of two $\epsilon$ tensors that define $G_{1L}$ and $G_{1T}$, provided $\Pi_{L,T}$ and $\epsilon_T$ are built with the same $\epsilon$.
We use $\epsilon_T^{\mu\nu}=\epsilon^{\mu\nu n\bar n}$, for which
\begin{equation}
\epsilon^{\mu\nu\alpha\beta}P_\alpha\,\varepsilon_{L\beta}=-M\,\epsilon_T^{\mu\nu}.
\end{equation}
The opposite ordering flips the sign of $G_{1L}$ and $G_{1T}$.
The spinor bilinears are written in terms of
\begin{equation}
  X^i\equiv\xi^\dagger\sigma^i\eta\ \ (\text{spin triplet}),\qquad
  S\equiv\xi^\dagger\eta\ \ (\text{spin singlet}),
\end{equation}
with $X^\mu\equiv L^\mu{}_iX^i, X\!\cdot\!P=0 $.
 
\subsection{Bilinears at  \texorpdfstring{$\mathcal O(q^0)$}{Oq0} and \texorpdfstring{$\mathcal O(q^1)$}{Oq1}}
\label{app:B_bilinears}
Any string of $\gamma$ matrices can be decomposed in the Clifford basis
$\{\mathbbm{1},\gamma^\mu,\sigma^{\mu\nu},\gamma^\mu\gamma_5,\gamma_5\}$, so the
following five bilinears give every Dirac structure needed in this work. For
the rest-frame spinors~\eqref{eq:B_spinors} one finds, up to $\mathcal O(q^2)$,
\begin{align}
  \bar u(p_1)\,v(p_2)&=2\,q\!\cdot\!X ,
  \label{eq:B_bil_1}\\
  \bar u(p_1)\gamma^\mu v(p_2)&=2m_Q\,X^\mu ,
  \label{eq:B_bil_V}\\
  \bar u(p_1)\sigma^{\mu\nu}v(p_2)&=i\big(P^\mu X^\nu-P^\nu X^\mu\big)
      +\frac{1}{m_Q}\,\epsilon^{\mu\nu\alpha\beta}P_\alpha q_\beta\,S ,
  \label{eq:B_bil_T}\\
  \bar u(p_1)\gamma^\mu\gamma_5v(p_2)&=P^\mu S
      +\frac{i}{m_Q}\,\epsilon^{\mu\nu\alpha\beta}P_\nu q_\alpha X_\beta ,
  \label{eq:B_bil_A}\\
  \bar u(p_1)\gamma_5v(p_2)&=2m_Q\,S .
  \label{eq:B_bil_5}
\end{align}
 
\subsection{Covariant spin projectors}
\label{app:B_proj}
 
Summing over the spins of the pair with Clebsch--Gordan coefficients,
\begin{equation}
  \sum_{s_1,s_2}\Big\langle\tfrac12s_1;\tfrac12s_2\Big|S\,S_z\Big\rangle
  v(p_2,s_2)\,\bar u(p_1,s_1)\equiv\Pi_S ,
  \qquad
  \bar u\,\Gamma\,v\ \longrightarrow\ \mathrm{Tr}\big[\Gamma\,\Pi_S\big] .
  \label{eq:B_Pi_def}
\end{equation}
In the Pauli language this is the replacement $\eta\,\xi^\dagger\to\mathbbm 1/\sqrt2$ for $S=0$ and
$\eta\,\xi^\dagger\to\boldsymbol\sigma\!\cdot\!\boldsymbol\varepsilon^*/\sqrt2$ for $S=1$, hence $X^i\to\sqrt2\,\varepsilon^{*i}$ and $S\to\sqrt2$. 
In covariant form
\begin{align}
  \Pi_1&=-\frac{1}{4\sqrt2\,E(E+m_Q)}\,(\slashed p_2-m_Q)\,\slashed\varepsilon^{*}\,
         (\slashed P+2E)\,(\slashed p_1+m_Q),
  \label{eq:B_Pi1}\\
  \Pi_0&=-\frac{1}{4\sqrt2\,E(E+m_Q)}\,(\slashed p_2-m_Q)\,\gamma_5\,
         (\slashed P+2E)\,(\slashed p_1+m_Q),
  \label{eq:B_Pi0}
\end{align}
with $P\cdot\varepsilon=0$. 
The relative sign between $\Pi_0$ and $\Pi_1$ is fixed by the replacement rule above. A different phase convention for $\eta$ changes only the global sign of $\Pi_0$.
 
At $q=0$, using $(\slashed P+2m_Q)(\slashed P/2+m_Q)=4m_Q(\slashed P/2+m_Q)$,
\begin{equation}
  \begin{aligned}
      \Pi_1\big|_{q=0}&=-\frac{1}{2\sqrt2\,m_Q}\Big(\frac{\slashed P}{2}-m_Q\Big)
     \slashed\varepsilon^{*}\Big(\frac{\slashed P}{2}+m_Q\Big),\\
  \Pi_0\big|_{q=0}&=-\frac{1}{2\sqrt2\,m_Q}\Big(\frac{\slashed P}{2}-m_Q\Big)
     \gamma_5\Big(\frac{\slashed P}{2}+m_Q\Big).
  \label{eq:B_Pi_q0}
  \end{aligned}
\end{equation}
Since $(\slashed P/2-m_Q)\Gamma=-\Gamma(\slashed P/2+m_Q)$ for $\Gamma=\slashed\varepsilon^*,\gamma_5$ and $(\slashed P/2+m_Q)^2=m_Q(\slashed P+2m_Q)$, these reduce to
\begin{equation}
  \Pi_1\big|_{q=0}=\frac{1}{2\sqrt2}\,\slashed\varepsilon^{*}(\slashed P+M),\qquad
  \Pi_0\big|_{q=0}=\frac{1}{2\sqrt2}\,\gamma_5(\slashed P+M).
  \label{eq:B_Pi_q0_short}
\end{equation}
To first order in $q$, the factors $E$ and $E+m_Q$ are even in $q$, and the middle factor obeys $(\slashed P+2m_Q)\slashed q=-2\slashed q(\slashed P/2-m_Q)$ because $P\cdot q=0$. 
Using also $P\cdot\varepsilon=0$ and $(\slashed P/2-m_Q)^2=-2m_Q(\slashed P/2-m_Q)$, one finds that the projector without the factor $(\slashed P+2E)$ is exact up to $\mathcal O(q^2)$:
\begin{equation}
  \Pi_1=-\frac{1}{2\sqrt2\,m_Q}\,\hat\Pi_1+\mathcal O(q^2),\qquad
  \hat\Pi_1\equiv\Big(\frac{\slashed P}{2}-\slashed q-m_Q\Big)
  \slashed\varepsilon^{*}\Big(\frac{\slashed P}{2}+\slashed q+m_Q\Big).
  \label{eq:B_Pi1_lin}
\end{equation}
The projector $\hat\Pi_1$ is the one that is differentiated with respect to $q$ in the P-wave calculation,
\begin{equation}
  \frac{\partial\hat\Pi_1^\alpha}{\partial q_\rho}\Big|_{q=0}=
  -\gamma^\rho\gamma^\alpha\Big(\frac{\slashed P}{2}+m_Q\Big)
  +\Big(\frac{\slashed P}{2}-m_Q\Big)\gamma^\alpha\gamma^\rho ,
  \label{eq:B_dPi1}
\end{equation}
where $\alpha$ is the spin index and $\rho$ the orbital index. 
In the ${}^3P_J^{[8]}$ rule of Eq.~\eqref{eq:spin_rule_3PJ}, the derivative acts on the full projected trace, including the quark propagator, whose momentum $l=-p_2-k=-\tfrac{P}{2}+q-k$ also depends on $q$:
\begin{equation}
  V^{\alpha\rho\mu\beta}=\frac{\partial}{\partial q_\beta}\left[\frac{\mathrm{Tr}\big[\hat\Pi_1^\mu\,\gamma^\alpha(\slashed l+m_Q)\gamma^\rho\big]}{l^2-m_Q^2}\right]_{q=0},
  \label{eq:B_V_def}
\end{equation}
where $\hat\Pi_1^\mu$ is $\hat\Pi_1$ with $\slashed\varepsilon^*$ replaced by $\gamma^\mu$. The derivative receives three contributions: from $\hat\Pi_1^\mu$, Eq.~\eqref{eq:B_dPi1}; from $\partial\slashed l/\partial q_\beta=\gamma^\beta$; and from the denominator, $l^2-m_Q^2=P\cdot k+k^2-q\cdot(P+2k)+\mathcal O(q^2)$. 

\bibliographystyle{utphys}
\bibliography{references}

\end{document}